\documentclass[12pt]{article}

\usepackage{graphicx}
\usepackage{latexsym}
\usepackage{amsfonts}
\usepackage{amssymb}
\usepackage{epsfig}
\usepackage{psfig}
\usepackage{psfrag}
\usepackage{dsfont}

\usepackage[sc,small]{caption2}
\setcaptionwidth{14cm}

\makeatletter
\@addtoreset{equation}{section}
\makeatother

\newcommand{\mbb}{\mathbb}
\newcommand{\beqn}{\begin{eqnarray}}
\newcommand{\eeqn}{\end{eqnarray}}
\newcommand{\be}{\begin{equation}}
\newcommand{\ee}{\end{equation}}
\newcommand{\non}{\nonumber \\}

\newcommand{\te}{{\tilde E}}

\newcommand{\ca}{{\cal A}}
\newcommand{\ce}{{\cal E}}
\newcommand{\ck}{{\cal K}}
\newcommand{\cn}{{\cal N}}
\newcommand{\cm}{{\cal M}}
\newcommand{\cp}{{\cal P}}
\newcommand{\cl}{{\cal L}}
\newcommand{\cf}{{\cal F}}

\newcommand{\cv}{{\cal V}}

\newcommand{\co}{{\cal O}}
\newcommand{\cs}{{\cal S}}
\newcommand{\cq}{{\cal Q}}
\newcommand{\ct}{{\cal T}}

\newcommand{\cz}{{\cal Z}}
\newcommand{\tr}{{\rm tr}}

\newcommand{\zba}[2]{[\!\!\begin{array}{c}{\scriptstyle#1}%
                        \\[-1.6mm]{\scriptstyle #2}\end{array}\!\!]}
\newcommand{\tht}{\vartheta}

\newcommand{\thbw}[2]{\vartheta[{ #1 \atop #2} ]}
\newcommand{\thba}[2]{\vartheta[\!\!\begin{array}{c}{\phantom{}\vspace{-.5mm}\scriptstyle#1}%
                        \\[-1.6mm]{\scriptstyle #2}\end{array}\!\!]}

\newcommand{\ba}[2]{[\!\!\begin{array}{c}{\scriptstyle#1}
                        \\[-1.6mm]{\scriptstyle #2}\end{array}\!\!]}

\newskip\humongous \humongous=0pt plus 1000pt minus 1000pt

\newif\ifdtup

\begin{document}

\title{}
\author{}
\date{}
\thispagestyle{empty}

\begin{flushright}
\vspace{-3cm}
{\small MIT-CTP-3671 \\
        NSF-KITP-2005-55 \\
        hep-th/0508043}
\end{flushright}
\vspace{1cm}

\begin{center}
{\bf\LARGE
String Loop Corrections to \\[.3cm]
K\"ahler Potentials in Orientifolds}

\vspace{1.5cm}

{\bf Marcus Berg}$^{\dag}$
{\bf,\hspace{.2cm} Michael Haack}$^{\dag}$
{\bf\hspace{.1cm} and\hspace{.2cm} Boris K\"ors}$^{*}$
\vspace{1cm}

{\it
$^{\dag}$Kavli Institute for Theoretical Physics, University of California \\
Santa Barbara, California 93106-4030, USA\\

$^*$Center for Theoretical Physics, Laboratory for Nuclear Science \\
and Department of Physics, Massachusetts Institute of Technology \\
Cambridge, Massachusetts 02139, USA \\

$^*$II. Institut f\"ur Theoretische Physik der Universit\"at Hamburg \\
Luruper Chaussee 149, D-22761 Hamburg, Germany\\

$^*$Zentrum f\"ur Mathematische Physik, Universit\"at Hamburg \\
Bundesstrasse 55, D-20146 Hamburg, Germany
}

\vspace{1cm}

{\bf Abstract}
\end{center}
\vspace{-.5cm}

We determine one-loop string corrections to K\"ahler
potentials in type IIB orientifold compactifications with either
$\cn=1$ or $\cn=2$ supersymmetry, including D-brane
moduli, by evaluating string scattering
amplitudes.

\clearpage

\tableofcontents

\vspace{3cm}

\section{Introduction}

Orientifolds represent an ideal laboratory to determine explicit loop
corrections to low energy effective actions in string theory, due to
their high degree of calculability.\footnote{See
\cite{Angelantonj:2002ct} and references therein for an introduction.}
In this paper we determine string one-loop corrections to K\"ahler
potentials in three different type IIB orientifold models, one with
$\cn = 2$ supersymmetry and two with $\cn =1$. The results could find
applications in various  contexts from cosmology to particle
phenomenology and we will exploit some of them in a companion paper
\cite{gg2}.

Our main motivation for calculating string corrections to K\"ahler
potentials is that they contribute to the scalar potential of the
low energy effective action. This can have effects on the vacuum
structure of the theory or on the dynamics of the scalar fields.
The price one pays for the advantage of concrete calculability in
orientifold models is the restriction to work at a special point
in moduli space, the orbifold point. Nevertheless, our results
give some important insights into the qualitative features arising
from string corrections to K\"ahler potentials in general.

For instance, the loop corrections introduce new dependence of the
K\"ahler potential on both the closed and open string moduli. Thus
they can have immediate bearing on the issue of moduli
stabilization. One interesting question in this context is whether
perturbative corrections to the K\"ahler potential could lead to
stabilization of the volume modulus without invoking non-perturbative
effects such as gaugino condensation \cite{Kachru:2003aw}. This
possibility was recently also mentioned by
\cite{vonGersdorff:2005bf}. To address this question one also 
has to take
other sources for corrections to the K\"ahler potential than
string loops  into account, for instance $\alpha'$ corrections at
string tree level. Some of these have been determined in
\cite{Becker:2002nn}. We will see that the one-loop corrections to
the K\"ahler potential in models with D3- and D7-branes are
suppressed for large values of the volume compared to the leading
tree level terms, but less suppressed than the
$\alpha'$ corrections of \cite{Becker:2002nn}. Unless the string
coupling is very small, they would thus generally be more
important in the large volume limit than the tree level
$\alpha'$ corrections. It is important to note that this is not in
contradiction with the fact that the leading $\alpha'$ corrections
in type IIB string theory arise at order $\co (\alpha'^3)$. The
corrections that we calculate only originate from world-sheets
that are not present with only oriented closed strings (they come
from D-branes and O-planes, i.e.\  annulus, M\"obius and Klein
bottle diagrams). We will come back
to this question in our companion paper \cite{gg2}.

A second interesting application of our results lies in the field of
brane inflation, initiated in
\cite{Dvali:1998pa,Dvali:2001fw,Burgess:2001fx,Kachru:2003sx}
where an open string scalar plays the role of the inflaton field.
Given the fact that the corrections depend on open string scalars,
they open up new possibilities to find regions in moduli space
where the mass of the inflaton takes a value that allows for slow
roll inflation. The corrections may or may not alleviate the
amount of fine tuning required to achieve that. 

In practice we determine the corrections to the K\"ahler metrics of
the scalar sigma-model by calculating 2-point functions of the
relevant scalars (which include the open string scalars),
following a similar calculation of  Antoniadis,
Bachas, Fabre, Partouche and Taylor, who considered a 2-point function
of gravitons in \cite{ABFPT}.\footnote{One
way to do this is to relax momentum conservation
\cite{Atick:1987gy,Minahan:1987ha,Poppitz:1998dj,Bain:2000fb,Antoniadis:2002cs}. We confirm
the validity of this prescription in our case in appendix
\ref{4ptcheck} by considering a 4-point function. In the case of 
\cite{ABFPT}, the result was confirmed by a 3-point function 
of gravitons in \cite{Antoniadis:2002tr}.}  
As the metrics on the
moduli spaces we consider are K\"ahler, it is convenient to use
vertex operators directly for the K\"ahler variables,
i.e.\ {\it K\"ahler structure adapted} vertex operators, cf.\
section \ref{vertexsec}. The K\"ahler variables for
the orientifold model with $\cn = 2$ supersymmetry were found in
\cite{ABFPT}. A crucial feature is that the definition of the
K\"ahler variables for the K\"ahler moduli\footnote{Notice the
twofold use of the term ``K\"ahler" here, because both the
compactification manifold and the moduli space are K\"ahler.}
involve a mixing between closed and open string scalars. A similar
kind of mixing between geometric and non-geometric scalar fields
in the definition of the appropriate K\"ahler variables is
familiar from other circumstances, like from compactifications of
the heterotic string (see for instance
\cite{Witten:1985xb}) or from $\cn=2$ compactifications of type II
theories \cite{Antoniadis:2003sw}.

The moduli dependence of our result for the $\cn=2$ orientifold is
in agreement with the one-loop K\"ahler metric given in
\cite{ABFPT} for the case of vanishing open string scalars, there
inferred from the one-loop correction to the Einstein-Hilbert
term.\footnote{The case with D9-brane Wilson line moduli was also
considered in \cite{Antoniadis:1997gu}. Their integral
representation for the K\"ahler metric (given in (2.4) of that
paper) is less straightforward to use in applications,
but it is consistent with the K\"ahler potential we find.} As
another check, we explicitly derive the one-loop correction to the
prepotential from the corrected K\"ahler potential. This confirms
that our result is consistent with supersymmetry.

Let us next summarize the content and organization of this paper.
We consider the same three orientifold models as in
\cite{Berg:2004ek}, i.e.\ the $\mbb T^4/\mathbb{Z}_2\times \mbb
T^2$ model with $\cn=2$ supersymmetry (in chapter \ref{z2}), the
$\mbb T^6/(\mathbb{Z}_2\times\mathbb{Z}_2)$ model (in chapter
\ref{z2z2}) and the $\mbb T^6/\mathbb{Z}_6'$ model (in chapter
\ref{z6}), both with $\cn=1$ supersymmetry. In section
\ref{treesection} we verify for the $\mbb T^4/\mathbb{Z}_2\times
\mbb T^2$ model that one can straightforwardly reproduce the tree
level sigma-model metrics by calculating 2-point functions on the
sphere and disk using the K\"ahler structure adapted vertex
operators (doing so, we built on previous work on disk amplitudes,
in particular
\cite{Garousi:1996ad,Hashimoto:1996bf,Garousi:1998fg,Garousi:2000ea,Lust:2004cx,Lust:2004fi}).
We then continue by determining a particular 2-point function at
one-loop that allows us to read off the one-loop K\"ahler potential.
We perform several checks on the result. In section
\ref{1loopkpot} we verify that the result is consistent with $\cn
= 2$ supersymmetry by determining the corresponding prepotential.
A second check is performed in appendix
\ref{other}, where we calculate five other 2-point functions for the
(vector multiplet) scalars of this model and show that the result is
consistent with the proposed K\"ahler potential. Given the fact that
the prepotential in $\cn =2$ theories only gets one-loop and
non-perturbative corrections, we conclude that our result holds
to all orders of perturbation theory in the $\mbb
T^4/\mathbb{Z}_2\times \mbb T^2$ case. The main result of chapter
\ref{z2} is given in formula (\ref{tollesK}).

We then continue in chapters \ref{z2z2} and \ref{z6} to generalize
this result to the $\cn=1$ cases of the $\mbb
T^6/(\mathbb{Z}_2\times\mathbb{Z}_2)$ and $\mbb T^6/\mathbb{Z}_6'$
models. The main results in these chapters can be found in
formulas (\ref{tollesK2}) and (\ref{K1z6}), respectively.

In section \ref{conclusions} we draw some conclusions and, in
particular, translate our results to the T-dual picture with D3-
and D7-branes.

Finally, we relegated some of the technical details to a series of
appendices.


\section{The $\cn=2$ orientifold $\mbb T^4/\mathbb{Z}_2\times \mbb T^2$}
\label{z2}

We first study the type IIB orientifold compactification with
$\cn=2$ supersymmetry on $\mathbb T^2\times$K3 at the orbifold
point $\mbb T^2\times \mbb T^4/\mbb Z_2$ described in
\cite{Bianchi:1990yu,Gimon:1996rq,Berkooz:1996iz}.
It is defined by world-sheet parity and the orbifold generator
$\Theta$ of $\mbb Z_2=\{ 1,\Theta\}$ which acts on the coordinates
of the $\mbb T^4$ by reflection. More precisely, it acts on the
complex coordinates along $\mbb T^6 = \mbb T^2_1\times\mbb
T^2_2\times\mbb T^2_3$ by multiplication with $\exp(2\pi i \vec
v)$, where
\beqn
\vec v = \Big(0, \frac12, -\frac12)\ .
\eeqn
The world-sheet parity operation $\Omega$ interchanges 
left- and right-moving fields on the closed
string world-sheet, see \cite{Angelantonj:2002ct} for a review.
The model contains orientifold O9- and O5-planes, the former
space-time filling, the latter localized at the fixed points of
the orbifold generator. In the same way there are D9-branes and
D5-branes wrapped on the $\mbb T^2$ and point-like on the
K3. We should stress that it is completely straightforward to translate the D9/D5
model into a model with D3- and D7-branes instead, by performing
six T-dualities along all the internal directions; we will come back to this in the
conclusions, sec.\ \ref{conclusions}.

\subsection{The classical Lagrangian\label{sugra}}
\label{classicalz2}

The relevant aspects of the effective action that describes the
low energy dynamics of the untwisted modes of this model have been
discussed in \cite{ABFPT}. The moduli of the K3 including all the blow-up
modes of the orbifold singularities fall into $\cn=2$
hypermultiplets and will not be important to us in the following.
The scalars that arise from reducing the ten-dimensional fields
along the $\mbb T^2$ reside in vector multiplets, and their moduli space
will be the subject of this section. Some of them arise from the closed string sector,
but in addition there are the
vector multiplets from the open string sectors of D9- and
D5-branes. For our present purposes we will only consider the
D9-brane scalars and set D5-brane scalars to zero, keeping the D5-brane
gauge fields.

To be more specific, we focus on the complex scalars $\{S,S',U,A_i\}$ that are defined as
\beqn\label{scalars}
&& S = \frac{1}{\sqrt{8\pi^2}} ( C + i e^{-\Phi} \sqrt{G} \cv_{\rm K3})\ ,  \non
&& U = \frac{1}{G_{44}} ( G_{45} + i \sqrt{G} )\ ,\quad
A^i = U a^i_4 - a^i_5
\eeqn
and
\beqn \label{sstrich}
S' &=& \frac{1}{\sqrt{8\pi^2}} ( C_{45} + i e^{-\Phi} \sqrt{G} ) +
\frac{1}{8\pi} \sum_i N_i (U (a_4^i)^2 - a_4^ia_5^i)
\non
&=& S_0' + \frac{1}{8\pi} \sum_i N_i A_i \frac{A_i-\bar A_i}{U-\bar U} \ .
\eeqn
We introduced the rescaled K3 volume in the string frame $\cv_{\rm
K3} = (4 \pi^2 \alpha')^{-2}{\rm vol (K3)}$, and used
the string frame metric on $\mathbb T^2$,
\beqn \label{T2metric}
G_{mn}  = \frac{\sqrt{G}}{U_2} \left(
\begin{array}{cc}
1 & U_1 \\
U_1 & |U|^2
\end{array}
\right) \  .
\eeqn
Moreover, $\Phi$ is the ten-dimensional dilaton, $C_{45}$ the
internal component of the RR 2-form $C_2$ along the torus and
$\partial_\mu  C= \frac12 \epsilon_{\mu\nu\rho\sigma}\partial^\nu
C^{\rho\sigma}$. Finally, we denoted the internal components of
the ten-dimensional abelian vectors of the D9-brane stack labelled
by $i$ as $a_m^i$ (which are defined to be dimensionless, i.e.
$\cf^i_{\mu m} = \partial_\mu a^i_m/\sqrt{\alpha'}$), and $N_i$
denotes the multiplicity of branes in the stack, i.e. the rank of
the respective $U(N_i)$ factor of the gauge group.\footnote{For
the open string scalars, we use the convention that $i,j$
enumerate the different stacks of D9-branes, whereas $m,n$ stand
for the internal components along the torus, i.e.\ $m,n \in
\{4,5\}$. Furthermore, we sometimes write the index $i,j$ upstairs
and sometimes downstairs, whichever is more convenient. We also do
not follow the Einstein summation rule for $i,j$, always writing
the summation explicitly.} Note that the definition
(\ref{sstrich}) of $S'$ contains a correction involving the open
string scalars, arising at disk level. We will review in a moment
why this is a good K\"ahler variable in the presence of open
string scalars.

The leading order interactions between the vector multiplets
(coupled to gravity) can be derived from the dimensional reduction
of ten-dimensional type I supergravity \cite{ABFPT}
\beqn \label{dimred}
\cs_{\rm SG} &=& \frac{1}{2\kappa^2_{10}} \int d^{10}x\sqrt{-g_{10}} \Big[ e^{-2\Phi}
\Big( R_{10} + 4\partial^I \Phi\partial_I \Phi \Big) - \frac12 |dC_2 - \frac{\kappa_{10}^2}{g_{10}^2} \omega_3|^2 \Big]
\\
&=&
\int d^{4}x\sqrt{-g} \frac{1}{\kappa_4^2} \Big[ \frac{1}{2} R
+ \frac{\partial_\mu S\partial^\mu \bar S}{(S-\bar S)^2}
+ \frac{\partial_\mu U\partial^\mu \bar U}{(U-\bar U)^2}
\non
&&
\hspace{5cm}
+ \frac{|\partial_\mu S_0'+\frac{1}{8 \pi} \sum_i N_i ( a^i_4\partial_\mu a^i_5-a^i_5\partial_\mu a^i_4)|^2
}{(S_0'-\bar S_0')^2} \Big] + \ \cdots \ ,
\nonumber
\eeqn
together with the Born-Infeld action that produces the kinetic terms
for the D9-brane gauge fields and scalars,
\beqn \label{dbi9}
\cs_{\rm BI} &=& \mu_9 \int d^{10}x\, e^{-\Phi} \Big[ -{\rm det}(
g_{IJ} + 2\pi\alpha' (\cf_{\rm D9})_{IJ}) \Big]^{1/2}
\\
&=&
\int d^4x \sqrt{-g} \Big[ \frac{1}{4\pi\kappa_4^2}
\frac{\sum_i N_i |U\partial_\mu a^i_4 - \partial_\mu a^i_5|^2}{(U-\bar U)(S_0'-\bar S_0')}
- \frac14 {\rm Im}(S)\, {\rm tr}\, \cf_{\rm D9}^2  \Big] + \ \cdots \ .
\nonumber
\eeqn
The ellipsis stands for higher derivative terms, $g_{IJ}$
denotes the ten-dimensional string frame metric and
$g_{\mu\nu}$ the four-dimensional Einstein frame metric.
All the non-abelian scalars are set to zero.
The kinetic terms for D5-brane gauge fields are
\beqn \label{dbi5}
\cs_{\rm BI} &=& \mu_5 \int_{\mbb R^{3,1}\times \mbb T^2}
 d^{6}x\, e^{-\Phi} \Big[ -{\rm det}( g_{IJ} + 2\pi\alpha' (\cf_{\rm D5})_{IJ}) \Big]^{1/2}
\\
&=&
\int d^4x \sqrt{-g} \Big[ - \frac14 {\rm Im}(S_0')\, {\rm tr}\, \cf_{\rm D5}^2 \Big] + \ \cdots \ .
\nonumber
\eeqn
The volume of the background torus is taken to be $4\pi^2 \alpha'
\sqrt G$ and the constants are
\beqn
\kappa_{10}^{2} = (4\pi^2\alpha')^3 \kappa_4^{2} = \frac{1}{4\pi}
(4\pi^2\alpha')^4 \ , \quad
\mu_p = \frac{2\pi}{\sqrt{2}} ( 4\pi^2\alpha')^{-(p+1)/2}\ , \quad
\frac{\kappa_{10}^2}{g_{10}^2}=\frac{\alpha'}{2\sqrt{2}}\ .
\eeqn
The gauge kinetic terms are completed by the Chern-Simons
action into the classical (i.e.\ disk level) holomorphic couplings
\beqn \label{gkin}
f^{(0)}_{\rm D9} = -i S\ , \quad
f^{(0)}_{\rm D5} = -i S_0'\ .
\eeqn
Let us make some comments on the numerical factors appearing in
(\ref{dimred}) and (\ref{dbi9}). We have made all scalars
dimensionless, hence the prefactors $\kappa_4^{-2} =
(\pi\alpha')^{-1}$ in (\ref{dimred}) and the first term of
(\ref{dbi9}). Moreover, the $S$ and $S'_0$ in (\ref{scalars}) and
(\ref{sstrich}) are defined such that the tree-level gauge kinetic
functions are given by (\ref{gkin}) without any numerical
factors. This leaves one with the unconventional prefactor
$(4\pi)^{-1}$ in (\ref{dbi9}) which matches precisely with the
relative factor in the reduction of the Chern-Simons corrected
kinetic term for the RR 2-form, i.e.\
\be \label{cscorr}
\frac{1}{\sqrt{8\pi^2}} \Big[ \partial_\mu C_{45} - \frac{\kappa_{10}^2}{g_{10}^2} (\omega_3)_{\mu
45} \Big] =
\frac{1}{\sqrt{8\pi^2}} \partial_\mu C_{45}
 + \frac{1}{2}\frac{1}{4\pi} \sum_i N_i \big( a^i_4 \partial_\mu a^i_5 - a^i_5
\partial_\mu a^i_4 \big) \ .
\ee
Putting the pieces together, the classical Lagrangian that follows
from the dimensional reduction of the leading order
ten-dimensional supergravity action reads
\beqn \label{dimred2}
\kappa_4^2 \cl_{4d} &=&
\frac{1}{2} R
+ \frac{\partial_\mu S\partial^\mu \bar S}{(S-\bar S)^2}
+ \frac{\partial_\mu U\partial^\mu \bar U}{(U-\bar U)^2}
+ \frac{|\partial_\mu S_0'+\frac{1}{8 \pi} \sum_i N_i ( a^i_4\partial_\mu  a^i_5-
                             a^i_5\partial_\mu  a^i_4)|^2}{(S_0'-\bar S_0')^2}
\non
&&
\hspace{0cm}
+ \frac{\sum_i N_i |U\partial_\mu  a^i_4 - \partial_\mu  a^i_5|^2}{4 \pi (U-\bar U)(S_0'-\bar S_0')}
- \frac14 \kappa_4^2 {\rm Im}(S)\, {\rm tr}\, \cf_{\rm D9}^2 - \frac14 \kappa_4^2 {\rm Im}(S_0')\, {\rm tr}\,
\cf_{\rm D5}^2 \ .
\eeqn
It is important to observe that in the kinetic term of $S_0'$,
there are terms of different dependence on the ten-dimensional
dilaton, and with different numbers of traces over gauge group
indices (i.e.\  factors of $N_i$). More
precisely, the cross term 
\[
\sum_i N_i ( a^i_4\partial_\mu a^i_5)\partial^\mu ({\rm Re}(S_0'))+\,
\cdots 
\]
has one trace and the dilaton dependence expected from open
string tree level (disk diagrams), while the term with only open
string scalars of the type 
\[
\sum_{i,j} N_i N_j (a^i_4\partial^\mu
a^i_5)(a^j_4\partial_\mu a^j_5)+\cdots
\]
has two traces and the
dilaton dependence of an open string one-loop diagram. Due to
these corrections to the kinetic term of $S_0'$ one has to use the
modified field $S'$ defined in (\ref{sstrich}) in order to make
the K\"ahler property of the scalar sigma-model metric explicit.
Thus the scalars $\{S, S', U, A_i\}$ are good K\"ahler
variables to use at the classical level and their (classical)
K\"ahler potential is
\beqn \label{kaeN2}
K^{(0)} &=&
 -\ln (S-\bar S) - \ln\big[ (S_0'-\bar S_0')(U-\bar U)\big]\non
&=&
 -\ln (S-\bar S) - \ln\Big[ (S'-\bar S')(U-\bar U) - \frac{1}{8\pi}
\sum_i N_i (A_i - \bar A_i)^2 \Big]\ .
\eeqn
As required by ${\cal N}=2$ supersymmetry,
this can be expressed by a prepotential, i.e.\ by
\beqn \label{precl}
\cf^{(0)}(S,S',U,A_i) = S \Big[ S'U - \frac{1}{8\pi} \sum_i N_i A_i^2 \Big] \ ,
\eeqn
via the standard formula for
the K\"ahler potential in special K\"ahler geometry
\beqn \label{n=2kaehler}
K = -\ln \Big[ 2 \cf - 2\bar\cf - \sum_\alpha ( \phi^\alpha - \bar\phi^{\bar \alpha} )
( \cf_\alpha + \bar \cf_{\bar \alpha} ) \Big] \ ,
\eeqn
$\phi^\alpha$ running over all scalars $\{ S,S',U,A_i \}$, and $\cf_\alpha =
\partial_{\phi^\alpha} \cf$. Since some terms in (\ref{dimred2}) go as $(e^{\Phi})^0$ in the string frame, which is characteristic of 
open string one-loop corrections, the moniker
``classical'' K\"ahler potential clearly 
has to be taken with a grain of
salt in the Type I context 
(this was also emphasized in \cite{Berg:2004ek}). We will see, however, in section \ref{treesection} that all
the kinetic terms resulting from (\ref{kaeN2}) can indeed be
derived by calculating just tree diagrams (i.e.\ sphere and disk)
if one uses appropriately defined K\"ahler structure adapted
vertex operators, cf.\ sec.\ (\ref{vertexsec}). We therefore
continue to call (\ref{kaeN2}) the tree-level (or classical)
K\"ahler potential.

Finally, note that in $\cn=2$ supersymmetry the gauge couplings
are also fixed by the prepotential and related to the K\"ahler
metric of the matter fields in the same vector multiplet. The open
string moduli are in general not the same as matter
fields\footnote{We adopt the convention of
\cite{Kaplunovsky:1994fg} who define matter fields as those that
are charged under the gauge group and, thus, whose vacuum
expectation values would break part of the gauge symmetry.} but at
points of enhanced gauge symmetry such as $A_i=0$ they should be
treated on the same footing. They are then expected to satisfy a
relation \cite{deWit:1995zg}
\beqn\label{km-gc}
2\pi i\, K_{A_i\bar A_i} \Big|_{A_i=0} = e^K {\rm Re}(f_{{\rm D9}_i}) \Big|_{A_i=0}\ ,
\eeqn
where the constant of proportionality has been fixed in a way that
$K_{A_i\bar A_i}^{(0)}$ and $f^{(0)}_{{\rm D9}}$
respect the condition up to terms of order $\co(e^{2\Phi})$.


\subsection{Classical K\"ahler metric}

The tree level K\"ahler potential in $\mbb T^4/\mathbb{Z}_2\times
\mbb T^2$ was already given in (\ref{kaeN2}). In the following we
will check and confirm it by performing explicit calculations of
string scattering amplitudes on sphere and disk world-sheet
topologies. This is a preliminary step before computing the
one-loop corrections that will prove useful to develop some new
techniques, in particular regarding choices of complex coordinates in the vertex
operators for the relevant moduli fields. In our calculation we
make use of results previously obtained in
\cite{Garousi:1996ad,Hashimoto:1996bf,Garousi:1998fg,Garousi:2000ea,Lust:2004cx,Lust:2004fi}.

To make contact with the K\"ahler potential (\ref{kaeN2}), we
explicitly write out the components of the K\"ahler metric that
result from it, immediately recasting them into the variables
$\{S'_0, U, A_i\}$, which makes the dilaton dependence of the
various terms obvious.\footnote{We do not consider $S$, because it
decouples at tree level, cf.\ (\ref{kaeN2}).} One finds
\beqn \label{km}
K^{(0)}_{S'\bar S'} &=& -\underbrace{\frac{1}{(S_0'-\bar S_0')^2}}_{\co(e^{2\Phi})} \ , \non
K^{(0)}_{U\bar U} &=& -\underbrace{\frac{1}{(U-\bar U)^2}}_{\co(1)}
 - \frac{1}{4 \pi} \underbrace{\frac{\sum_i N_i (A_i-\bar A_i)^2}{(U-\bar U)^3(S_0'-\bar S_0')}}_{\co(e^\Phi)}
 - \frac{1}{64 \pi^2} \underbrace{\frac{(\sum_i N_i (A_i-\bar A_i)^2)^2}{(U-\bar U)^4(S_0'-\bar S_0')^2}}_{\co(e^{2\Phi})} \ , \non
K^{(0)}_{A_i\bar A_j} &=& - \frac{1}{4 \pi} \underbrace{\frac{N_i\delta_{ij}}{(U-\bar U)(S_0'-\bar S_0')}}_{\co(e^{\Phi})}
 - \frac{1}{16 \pi^2} \underbrace{\frac{N_i (A_i-\bar A_i) N_j (A_j-\bar A_j)}{(U-\bar U)^2(S_0'-\bar S_0')^2}}_{\co(e^{2\Phi})} \ , \non
K^{(0)}_{U\bar S'} &=& - \frac{1}{8\pi}
 \underbrace{\frac{\sum_i N_i (A_i-\bar A_i)^2}{(U-\bar U)^2(S_0'-\bar S_0')^2}}_{\co(e^{2\Phi})} \ , \non
K^{(0)}_{U\bar A_i} &=& \frac{1}{4 \pi} \underbrace{\frac{N_i (A_i-\bar A_i)}{(U-\bar U)^2(S_0'-\bar S_0')}}_{\co(e^\Phi)}
 + \frac{1}{32 \pi^2} \underbrace{\frac{N_i (A_i -\bar A_i) \sum_j N_j (A_j-\bar A_j)^2}{(U-\bar U)^3(S_0'-\bar S_0')^2}}_{\co(e^{2\Phi})} \ , \non
K^{(0)}_{A_i \bar S'} &=& \frac{1}{4 \pi} \underbrace{\frac{N_i (A_i-\bar A_i)}{(U-\bar U)(S_0'-\bar S_0')^2}}_{\co(e^{2\Phi})}\ . \label{ktree}
\eeqn
We have indicated to which order in the dilaton expansion the
respective terms contribute. To read off the order in a
perturbative expansion of the effective Lagrangian, one
has to take into account that for each derivative with respect to
$\bar S'$ one also has a term $\partial_\mu \bar S'$ in the
kinetic term that compensates one power of $e^{-\Phi}$. Another
indicator for the order in perturbation theory of a given term is
the number of traces (factors of $N_i$) that appear. As mentioned
earlier, there are not only terms of sphere and disk order but
also a number of terms that appear to originate from annulus
diagrams, i.e.\ at order $e^{2\Phi}$ with two traces. 
Amidst this barrage of confusion, there is hope:
we will show that the full expression can be
obtained from a purely tree level
(sphere $+$
disk diagram) computation by using 
{\it K\"ahler structure adapted}
vertex operators. They produce the additional factors of the string
coupling and involve traces over gauge group indices.

In order to do so we will calculate a number of scattering
amplitudes 
and show that they are consistent with the K\"ahler
metric in (\ref{km}). Concretely, we choose $K^{(0)}_{A_i\bar
A_j}$ as an example and confirm the presence of both terms in this
metric component. The other cases could be treated analogously.
Since all tree-level 2-point functions vanish, we will have to
calculate 3-point functions. The first step, though, is to
determine the vertex operators.


\subsection{Vertex operators for moduli fields}
\label{vertexsec}

The correct vertex operators to represent the K\"ahler variables
$S', U, A$ can be determined by expressing the world-sheet action
in terms of these fields and taking its variations. We write the
relevant part of the bosonic world-sheet action
\beqn \label{wsaction}
\cs_{\rm ws} ~=~ \frac{1}{2\pi\alpha'} \int_\Sigma d^2z\ G_{mn} \partial X^m \bar\partial X^n
  - \frac{i}{\sqrt{\alpha'}} \sum_{B} \int_{(\partial\Sigma)_B} d
\theta\ q_B\, a^{s[B]}_m \dot X^m +\, \cdots\ ,
\eeqn
where the new label $B$ enumerates the components of the boundaries ($B$ takes
values in the empty set for the sphere and Klein bottle, $B \in \{ 1
\}$ for the disk and M\"obius strip, and $B \in \{ 1,2 \}$ for the
annulus),  $s[B]$ is the label for the stack of D-branes on which
the boundary component $B$ ends, and thus $\vec{a}^{\, s[B]}
=(a^{s[B]}_4,a^{s[B]}_5)$ denotes the (dimensionless) Wilson line
modulus on the $s[B]$th stack of branes on which the $B$th
boundary ends. Finally $q_B$ takes values $(q_1,q_2)=(1,-1)$ for
the annulus, distinguishing the two possible orientations (for the
M\"obius strip one would only have $q_1 = 1$). The metric
$G_{mn}$ on the two-torus was introduced in (\ref{T2metric}).

Let us explain our notation for the Wilson lines in more detail.
The conventions are such that there are 32 D9-branes labelled by
the Chan-Paton (CP) indices. This means that the Wilson lines of
all the D9-branes are collectively written as $32\times 32$
dimensional matrices $W$. All fields are then subject to
projections onto invariant states under the world-sheet parity
$\Omega$ and under $\Theta$,
which act on the $W$ via the gamma-matrices $\gamma_{\Omega}$ and
$\gamma_{\Theta}$ which we define later. They identify the upper
and lower $16\times 16$ blocks in $W$ up to sign, and project out
the off-diagonal blocks. In the end, the invariant Wilson lines
are described by a single unconstrained $16\times 16$ matrix which
is just right to collect all the degrees of freedom of the adjoint
representation of the maximal $U(16)$ gauge group on the
D9-branes. When some of the Wilson line scalars take non-trivial
vacuum expectation values, the D9-branes are separated into stacks
labelled by $i$ and the gauge group is broken up into
factors\footnote{Similarly, one can also break the D5-brane gauge
group $U(16)$ into factors $U(N_a)$ with $\sum_a N_a=16$ but we
are only dealing with the D9-brane Wilson line moduli explicitly.}
\beqn \label{gauge}
U(16) ~\longrightarrow~ \prod_i U(N_i) \ ,
\eeqn
with
\be
\sum_i N_i = 16\ .
\ee
We then break up all matrices into $(2N_i)\times (2N_i)$ blocks,
denoted $\gamma_{\Omega i}$, $\gamma_{\Theta i}$ and $W_i$ for
instance. We are only interested in the scalars that behave like
moduli in this situation, i.e. the abelian $U(1)_i$ factors in
$U(N_i)=SU(N_i)\times U(1)_i$. These are described by matrices
\beqn \label{Wmatrix}
W_i &=& {\rm diag}( {\bf 1}_{N_i} , -{\bf 1}_{N_i}) \oplus {\bf
0}_{32-2N_i} \ ,
\eeqn
and the matrix valued Wilson line vector along the two circles of
the $\mbb T^2$ is written as $\vec a^{\, i} W_i$, which has the
property that it commutes with all matrices $\gamma$. The
normalization is such that\footnote{A different normalization would lead to 
a rescaling of the scalars $A^i$.}
\beqn
{\rm tr}( W_i W_j ) = 2N_i \delta_{ij}\ .
\eeqn
Expressed in terms of the K\"ahler coordinates $A^i$ and $U$, the $\vec a^i$ read
\beqn
a_4^i = \frac{A^i-\bar A^i}{U-\bar U} \ , \quad
a_5^i = \frac{A^i \bar U-\bar A^i U}{U-\bar U} \ .
\eeqn
In order to proceed it is also useful to define the ordinary
volume modulus of the $\mbb T^2$ as $T = \sqrt{8\pi^2} ({\rm
Re}(S_0') + i e^\Phi {\rm Im}(S_0'))$ such that $T_2 = \sqrt G$.
For the moment, we keep $G_{mn}$ and $a_m^i$ as functions of
$T_2$, and introduce $S'$ later, by rewriting $T_2$ as a function
of $S',U,A^i$ and their conjugates. Then one has
\beqn
G_{mn} \partial X^m \bar\partial X^n &=&
 \partial\bar Z \bar\partial Z + \partial Z \bar\partial \bar Z ~\equiv~ f_1(T_2,U) \ , \non
\partial_{T_2}  f_1(T_2,U) &=& \frac{1}{T_2} f_1(T_2,U) \ , \non
\partial_U  f_1(T_2,U) &=& -\frac{2}{U-\bar U} \partial Z\bar\partial Z\ ,
\eeqn
and
\beqn
a_m^{s[B]} \dot X^m &=& \frac{\sqrt 2 }{(U-\bar U)^{1/2} (T-\bar
T)^{1/2}} ( A^{s[B]} \dot Z - \bar A^{s[B]} \dot{\bar Z} )
 ~\equiv~ f_2(U,A^{s[B]}) \ , \non
\partial_{A^i}  f_2(U,A^{s[B]}) &=& \frac{\sqrt 2\delta_{is[B]}}{(U-\bar U)^{1/2} (T-\bar T)^{1/2}} \dot Z \ , \non
\partial_U  f_2(U,A^{s[B]}) &=& - \frac{\sqrt 2(A^{s[B]}-\bar A^{s[B]})}{(U-\bar U)^{3/2} (T-\bar T)^{1/2}} \dot Z\ ,
\eeqn
where, as in \cite{Lust:2004cx}, we defined complexified
coordinates
\beqn \label{zpsi}
Z = \sqrt{\frac{T_2}{2 U_2}} (X^{4} + \bar
U X^{5}) \quad , \quad
\bar Z = \sqrt{\frac{T_2}{2 U_2}} (X^{4}
+ U X^{5})\ , \non
\Psi = \sqrt{\frac{T_2}{2 U_2}} (\psi^{4} + \bar U
\psi^{5}) \quad , \quad
\bar \Psi = \sqrt{\frac{T_2}{2 U_2}} (\psi^{4}
+ U \psi^{5})\ .
\eeqn
To introduce $S'$ we also have to express $e^\Phi$ in terms of
$T,U,A^i$. According to \cite{ABFPT}, there is another scalar
modulus given by the six-dimensional dilaton
\beqn
e^{-2\Phi_6} = e^{-2\Phi} \cv_{\rm K3}
\eeqn
which is part of a hypermultiplet. With this, we can express
\beqn
T_2 = \sqrt G= -i \pi \sqrt{2} \Big[ \frac{(S_0'-\bar S_0')(S-\bar
S)}{e^{-2\Phi_6}} \Big]^{1/2}\ .
\eeqn
Inserting
\be
S_0 ' - \bar S_0 '= S'-\bar S' - \frac{1}{8 \pi}  \frac{\sum_i N_i (A_i-\bar A_i)^2}{U-\bar U}
\ee
and using
\be
\Big[ \frac{S - \bar{S}}{(S_0'-\bar S_0') e^{-2\Phi_6}} \Big]^{1/2} = e^{\Phi}\ ,
\ee
one arrives at
\beqn
\partial_U T_2 &=& -i \frac{1}{8 \sqrt{2}}  \frac{\sum_i N_i (A_i-\bar A_i)^2}{(U-\bar U)^2}e^{\Phi}  \ , \non
\partial_{A_i} T_2 &=& i \frac{1}{4 \sqrt{2}} \frac{N_i(A_i-\bar A_i)}{U-\bar U}e^{\Phi}  \ , \non
\partial_{S_2'} T_2 &=& \pi \sqrt{2} e^{\Phi}   \ .
\eeqn
Now we can compute the vertex operators from differentiating the full world-sheet action, obtaining
\beqn
\frac{\delta \cs_{\rm ws}}{\delta S_2'} &=& \partial_{S_2'} T_2\,  \partial_{T_2} \cs_{\rm ws} \non
                          &=& \frac{1}{2\pi\alpha'} \int_\Sigma d^2z\
\frac{i}{S_0'-\bar S_0'}
 ( \partial \bar Z \bar \partial Z + \partial Z \bar \partial \bar Z )\ ,
\non
\frac{\delta \cs_{\rm ws}}{\delta U} &=& \partial_U T_2\, \partial_{T_2} \cs_{\rm ws} + \partial_U \cs_{\rm ws} \non
              &=& \frac{1}{2\pi\alpha'} \int_\Sigma d^2z\ \Big[ \frac{1}{16 \pi} \frac{\sum_i N_i (A_i-\bar A_i)^2}{(U-\bar U)^2( S_0'-\bar S_0')}
  ( \partial \bar Z \bar \partial Z + \partial Z \bar \partial \bar Z )
 - \frac{2}{U-\bar U} \partial Z\bar \partial Z \Big]
\non
&&
 + \frac{i}{\sqrt{\alpha'}} \sum_B \int_{(\partial\Sigma)_B}d \theta\  q_B \frac{\sqrt 2 (A_{s[B]}-\bar A_{s[B]})}{(U-\bar U)^{3/2} (T-\bar T)^{1/2}} \dot Z \ ,
\non
\frac{\delta \cs_{\rm ws}}{\delta A_j} &=& \partial_{A_j} T_2\, \partial_{T_2} \cs_{\rm ws} + \partial_{A_j} \cs_{\rm ws}
\non
                         &=& \frac{1}{2\pi\alpha'} \int_\Sigma d^2z\
 \Big[ - \frac{1}{8 \pi} \frac{N_j(A_j-\bar A_j)}{(U-\bar U)( S_0'-\bar S_0')}
  ( \partial \bar Z \bar \partial Z + \partial Z \bar \partial \bar Z ) \Big]
\non
&&
 - \frac{i}{\sqrt{\alpha'}} \sum_B \int_{(\partial\Sigma)_B} d \theta\ q_B  \frac{\sqrt 2 \delta_{j s[B]}}{(U-\bar U)^{1/2} (T-\bar T)^{1/2}} \dot Z \ .
\eeqn
Let us define a set of ``building block'' 
vertex operators, in the $0$ picture
\beqn \label{nakedvo}
V_{Z\bar Z}^{(0,0)} &=& -\frac{2}{\alpha'} \int_\Sigma d^2z\
\big[ i \partial \bar Z + \frac{1}{2} \alpha' (p \cdot \psi) \bar \Psi \big]
\big[ i \bar \partial Z + \frac{1}{2} \alpha' (p \cdot \tilde \psi) \tilde \Psi \big]
 e^{ipX}
\non
&& - \frac{2}{\alpha'} \int_\Sigma d^2z\
                                            \big[ i \partial Z + \frac{1}{2} \alpha' (p \cdot \psi) \Psi \big]
                                            \big[ i \bar \partial \bar Z + \frac{1}{2} \alpha' (p \cdot \tilde \psi) \bar {\tilde \Psi} \big] e^{ipX} \ ,
\non
V_{ZZ}^{(0,0)} &=& -\frac{2}{\alpha'} \int_\Sigma d^2z\
                                            \big[i \partial Z + \frac{1}{2} \alpha' (p \cdot \psi) \Psi \big]
                                             \big[i \bar \partial Z + \frac{1}{2} \alpha' (p \cdot \tilde \psi) \tilde \Psi \big]
                                             e^{ipX} \ ,
\non
V_{\bar Z \bar Z}^{(0,0)} &=& -\frac{2}{\alpha'} \int_\Sigma d^2z\
                                            \big[i \partial \bar Z + \frac{1}{2} \alpha' (p \cdot \psi) \bar \Psi \big]
                                             \big[i \bar \partial \bar Z + \frac{1}{2} \alpha' (p \cdot \tilde \psi) \bar{\tilde \Psi} \big]
                                             e^{ipX} \ ,
\non
V_Z^{(0)B} &=& \frac{1}{\sqrt{2 \alpha'}} \lambda_{s[B]} \int_{(\partial\Sigma)_B}d \theta\ \big[i \dot Z + 2\alpha' (p \cdot \psi) \Psi \big] e^{ipX} \ ,
\non
V_{\bar Z}^{(0)B} &=&  \frac{1}{\sqrt{2 \alpha'}} \lambda_{s[B]}^\dagger \int_{(\partial\Sigma)_B}d \theta\ \big[i \dot{\bar Z} + 2\alpha' (p \cdot \psi) \bar \Psi \big] e^{ipX}
\; ,
\eeqn
and in the $(-1)$ picture
\beqn \label{nakedvo2}
V_{Z\bar Z}^{(-1,-1)} &=&  \int_\Sigma d^2z\ e^{-\phi - \tilde \phi} \bar \Psi \tilde \Psi e^{ipX}
+ \int_\Sigma d^2z\  e^{-\phi - \tilde \phi} \Psi \bar {\tilde \Psi}
                                           e^{ipX} \ ,
\non
V_{ZZ}^{(-1,-1)} &=&  \int_\Sigma d^2z\ e^{-\phi - \tilde \phi} \Psi \tilde \Psi e^{ipX} \ ,
\non
V_{\bar Z \bar Z}^{(-1,-1)} &=&  \int_\Sigma d^2z\ e^{-\phi - \tilde \phi} \bar \Psi \bar{\tilde \Psi} e^{ipX} \ ,
\non
V_Z^{(-1)B} &=& \lambda_{s[B]} \int_{(\partial\Sigma)_B}d \theta\ e^{-\phi} \Psi e^{ipX} \ ,
\non
V_{\bar Z}^{(-1)B} &=& \lambda_{s[B]}^\dagger \int_{(\partial\Sigma)_B}d \theta\ e^{-\phi} \bar \Psi e^{ipX}\ .
\eeqn
We then define the {\it K\"ahler structure adapted}
vertex operators corresponding to the K\"ahler coordinates
\beqn \label{vops}
V_{S'_2}^{(l,m)} &=& g_c \alpha'^{-2} \frac{i}{S_0'-\bar S_0'} V_{Z\bar Z}^{(l,m)} \ ,
\non
V_{U}^{(l,m;n)} &=& g_c \alpha'^{-2} \Big( \frac{1}{16 \pi} \frac{\sum_i N_i (A_i-\bar A_i)^2}{(U-\bar U)^2( S_0'-\bar S_0')} V_{Z\bar Z}^{(l,m)}
 -  \frac{2}{U-\bar U} V_{ZZ}^{(l,m)} \non
&& \hspace{3cm}
+ \sum_B \frac{4 \pi (A_{s[B]}-\bar A_{s[B]})}{(U-\bar U)^{3/2} (T-\bar T)^{1/2}} V_Z^{(n)B} \Big)\ ,
\non
V_{\bar U}^{(l,m; n)} &=& g_c \alpha'^{-2} \Big(- \frac{1}{16 \pi} \frac{\sum_i N_i (A_i-\bar A_i)^2}{(U-\bar U)^2( S_0'-\bar S_0')} V_{Z\bar Z}^{(l,m)}
 +  \frac{2}{U-\bar U} V_{\bar Z \bar Z}^{(l,m)} \non
&& \hspace{3cm}
- \sum_B \frac{4 \pi (A_{s[B]}-\bar A_{s[B]})}{(U-\bar U)^{3/2} (T-\bar T)^{1/2}} V_{\bar Z}^{(n)B} \Big) \ ,
\non
V_{A_j}^{(l,m;n)} &=& g_o \alpha'^{-2} \Big(- \frac{1}{8 \pi^2} \frac{N_j(A_j-\bar A_j)}{(U-\bar U)( S_0'-\bar S_0')} V_{Z\bar Z}^{(l,m)}
\non
&& \hspace{3cm}
 - \sum_B \frac{4 \delta_{j s[B]}}{(U-\bar U)^{1/2} (T-\bar T)^{1/2}} V_Z^{(n)B} \Big) \ ,
\non
V_{\bar A_j}^{(l,m;n)} &=& g_o \alpha'^{-2} \Big( \frac{1}{8 \pi^2} \frac{N_j(A_j-\bar A_j)}{(U-\bar U)( S_0'-\bar S_0')} V_{Z\bar Z}^{(l,m)}
\non
&& \hspace{3cm}
 + \sum_B \frac{4 \delta_{j s[B]}}{(U-\bar U)^{1/2} (T-\bar T)^{1/2}} V_{\bar Z}^{(n)B} \Big)\ ,
\eeqn
where the explicit factors of $\alpha'$ are chosen such that the
vertex operators are dimensionless and  we allowed for different
superghost pictures for the closed and open string parts of the
vertex operators. In practice we will always use the vertex
operators for $l=m=n \in \{0,-1\}$. For $S'$ we only write down
the vertex operator of the imaginary part $S'_2$ because the real
part is given by an RR-field and does not appear in the world
sheet action (\ref{wsaction}). Note, moreover, that the charges
$q_B$ have been absorbed in the definition of the Chan-Paton
matrices $\lambda_{s[B]}$ appearing in the building block vertex operators
(\ref{nakedvo}) and (\ref{nakedvo2}).

A  striking feature of the expressions (\ref{vops}) is the fact
that the vertex operators for the K\"ahler coordinates are
combinations of open and closed string vertex operators (except
for $S_2'$) and thus contribute all in string diagrams with or
without boundaries. For instance, the vertex operator for an open
string scalar $A_i$ can appear not only  in a disk diagram 
but also in a
sphere diagram at tree-level via the term proportional to
$V_{Z\bar Z}^{(l,m)}$. Furthermore, the coefficients involve
powers of the dilaton and thus the naive counting of the order (in
an expansion in the dilaton) at which a certain diagram
contributes is no longer valid when the K\"ahler structure
adapted vertex operators are inserted. The reason for this is, of
course, the redefinition of $S'$ at disk level (\ref{sstrich}).
This feature of the vertex operators (\ref{vops}) allows us to
derive even those terms in (\ref{km}) from tree level diagrams
that usually (i.e.\ when using unadapted vertex operators) would
arise only at one-loop level.


\subsection{Tree level diagrams}
\label{treesection}

\begin{figure}[t]
\begin{center}
\psfrag{Ai}[bc][bc][1][0]{$A_i$}
\psfrag{Ajb}[bc][bc][1][0]{$\bar{A}_j$}
\psfrag{Z}[bc][bc][1][0]{$Z$}
\psfrag{g}[bc][bc][1][0]{$g$}
\psfrag{+}[bc][bc][1][0]{$+$}
\psfrag{=}[bc][bc][1][0]{$=$}
\psfrag{Zb}[bc][bc][1][0]{$\bar{Z}$}
\psfrag{ZZb}[bc][bc][1][0]{$Z\bar{Z}$}
\psfrag{K}[bc][bc][0.5][0]{{\sc K}}
\includegraphics[width=0.7\textwidth]{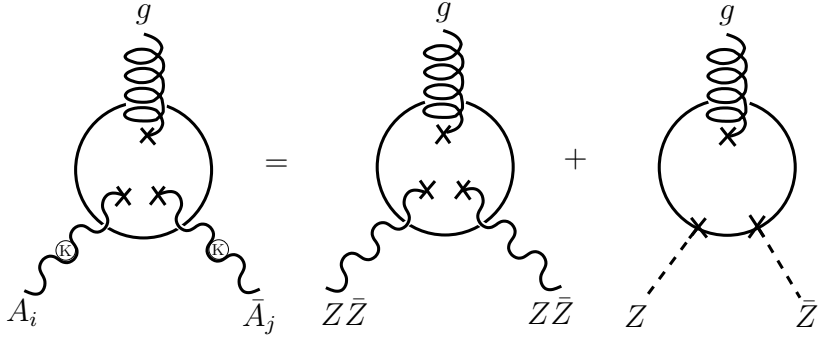}
\caption{Computing
a disk 3-point function using K\"ahler structure adapted
vertex operators (denoted by 
 $\bigcirc$\hspace{-2.9mm}{\tiny {\sc K}} ).
The adapted vertex
operators contain both open and closed vertex operators.}
\label{fig:disk}
\end{center}
\vspace{-5mm}
\end{figure}

Now we are in a position to calculate 3-point functions on the
sphere and disk. As an example we consider the case of 3-point
functions with one graviton and two open string scalars, from
which one can read off the sigma-model metric $K^{(0)}_{A_i \bar
A_j}$ of (\ref{ktree}). We thus obtain on the sphere
\beqn \label{3ptsphere}
\langle
V_{g}^{(0,0)} V_{A_i}^{(-1,-1;-1)} V_{\bar A_j}^{(-1,-1;-1)}
\rangle_{\rm sphere}
&=& \non
&& \hspace{-4cm}
- \frac{g_o^2 \alpha'^{-4}}{64 \pi^4} \frac{N_i (A_i  -\bar A_i) N_j (A_j  -\bar A_j)}{(U-\bar U)^2( S_0'-\bar S_0')^2}
 \langle V_{g}^{(0,0)}V_{Z\bar Z}^{(-1,-1)}V_{Z\bar Z}^{(-1,-1)}\rangle_{\rm sphere} \ , \non
\eeqn
whereas on the disk we have to calculate\footnote{As there is just one
boundary on the disk we omit the index $B$ in this section.}
(see fig.\ \ref{fig:disk})
\beqn \label{3ptdisk}
\langle
V_{g}^{(-1,-1)} V_{A_i}^{(0,0;0)} V_{\bar A_j}^{(0,0;0)}
\rangle_{\rm disk}
&=&\non
&& \hspace{-3cm}
- \frac{g_o^2\alpha'^{-4}}{64 \pi^4} \frac{N_i (A_i  -\bar A_i) N_j (A_j  -\bar A_j)}{(U-\bar U)^2( S_0'-\bar S_0')^2}
 \langle V_{g}^{(-1,-1)} V_{Z\bar Z}^{(0,0)}V_{Z\bar Z}^{(0,0)}\rangle_{\rm disk} \non
&& \hspace{-3cm}
- \frac{16 g_o^2 \alpha'^{-4} \delta_{ij}}{(U-\bar U)(T-\bar T)} \langle  V_{g}^{(-1,-1)} V_Z^{(0)} V_{\bar Z}^{(0)} \rangle_{\rm disk} \ .
\eeqn
The graviton vertex operators are given by
\beqn \label{gravitonvo}
V_g^{(0,0)} & = & -\frac{2g_c}{\alpha'} \epsilon_{\mu \nu} \int_\Sigma d^2z\
                                            \big[i \partial X^\mu + \frac{1}{2} \alpha' (p \cdot \psi) \psi^\mu \big]
                                             \big[i \bar \partial X^\nu + \frac{1}{2} \alpha' (p \cdot \tilde \psi) \tilde \psi^\nu \big]
                                             e^{ipX}\ , \non
V_g^{(-1,-1)} & = & g_c  \epsilon_{\mu \nu} \int_\Sigma d^2z\ e^{-\phi - \tilde \phi} \psi^\mu \tilde \psi^\nu e^{ipX} \ ,
\eeqn
for a symmetric, transverse and traceless polarization tensor
$\epsilon_{\mu \nu}$. Moreover, we use the same world-sheet
correlators as in \cite{Polchinski:rr}, i.e.
\beqn \label{xx}
\langle X^M(z_1) X^N(z_2) \rangle &=& -\frac{\alpha'}{2} g^{MN} \ln |z_{12}|^2\ , \non
\langle \psi^M(z_1) \psi^N(z_2) \rangle &=& g^{MN} z_{12}^{-1} \quad  , \non
\langle \tilde \psi^M(\bar z_1) \tilde \psi^N(\bar z_2) \rangle &=& g^{MN} \bar z_{12}^{-1}\ , \non
\langle e^{-\phi}(z_1) e^{-\phi}(z_2) \rangle &=& z_{12}^{-1}\quad  ,
\non 
 \langle e^{-\tilde \phi}(\bar z_1) e^{-\tilde \phi}(\bar z_2) \rangle &=& \bar{z}_{12}^{-1}\ ,
\eeqn
where $M,N$ can stand either for the external coordinates or the torus directions and $z_{12}=z_1-z_2$.
On the disk there are further contributions coming from
\beqn
\langle \bar \partial X^M(\bar z_1) \partial X^N(z_2) \rangle_{\rm disk} &=& -\frac{\alpha'}{2} g^{MN} (\bar z_1 - z_2)^{-2}\ , \non
\langle \tilde \psi^M(\bar z_1) \psi^N(z_2) \rangle_{\rm disk} &=& g^{MN} (\bar z_1- z_2)^{-1}\ .
\eeqn
Notice also that the correlators for the complexified variables
(\ref{zpsi}) follow as \cite{Lust:2004cx}
\beqn
\langle \partial Z(z_1) \partial \bar Z(z_2) \rangle &=&
-\frac{\alpha'}{2} z_{12}^{-2} \quad , \non
\langle \bar \partial Z(\bar z_1) \bar \partial \bar Z(\bar z_2) \rangle &=&
-\frac{\alpha'}{2} \bar z_{12}^{-2} \ , \\
\langle \bar \partial Z(\bar z_1) \partial \bar Z(z_2) \rangle &=&
\langle \bar \partial \bar Z(\bar z_1) \partial Z(z_2) \rangle ~=~
-\frac{\alpha'}{2} (\bar z_1 - z_2)^{-2}\ , \nonumber
\eeqn
and
\beqn
\langle \Psi(z_1) \bar \Psi(z_2) \rangle &=& z_{12}^{-1} \quad  , \non
\langle \tilde \Psi(\bar z_1) \bar{\tilde \Psi}(\bar z_2) \rangle &=&
\bar z_{12}^{-1} \ , \non
\langle \tilde \Psi(\bar z_1) \bar \Psi(z_2) \rangle &=&
\langle \bar{\tilde \Psi}(\bar z_1) \Psi(z_2) \rangle ~=~
(\bar z_1 - z_2)^{-1}
\ .
\eeqn
All other correlators, i.e.\ those for purely holomorphic or
anti-holomorphic fields, vanish. Now we are ready to
calculate (\ref{3ptsphere}) and (\ref{3ptdisk}). Fixing
the positions of the three vertex operators in (\ref{3ptsphere})
at $0,1$ and $\infty$, respectively, and including the
corresponding ghost factor leads to (cf.\
\cite{Polchinski:rr})
\be \label{As2}
\langle
V_{g}^{(0,0)} V_{A_i}^{(-1,-1;-1)} V_{\bar A_j}^{(-1,-1;-1)}  \rangle_{\rm sphere} \sim
\frac{g_c g_o^2 \alpha'^{-3} N_i (A_i  -\bar A_i) N_j (A_j  -\bar A_j)}{(U-\bar U)^2( S_0'-\bar S_0')^2} \epsilon_{\mu \nu} \, p_{23}^\mu \, p_{23}^\nu \ ,
\ee
where $p_{23}^\mu = p_2^\mu - p_3^\mu$. This result confirms the presence of the second term of
$K^{(0)}_{A_i \bar A_j}$ of (\ref{ktree}). We do not bother about determining the overall factor here, because the main purpose of this exercise is to show that the modification of the vertex operators is essential for deriving the terms with the right moduli dependence in the sigma-model metrics (\ref{ktree}).

The first term of (\ref{3ptdisk}) actually does not contribute to the sigma-model metric. However, in order  to see this, it is much easier to consider the 2-point function
\be \label{2ptdisk}
\langle
V_{A_i}^{(-1,-1;-1)} V_{\bar A_j}^{(0,0;0)}
\rangle_{\rm disk}
=- \frac{g_o^2 \alpha'^{-4}}{64 \pi^2} \frac{N_i (A_i  -\bar A_i) N_j (A_j  -\bar A_j)}{(U-\bar U)^2( S_0'-\bar S_0')^2}
 \langle V_{Z\bar Z}^{(-1,-1)}V_{Z\bar Z}^{(0,0)}\rangle_{\rm disk}\ .
\ee
This does not vanish automatically due to the infinite volume of an
unfixed conformal Killing group, as in the case of the closed string
2-point function on the sphere or the open string 2-point function on
the disk. However, it would vanish on-shell. 
To deal with this, one can attempt to relax momentum conservation.
Concretely, setting 
$\delta= p_1 \cdot p_2 \neq 0$ and expanding the
result to linear order in $\delta$, one can try to read off
the metric as the
coefficient of $\delta$. A similar procedure 
was used in 
\cite{Atick:1987gy,Poppitz:1998dj} to investigate the generation of
one-loop mass terms of scalar fields in heterotic and type I theories, in \cite{Bain:2000fb} to compute 
anomalous dimensions and K\"ahler metric renormalization
for matter fields in orientifolds,
and in \cite{Antoniadis:2002cs} to calculate the mass terms for
anomalous $U(1)$ gauge fields in type I compactifications.\footnote{A similar method
was also used in \cite{Dixon:1990pc} and in
\cite{Minahan:1987ha} for calculations of 3-point functions at one-loop in the heterotic
string. As pointed out in \cite{Minahan:1987ha}, the reason why momentum-conservation relaxation
works particularly well for 2-point functions is
that there is 
little ambiguity in how to go off-shell. For further
discussion, see also chapter 13 of \cite{Kiritsis:1997hj}.}  The
relevant amplitude was already  calculated in appendix A.2 of
\cite{Lust:2004cx} with the result that there are no contributions to
linear order in $\delta$.\footnote{One 
more observation is worth making. In the case of D$p$-branes
with $p<9$, the amplitude
contains a term proportional to $s/t$, where $s$
is the square of the momenta parallel to
the brane, $s=2
(p_1^{||})^2=2 (p_2^{||})^2$, and $t=p_1 \cdot p_2$. 
This $t$-channel pole arises when the
two vertex operators come together, and the coefficient of
$s/t$ must be proportional to the sigma-model metric at 
sphere level,
cf.\ figure 2 in \cite{Hashimoto:1996bf}. 
Comparing the moduli
dependence in (\ref{3ptsphere}) and (\ref{2ptdisk}) we see
that this proportionality also holds in our case.}

The second term  of (\ref{3ptdisk}) does, however, contribute to
the sigma-model metric of the open string scalars. Again, we do
not keep track of the exact prefactors. The result can be easily
extracted from section 5 of \cite{Hashimoto:1996bf} (in particular
from their formulas (29) and (30)). Using that all the momenta are
external and the polarization of the open string scalars is along
the untwisted internal 
2-torus, all contractions between momenta and scalar
polarization can be discarded. The same is true for contractions
between polarizations of the scalars and the graviton (which has
external indices like the momenta). Finally, using momentum
conservation (which does not need to be relaxed for this 3-point function) and tracelessness and transversality of the
graviton polarization tensor, it is easy to see that the only
contribution in our case is of the form
\be
\langle
V_{g}^{(-1,-1)} V_{A_i}^{(0,0;0)} V_{\bar A_j}^{(0,0;0)}
\rangle_{\rm disk} \sim \frac{g_c g_o^2 \alpha'^{-3} N_i
\delta_{ij}}{(U-\bar U)(T-\bar T)} \epsilon_{\mu \nu} \, p_{23}^\mu \,
p_{23}^\nu \ ,
\ee
where the factor $N_i\delta_{ij}$ comes from the trace over CP
labels. Being a disk diagram, the above correlator gets an additional
factor $e^{\Phi}$ relative to the sphere after performing the Weyl
rescaling to Einstein frame, which promotes $T-\bar T$ to
$S_0'-\bar S_0'$. This confirms the presence of the first term of
$K^{(0)}_{A_i \bar A_j}$ of (\ref{ktree}).


\subsection{One-loop diagrams}
\label{1loopsection}

After establishing the correct form for the vertex operators and
clarifying the status of the tree-level action we now come to the main
purpose of this paper: the calculation of one-loop corrections
from Klein bottle, annulus
and M\"obius strip diagrams. We have computed all the relevant 2-point
functions which allows us 
to read off the correction terms for all the
components of the K\"ahler metric whose classical terms are given in
(\ref{km}). To determine the K\"ahler potential it turns out to be
sufficient to compute a single such component and integrate. For the sake
of transparency, we have chosen to present
the simplest 2-point function in this section, 
and collected all the other
amplitudes in appendix
\ref{other}. There we will show that the other correlation functions are
consistent with the K\"ahler potential derived from the one
presented in this section.

\begin{figure}[h]
\begin{center}
\psfrag{Sp}[bc][bc][1][0]{$S_2'$}
\psfrag{K}[bc][bc][0.6][0]{{\sc K}}
\includegraphics[width=0.7\textwidth]{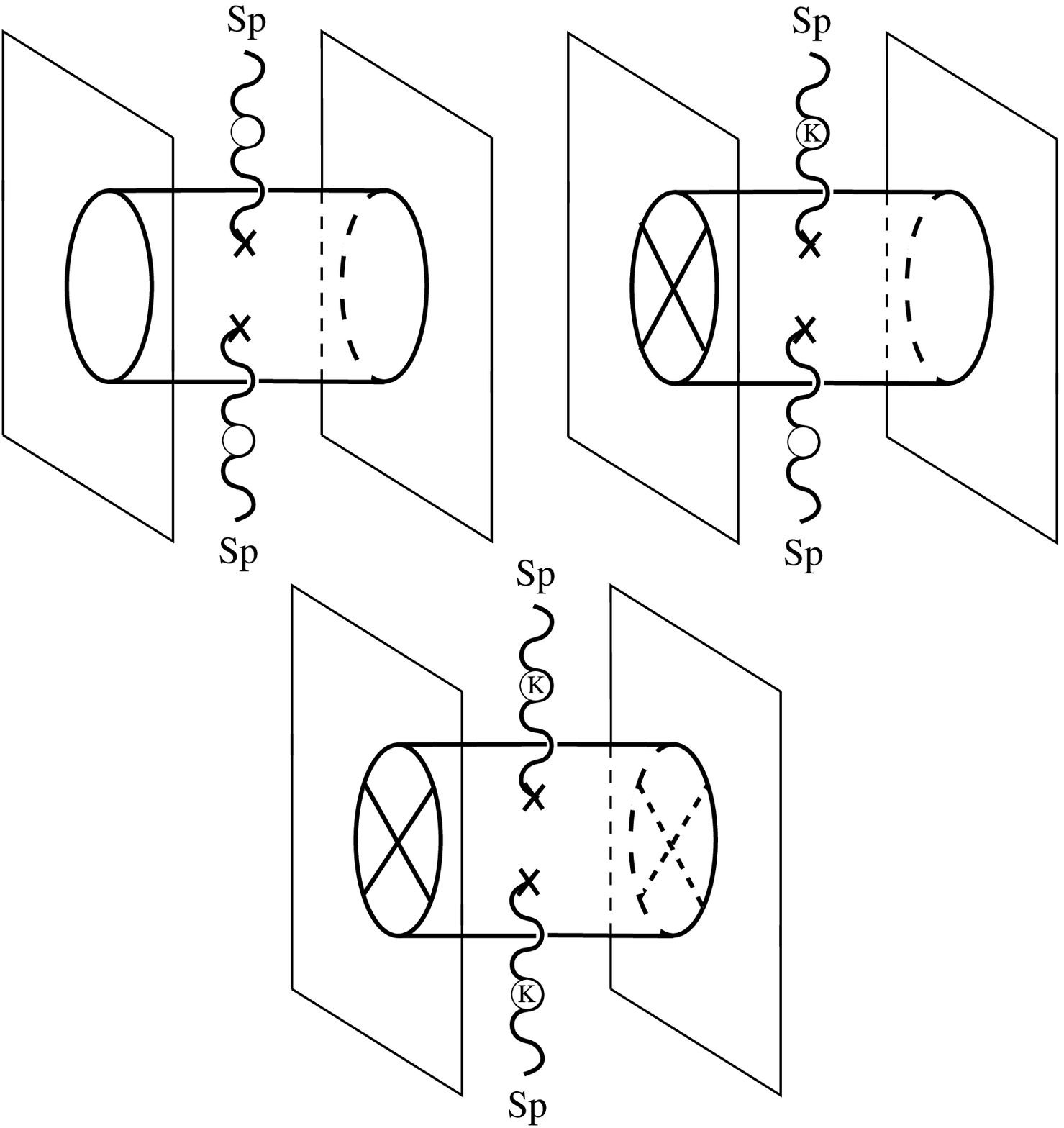}
\caption{$\langle
V_{S_2'}V_{S_2'}\rangle_{\sigma}$ for $\sigma=\ca,\cm,\ck$,
with K\"ahler adapted vertex operators.}
\label{fig:2pt}
\end{center}
\end{figure}

In this section we calculate the 2-point function for $S_2'$ on
all one-loop surfaces: the torus, Klein bottle, annulus
and M\"obius strip. This is sufficient to determine the K\"ahler
potential on the moduli space of 
vector moduli, as we will see in
section \ref{1loopkpot}.
The calculation again necessitates the use 
of the off-shell prescription
introduced in the last section for the 2-point function on the disk
(\ref{2ptdisk}). More precisely, now we want to compute
\be \label{2pts2}
\langle V_{S_2'} V_{S_2'} \rangle = - g_c^2 \alpha'^{-4} \frac{1}{(S_0'-\bar S_0')^2}
   \sum_\sigma \langle V_{Z\bar Z}^{(0,0)}V_{Z\bar Z}^{(0,0)} \rangle_\sigma\ ,
\ee
where the index $\sigma$ is used to label different types of
amplitudes. This 2-point function is depicted in figure
\ref{fig:2pt}.
We enumerate the contributions of the various diagrams
symbolically in the form
\beqn
\ck + \ca + \cm = \sum_{k=0,1} \Big[ \ck^{(k)}_{1} + \ck^{(k)}_{\rm \Theta} +
\ca_{99}^{(k)} + \ca_{55}^{(k)} + \ca_{95}^{(k)} + \ca_{59}^{(k)} +
\cm_9^{(k)} + \cm_5^{(k)} \Big]
\ ,
\eeqn
where warn the reader that we do not mean partition functions, 
but the above correlator $\langle V_{Z\bar
Z}^{(0,0)}V_{Z\bar Z}^{(0,0)} \rangle_\sigma$ of (\ref{2pts2})
evaluated on these world-sheets. There is no contribution from the
torus \cite{ABFPT}. The upper index $k$ stands for the power of
$\Theta$ inserted in the trace (coming from the orbifold projector
$\cp = \frac12 (1+\Theta)$). The Klein bottle contains a sum over
all 16 twisted sectors at the fixed points of $\Theta$ on the
$\mbb T^4$. When we break up the D9-branes (and potentially
D5-branes) into stacks, we use the notation
\beqn \label{actually}
&&
\ca_{99}^{(k)} =\sum_{i,j} \ca_{ij}^{(k)} \ , \quad
\ca_{95}^{(k)} =\sum_{i,a} \ca_{ia}^{(k)} \ , \quad
\ca_{55}^{(k)} =\sum_{a,b} \ca_{ab}^{(k)} \ , \non
&&
\cm_9^{(k)} =\sum_{i} \cm_{i}^{(k)} \ , \quad
\cm_5^{(k)}  =\sum_{a} \cm_{a}^{(k)} \ .
\eeqn
In this notation $\sigma$ takes values $\sigma \in\{ (ij), (ab), (ia),
(ai)\}$ for annulus diagrams,  $\sigma \in\{(i), (a)\}$ for M\"obius diagrams,
and $\sigma\in\{ (\Theta),(1) \}$ for the twisted and untwisted
sectors of the Klein bottle, respectively.
Each diagram can have a 
different (and characteristic) dependence on 
the Wilson lines.
We therefore also introduce a symbolic notation for the Wilson lines,
writing $\vec {\bf A}_\sigma$ for the matrix valued Wilson lines,
which now also become tensor valued in the case of annulus diagrams,
where they can appear at both of the two ends individually. We will
provide a list of these expressions later in table \ref{tolletabelle}.

The correlator in (\ref{2pts2}) is now expanded in powers of $\delta = p_1
\cdot p_2$, and only the linear term is relevant for the K\"ahler
metric. The sum over
spin structures causes the amplitudes to vanish unless
we contract at least four of the world
sheet fermions appearing in $V_{Z \bar Z}^{(0,0)}$. This already
gives the momentum dependence we want
and we set $p=0$ elsewhere, 
as in \cite{ABFPT}.\footnote{In general, a shortcut
like this can be invalidated by poles from the integration
over vertex operator positions.
We check the validity of this procedure in
one example, by calculating a 4-point function in appendix
\ref{4ptcheck}.} Doing this we find that (\ref{2pts2}) can be
expressed in the form
\beqn
\langle V_{Z \bar Z}^{(0,0)}V_{Z \bar Z}^{(0,0)} \rangle_\sigma &=& -  V_4
\frac{(p_1 \cdot p_2) \sqrt{G}}{8 (4 \pi^2 \alpha')^2} \int_0^\infty \frac{dt}{t^4}
  \int_{\cf_\sigma} d^2\nu_1 d^2\nu_2  \label{ZZbZZb} \\
&& \hspace{-3cm}
\times \sum_{k=0,1} \sum_{\vec n=(n,m)^T}
\tr \Bigg[
e^{-\pi \vec{n}^{T} G \vec{n} t^{-1}}
e^{2 \pi i \vec{\bf A}_\sigma \cdot \vec{n}} \sum_{{\alpha\beta}\atop{\rm even}} \frac{\thba{\alpha}{\beta}(0,\tau)}{\eta^3(\tau)}
 \gamma_{\sigma,k} \cz_{\sigma,k}^{\rm int} \zba{\alpha}{\beta} \non
&& \hspace{-1.5cm}
 \times
\Big[ \langle \bar \partial Z(\bar \nu_1)
\bar \partial \bar{Z}(\bar \nu_2) \rangle_\sigma \langle \bar \Psi(\nu_1)
\Psi(\nu_2) \rangle_\sigma^{\alpha,\beta} \langle \psi(\nu_1)
\psi(\nu_2) \rangle_\sigma^{\alpha,\beta} \non
&& \hspace{-1cm}
+ \langle \bar \partial Z(\bar \nu_1)
\partial \bar Z(\nu_2) \rangle_\sigma \langle \bar \Psi(\nu_1)
{\tilde{\Psi}}(\bar \nu_2) \rangle_\sigma^{\alpha,\beta} \langle \psi(\nu_1) \tilde
\psi(\bar \nu_2) \rangle_\sigma^{\alpha,\beta} + {\rm c.c.} \Big]
\Bigg] + {\cal O} (\delta^2)\ . \nonumber
\eeqn
Many comments are in order here.
$V_4$ denotes the regulated volume of the four-dimensional
spacetime. We changed variables on the
world-sheet and took
all vertex operators (\ref{nakedvo}) to depend on
the coordinate $\nu$, which is related to $z$ by $z=e^{-i
\nu}$. This choice coincides with the convention of \cite{Polchinski:rr} but differs from
the one of \cite{ABFPT} by a factor of $2 \pi$ in the exponent;
cf.\ appendix \ref{app:toolbox} for more details on our world-sheet
conventions, in particular figure \ref{fig:surfaces}. The sum
over bosonic zero modes has been made explicit, since there is
also an implicit dependence on $m,n$ 
in the bosonic correlators: this arises from the
classical piece in the split into zero modes and fluctuations. 
That is, $Z(\nu) = Z_{\rm class}(\nu) +
Z_{\rm qu}(\nu)$, where the classical part is given by
\be \label{zeromode}
Z_{\rm class} = \sqrt{\alpha'} \sqrt{\frac{T_2}{2 U_2}} \frac{(n + m
\bar U)}{\tau_2} {\rm Im}(\nu) \, c_\sigma\ , \quad c_\sigma = \left\{
\begin{array}{ll}
1& {\rm for}\ \ca \ , \cm \\
2&{\rm for}\  \ck
\end{array} \right.\ .
\ee
These zero modes have the right periodicity under ${\rm Im}(\nu)
\rightarrow {\rm Im}(\nu) + 2 \pi \tau_2$ (for $\ca , \cm$) or ${\rm
Im}(\nu) \rightarrow {\rm Im}(\nu) + \pi \tau_2$ (for $\ck$),
i.e.\ $X^4 \rightarrow X^4 + 2 \pi n \sqrt{\alpha'}$ and $X^5
\rightarrow X^5 + 2 \pi m \sqrt{\alpha'}$. The zero modes do not
have any analogue at tree level; one needs a non-trivial
1-cycle on the world-sheet. They do  play an
important role in calculating the moduli dependence of one-loop
corrections to the gauge couplings in the heterotic string
performed in \cite{Dixon:1990pc} and reviewed in, for instance,
\cite{Kiritsis:1997hj}.

The internal partition function is abbreviated
$\cz_{\sigma,k}^{\rm int}$ for the diagram $\sigma$ with insertion
$\Theta^k$ and carries a label $\alpha$, $\beta$ for the spin
structure.\footnote{To be more precise, we should say that
$\cz_{\sigma,k}^{\rm int} \zba{\alpha}{\beta}$ gives the internal
partition function without the contribution from the zero modes
that we split off.} For the annulus and M\"obius strip diagrams it
is \cite{Gimon:1996rq,Bachas:1996zt,ABD}
\beqn\nonumber
\cz_{\sigma,k}^{\rm int} \zba{\alpha}{\beta} ~=~
\eta_{\alpha\beta} \frac{\thba{\alpha}{\beta}(0,\tau)}{\eta^3(\tau)}
\frac{\thba{\alpha+h}{\beta+g}(0,\tau)\thba{\alpha-h}{\beta-g}(0,\tau)}
     {\thba{1/2+h}{1/2+g}(0,\tau)\thba{1/2-h}{1/2-g}(0,\tau)}
\times \left\{
\begin{array}{cl}
[-(2\sin(\pi g))^2] & {\rm for}\ h=0 \\
1 & {\rm for}\ h=\frac12
\end{array} \right.
\nonumber
\eeqn
where $(g,h)$ take values $h=0$ for M\"obius strip and the annulus
diagrams with pure NN (N for Neumann) or DD (D for Dirichlet)
boundary conditions whereas $h=\frac12$ for ND, DN for annulus
diagrams with boundary conditions. Further, $g=\frac12$ whenever
there is a reflection acting on the world-sheet oscillators in the
trace, $g=0$ otherwise. Note that in the M\"obius strip $\Omega$
acts on D directions with an additional reflection compared to N
directions. To make sense of this formula also for $g=h=0$ one has
to use
\be\label{lim}
\lim_{\epsilon \rightarrow 0} \Bigg[ \frac{2 \sin (\pi \epsilon)}{\thba{1/2}{1/2 + \epsilon}(0,\tau)} \Bigg]
=  -\frac{1}{\eta^3(\tau)}\ .
\ee
The internal partition function of the Klein bottle is given by
\cite{Gimon:1996rq,Bachas:1996zt,Aldazabal:1998mr}
\beqn
\cz_{\sigma,k}^{\rm int} \zba{\alpha}{\beta} ~=~
\eta_{\alpha\beta} \frac{\thba{\alpha}{\beta}(0,\tau)}{\eta^3(\tau)}
\frac{\thba{\alpha+h}{\beta+2g}(0,\tau)\thba{\alpha-h}{\beta-2g}(0,\tau)}
     {\thba{1/2+h}{1/2+2g}(0,\tau)\thba{1/2-h}{1/2-2g}(0,\tau)}
\times \left\{
\begin{array}{cl}
[-(2\sin(2\pi g))^2] & {\rm for}\ h=0 \\
16 & {\rm for}\ h=\frac12
\end{array} \right.
\nonumber
\eeqn
where $h=0$ or $\frac12$ for untwisted and twisted sectors, the
factor of $16$ coming from the $16$ fixed points of the orbifold
action. In the presence of a reflection in the trace $g=\frac12$,
otherwise $g=0$, but this does not have any effect, since only
$2g$ appears.

The matrices $\gamma_{\sigma,k}$ stand for the operation on CP labels of the
operators that appear in the trace and are given in table \ref{tolletabelle} for the
different sectors. They are given by
\cite{Aldazabal:1998mr,ABD}
\beqn \label{gammas}
\gamma_{i} &=& {\bf 1}_{2N_i} \oplus {\bf 0}_{32-2N_i} \ ,  \quad
\gamma_{a} ~=~ {\bf 1}_{2N_a} \oplus {\bf 0}_{32-2N_a} \ , \non
\gamma_{\Theta i} &=& {\rm diag}( i{\bf 1}_{N_i} , -i{\bf 1}_{N_i})
\oplus {\bf 0}_{32-2N_i} \ , \quad
\gamma_{\Theta a} ~=~ {\rm diag}( -i{\bf 1}_{N_a} , i{\bf 1}_{N_a}) \oplus {\bf 0}_{32-2N_a} \ , \non
\gamma_{\Omega i} &=& \sigma_{1N_i} \oplus {\bf 0}_{32-2N_i} \ , \quad
\gamma_{\Omega a} ~=~  \sigma_{2N_a} \oplus {\bf 0}_{32-2N_a} \ , \non
\gamma_{\Omega \Theta i} &=& \sigma_{2N_i} \oplus {\bf 0}_{32-2N_i} \ , \quad
\gamma_{\Omega \Theta a} ~=~  \sigma_{1N_a} \oplus {\bf 0}_{32-2N_a}
\eeqn
with
\beqn \label{sigmas}
\sigma_{1N_i} = \left( \begin{array}{cc}
{\bf 0}_{N_i} & {\bf 1}_{N_i} \\
{\bf 1}_{N_i} & {\bf 0}_{N_i}
\end{array}
\right) \ , \quad
\sigma_{2N_i} =
\left(
\begin{array}{cc}
{\bf 0}_{N_i} & i{\bf 1}_{N_i} \\
-i{\bf 1}_{N_i} & {\bf 0}_{N_i}
\end{array}
\right) \ .
\eeqn
In order to calculate the amplitude (\ref{ZZbZZb}), we need the
correlators (see also appendix \ref{app:toolbox} for a
discussion)\footnote{The derivative in $\vartheta_1'(0,\tau)$
is with respect to the first argument and not with respect to $\nu$.}
for bosons on the torus
\beqn \label{correl}
\langle Z_{\rm qu}(\nu_1) \bar Z_{\rm qu}(\nu_2) \rangle_{\ct} &=& -\frac{\alpha'}{2} \ln \Big|
   \frac{2 \pi}{\vartheta_1'(0,\tau)} \vartheta_1\Big(\frac{\nu_1-\nu_2}{2 \pi},\tau \Big) \Big|^2
 + \alpha'  \frac{({\rm Im}(\nu_1-\nu_2))^2}{4 \pi {\rm Im} (\tau)}
\ ,
\eeqn
and for fermions on any world-sheet
\beqn
\langle \Psi(\nu_1) \bar \Psi(\nu_2) \rangle_{\sigma}^{\alpha,\beta} &=&
\langle \psi(\nu_1) \psi(\nu_2) \rangle_{\sigma}^{\alpha,\beta} ~=~
\frac{1}{2 \pi} \frac{\thba{\alpha}{\beta}\Big(\frac{\nu_1-\nu_2}{2\pi},\tau\Big) \vartheta_1'(0,\tau)}
      {\vartheta_1\Big(\frac{\nu_1-\nu_2}{2
\pi},\tau\Big)\thba{\alpha}{\beta}(0,\tau)} \ .
\eeqn
All correlators are obtained from correlators on the torus 
via the method of images \cite{ABFPT}, and the remaining ones are
\beqn
\langle Z(\nu_1) \bar Z(\nu_2) \rangle_\sigma &=& \langle Z(\nu_1) \bar Z(\nu_2) \rangle_{\ct}
 + \langle Z(\nu_1) \bar Z(I_\sigma(\nu_2)) \rangle_{\ct} \ , \non
\langle \tilde \Psi(\bar \nu_1) \bar \Psi(\nu_2) \rangle_{\sigma}^{\alpha,\beta} &=&
\langle \tilde \psi(\bar \nu_1) \psi(\nu_2) \rangle_{\sigma}^{\alpha,\beta} ~=~
i \langle \Psi(I_\sigma(\nu_1)) \bar \Psi(\nu_2) \rangle_{\sigma}^{\alpha,\beta}\ , \non
\langle \tilde \Psi(\bar \nu_1) \bar{\tilde \Psi}(\bar \nu_2) \rangle_{\sigma}^{\alpha,\beta} &=&
\langle \tilde \psi(\bar \nu_1) \tilde \psi(\bar \nu_2) \rangle_{\sigma}^{\alpha,\beta} ~=~
\overline{\langle \Psi(\nu_1) \bar \Psi(\nu_2)
\rangle_{\sigma}^{\alpha,\beta}} \ .
\eeqn
where the involution $I_\sigma(\nu)$ defines the type of world-sheet
by identifying points on the torus. The involutions for the three
types of one-loop diagrams are\footnote{Compare also figure \ref{fig:surfaces}.}
\beqn \label{is}
I_{{\cal A}}(\nu) = I_{{\cal M}}(\nu) = 2 \pi - {\bar \nu} \ , \quad
I_{{\cal K}}(\nu) = 2 \pi - {\bar \nu} + \pi \tau \ .
\eeqn
Again the purely holomorphic or anti-holomorphic correlators vanish.
One can then use
\beqn \label{simple}
\sum_{\alpha, \beta \atop {\rm even}} \eta_{\alpha\beta} \thba{\alpha}{\beta}(\nu,\tau)^2
 \thba{\alpha+h}{\beta+g}(0,\tau)\thba{\alpha-h}{\beta-g}(0,\tau)
= \vartheta_1(\nu,\tau)^2 \thba{1/2+h}{1/2+g}(0,\tau)\thba{1/2-h}{1/2-g}(0,\tau)
\eeqn
to evaluate the sum over spin structures. In doing so, we define
\beqn \label{spinstructuresums}
\cq_{\sigma,k} &=&
\sum_{{\alpha\beta}\atop{\rm even}} \frac{\thba{\alpha}{\beta}(0,\tau)}{\eta^3(\tau)}
 \cz_{\sigma,k}^{\rm int} \zba{\alpha}{\beta}
\langle \Psi(\nu_1)\bar \Psi(\nu_2) \rangle_\sigma^{\alpha,\beta} \langle\psi(\nu_1)\psi(\nu_2)\rangle_\sigma^{\alpha,\beta}
\non
&=& \left\{
\begin{array}{cl}
     \left. \begin{array}{cc}
     -(2\sin(\pi g))^2 & {\rm for}\ h=0 \\
     1 & {\rm for}\ h=\frac12
     \end{array} \right\} & {\rm for}\ \ca , \cm \\
\hspace{-2.28cm} 0 & {\rm for}\ \ck_{1} \\
\hspace{-2.35cm} 16 & {\rm for}\ \ck_{\Theta}
\end{array}
\right. \ .
\eeqn
Note that the dependence on $\nu_i$ and $\tau$ drops out.
Furthermore, $\cq_{\sigma,k}$ is zero for the $\cn=4$ subsector of the
theory, which means that
\beqn
\ck^{(0)}_{1} = \ca_{99}^{(0)} = \cm_9^{(0)} = 0 \ , \quad
\ck^{(1)}_{1} = \ca_{55}^{(0)} = \cm_5^{(1)} = 0 \ .
\eeqn
These are the contributions of the original type I theory
compactified on a torus, and its T-dual along the $\mbb T^4$.
Furthermore, since the matrix $\gamma_{\Theta a}$ is traceless,
the annulus diagrams with $k=1$ and at least one boundary on a
D5-brane, i.e.\  $\ca_{55}^{(1)} + \ca_{95}^{(1)} +
\ca_{59}^{(1)}$ also do not contribute. This leaves us with the
non-vanishing contribution from diagrams
\beqn \label{contribu}
\ck^{(0)}_{\Theta} + \ck^{(1)}_{\Theta} + \ca_{99}^{(1)} +
\ca_{95}^{(0)} + \ca_{59}^{(0)} + \cm_9^{(1)} + \cm_5^{(0)} \ .
\eeqn
To be explicit, in table \ref{tolletabelle} we list all the quantities
$\cq_{\sigma,k}$, $\gamma_{\sigma,k}$ and the Wilson lines $\vec {\bf
A}_\sigma$ that are relevant to compute the contributions.
\begin{table}
\beqn
\begin{array}{|c|l|c|c|c|c|}
\hline
 & {\rm Sector} & {\rm Insertion} & \cq_{\sigma,k} & \gamma_{\sigma,k} & \vec {\bf A}_\sigma \\
\hline
\hline
\ck      & \sigma = (1) &  k=0,1 & 
                                            0 & & 
                                                  \vec{0} \\
         & \sigma = (\Theta) &  k=0,1
         & 
           16 & & \vec{0} \\
\hline
\ca_{99} & \sigma = (ij) & k=0 & 
                                 0
         & 
           \gamma_{i} \otimes \gamma^{-1}_{j}
         & 
           (\vec a_i W_i \otimes {\bf 1}_{32}) \oplus  ({\bf 1}_{32} \otimes (- \vec a_jW_j))
         \\
         & \sigma = (ij) & k=1
         & 
           -4
         & 
           \gamma_{\Theta i} \otimes \gamma^{-1}_{\Theta j}
         & (\vec a_i W_i \otimes {\bf 1}_{32}) \oplus  ({\bf
             1}_{32} \otimes (- \vec a_jW_j)) \\
\ca_{55} & \sigma = (ab) & k=0 & 
                                 0
         & 
           \gamma_{a} \otimes \gamma^{-1}_{b}
         & 
           \vec{0} \\
         & \sigma = (ab) & k=1 & 
                                 -4
         & 
           \gamma_{\Theta a} \otimes \gamma^{-1}_{\Theta b}
         & \vec{0} \\
\ca_{95} & \sigma = (ia) & k=0 & 
                                 1
         & 
           \gamma_{i} \otimes \gamma^{-1}_{a}
         & 
           \vec a_i W_i \otimes {\bf 1}_{32} \\
         & \sigma = (ia) & k=1 & 
                                 1
         & 
           \gamma_{\Theta i} \otimes \gamma^{-1}_{\Theta a}
         & \vec a_i W_i \otimes {\bf 1}_{32} \\
\hline
\cm_9    & \sigma = (i) & k=0 & 
                                0
         & 
           -\gamma^T_{\Omega i}\gamma^{-1}_{\Omega i}
         & 
           2 \vec a_iW_i \\
         & \sigma = (i) & k=1 & 
                                -4
         & 
           -\gamma^T_{\Omega\Theta i}\gamma^{-1}_{\Omega\Theta i} & 2 \vec a_iW_i \\
\cm_5    & \sigma = (a) & k=0 & 
                                -4
         & 
           -\gamma^T_{\Omega a}\gamma^{-1}_{\Omega a} & 
                                                         \vec{0}  \\
         & \sigma = (a) & k=1 & 
                                0
         & 
           -\gamma^T_{\Omega\Theta  a}\gamma^{-1}_{\Omega\Theta  a} & \vec{0}  \\
\hline
\end{array}
\nonumber
\eeqn
\caption{Information needed to compute the amplitudes.} \label{tolletabelle}
\end{table}

This leaves the correlators of the bosonic fields as the only
piece that depends on the positions $\nu_i$ of the vertex
operators. The integral is evaluated via
\beqn \label{torustrick}
&& \int_{\cf_\sigma} d^2\nu_1 d^2\nu_2
\Big[ \langle \bar \partial Z(\bar \nu_1)
\bar \partial \bar{Z}(\bar \nu_2) \rangle_\sigma
- \langle \bar \partial Z(\bar \nu_1)
\partial \bar Z(\nu_2) \rangle_\sigma + {\rm c.c.} \Big] \\
&&\hspace{5cm} = -2 \pi^4 c_\sigma^2 \frac{T_2}{U_2} |n + mU|^2 \alpha' + \pi^3 c_\sigma^2 t \alpha'
\non
&&\hspace{5cm} = \left\{
\begin{array}{ll}
      -2 \pi^4 \frac{T_2}{U_2} |n + mU|^2 \alpha' + \pi^3 \alpha' t & {\rm for}\ \ca , \cm \\[.1cm]
      -8 \pi^4 \frac{T_2}{U_2} |n + mU|^2  \alpha' + 4 \pi^3 \alpha' t& {\rm for}\ \ck
\end{array}
\right. \ , \nonumber
\eeqn
where the first contribution comes from the zero modes given in
formula (\ref{zeromode}), where also the constants $c_\sigma$ are
introduced.\footnote{For reference, we also collect these in
(\ref{constants}), together with other constants that will be
introduced shortly.} In order to evaluate the quantum part of
(\ref{torustrick}), i.e.\ the second contribution, we made use of
the fact that a function $f(\nu)$ that is periodic on the
covering torus satisfies \cite{ABFPT}
\be \label{abfpttrick}
\int_{\cf_\sigma} d^2\nu\, \Big[\partial_\nu f(\nu) - \partial_{\bar \nu} f
(I_\sigma (\nu))\Big] = \int_{\cf_{\ct}} d^2\nu\, \partial_\nu f (\nu) = 0\
.
\ee
To evaluate the trace and KK sum, it is useful to regularize the
integral over the $t$ with a UV cutoff $\Lambda$  and introduce new variables
\be
l = \frac{1}{e_\sigma t} =  \left\{
\begin{array}{ll}
      1/t & {\rm for}\ \ca \\
      1/(4t) & {\rm for}\ \cm , \ck \\
\end{array}
\right. \ ,
\ee
where we introduced yet another constant $e_\sigma$, also listed in
(\ref{constants}). With this we get
\beqn\label{intKK}
&& \int_{1/(e_\sigma \Lambda^2)}^\infty  \frac{dt}{t^4}
\sum_{\vec n=(n,m)^T} \Bigg[ e^{-\pi \vec{n}^{T} G \vec{n} t^{-1}}
e^{2 \pi i \vec{\bf A}_\sigma \cdot \vec{n}} \Big[
-2 \pi^4 c_\sigma^2 \frac{T_2}{U_2} |n + mU|^2 \alpha' + \pi^3
c_\sigma^2 t \alpha' \Big] \Bigg] \non
&&
\hspace{0cm} = \int_0^{\infty} dl \sum_{\vec n=(n,m)^T} \!\!\!\!\!\! ' \; \; \;  \Bigg[ e^{-\pi \vec{n}^{T} G \vec{n} e_\sigma l}
e^{2 \pi i \vec{\bf A}_\sigma \cdot \vec{n}} \Big[  -2 \pi^4
c_\sigma^2 e_\sigma^3 l^2 \frac{T_2}{U_2} |n+m U|^2 \alpha' + \pi^3
c_\sigma^2 e_\sigma^2 l  \alpha' \Big] \Bigg] \non
&&
\hspace{.5cm} +~ \pi^3 \alpha' c_\sigma^2 e_\sigma^2 \int_0^{\Lambda^2} dl\, l ~+~ \ldots \non
&&
\hspace{0cm} = \frac12 \pi^3 \alpha'c_\sigma^2 e_\sigma^2 \Lambda^4  -
3 \pi \alpha' c_\sigma^2T_2^{-2} E_2 ( {\bf A}_\sigma, U) ~+~ \ldots \ ,
\eeqn
where the prime at the sum indicates that one has to leave out the term with $(n,m)=(0,0)$.
Terms that go to zero in the limit $\Lambda\rightarrow \infty$ have
been dropped (indicated by the ellipsis).
Moreover, we abbreviated
\beqn \label{eisen}
E_s ( A, U) &=& \sum_{\vec n=(n,m)^T} \!\!\!\!\!\! ' \; \; \; \frac{U_2^s e^{2\pi i \vec n\cdot \vec a}}{|n+mU|^{2s}}  \\
&=& \sum_{\vec n=(n,m)^T} \!\!\!\!\!\! ' \; \; \; \frac{U_2^s}{|n+mU|^{2s}}
 \exp\left[ 2\pi i \frac{A(n+m\bar U) - \bar A (n+mU)}{U-\bar U} \right]\ . \nonumber
 \eeqn
This function reduces to the ordinary non-holomorphic Eisenstein series $E_s(U)$
when $A=0$. In (\ref{intKK}) we have used a boldface $\bf A_\sigma$
(without vector arrow) referring to matrix valued open string scalars, i.e.\ $\bf A_\sigma$
is defined by replacing $\vec a_i$ by $A_i$ everywhere in table \ref{tolletabelle} (recall the
relation between them, given in (\ref{scalars})).
The terms involving the UV cutoff $\Lambda$ drop out after summing
over all diagrams due to tadpole cancellation. We
are thus left with the expression
\be \label{final}
\langle V_{Z\bar Z}^{(0,0)}V_{Z\bar Z}^{(0,0)} \rangle_\sigma
=(p_1 \cdot p_2) \alpha' \frac{V_4}{(4 \pi^2 \alpha')^2} \frac{3 c_\sigma^2 \pi}{8T_2}
\sum_{k} \tr \Big[ E_2({\bf A}_\sigma,U) \gamma_{\sigma,k}
\cq_{\sigma,k} \Big]  + {\cal O} (\delta^2) \ .
\ee

We can then evaluate the relevant traces for all the diagrams that
appear in (\ref{contribu}) and find
\beqn \label{sumE}
\sum_{\sigma} c_\sigma^2
 \sum_{k=0,1} \tr \Big[
E_2({\bf A}_\sigma,U) \gamma_{\sigma,k} \cq_{\sigma,k} \Big]
&=&
4 E_2(0,U) \Big[ \cq_{(\Theta),0} + \cq_{(\Theta),1} \Big] \non
&& \hspace{-4cm}
+
\sum_{i,j} \tr \Big[ E_2({\bf A}_{(ij)},U) \gamma_{(ij),1} \cq_{(ij),1} \Big] +
2 \sum_{i,a} \tr \Big[ E_2({\bf A}_{(i)},U) \gamma_{(ia),0} \cq_{(ia),0} \Big] \non
&& \hspace{-4cm} +
\sum_{i} \tr \Big[ E_2({\bf A}_{(i)},U) \gamma_{(i),1} \cq_{(i),1} \Big] +
\sum_{a} \tr \Big[ E_2(0,U) \gamma_{(a),0} \cq_{(a),0} \Big]
\non
&& \hspace{-4.5cm} =
-~4 \sum_{i,j} N_iN_j \big[  E_2(A_i - A_j,U) +
     E_2(-A_i + A_j,U) \nonumber \\[-.3cm]
&& \hspace{-1cm}
    -  E_2(A_i + A_j,U)
    - E_2(-A_i - A_j,U) \big] \nonumber \\[.3cm]
&& \hspace{-4cm}
+~ 2\cdot 32 \sum_i N_i \big[  E_2(A_i,U) + E_2(-A_i,U)\big] \non
&& \hspace{-4cm}
-~ 4\sum_i N_i \big[  E_2(2A_i,U) + E_2(-2A_i,U)\big] \ ,
\eeqn
where the contributions from the Klein bottle and the 5-brane M\"obius strip diagrams,
which involve $E_2(0,U)$,  just cancel out.
We denote this quantity as
\beqn \label{e2au_def}
\ce_2(A_i,U) = \sum_{\sigma} c_\sigma^2
\sum_{k=0,1} \tr \Big[
E_2({\bf A}_\sigma,U) \gamma_{\sigma,k} \cq_{\sigma,k} \Big] \ .
\eeqn
Putting everything together we end up with
\be \label{s2stringframe}
\langle V_{S_2'} V_{S_2'} \rangle = - i (p_1 \cdot p_2) \alpha' \frac{V_4}{(4 \pi^2 \alpha')^2} \frac{3 g_c^2 \alpha'^{-4} e^{-\Phi}}{8 \sqrt{2} } \frac{1}{(S_0'-\bar S_0')^3}  \ce_2(A_i,U)\ .
\ee
From this we can read off the one-loop correction to the kinetic
term of $S'$ after performing a Weyl rescaling to the Einstein
frame. In the one-loop term (\ref{s2stringframe}) this just leads
to an additional factor of
\be \label{Weyl}
\frac{e^{2 \Phi}}{\cv_{\rm K3}\sqrt{G}} = \frac{i e^{\Phi}}{\sqrt{2} \pi (S-\bar S)}\ .
\ee
There is one further complication,
the one-loop correction to the Einstein-Hilbert
term calculated in \cite{ABFPT}. In the case at hand, i.e.\ in
the presence of Wilson line moduli, the corrected Einstein-Hilbert
term is
\be \label{eh1loop}
\frac12 \Big(e^{-2\Phi} \cv_{\rm K3} \sqrt{G} + \frac{\tilde c}{\sqrt{G}} \ce_2(A_i,U) \Big) R\ ,
\ee
with $\tilde c$ a constant, whose value is not important for
us at the moment. Due to the presence of a kinetic term for $S'$
already at sphere level this produces extra corrections to the
one-loop corrected kinetic term in Einstein frame. Performing a
Weyl rescaling to the four-dimensional Einstein frame, i.e.\
\be
g_{\mu \nu} ~\stackrel{\rm Weyl}{\longrightarrow}~ \Big(e^{-2\Phi}  \cv_{\rm K3} \sqrt{G}
 + \frac{\tilde c}{\sqrt{G}}\ce_2(A_i,U) \Big)^{-1}
g_{\mu \nu}\ ,
\ee
and expanding in the dilaton, the term proportional to $\ce_2(A_i,U)$
contributes to the sphere level kinetic term, cf.\ the first line of (\ref{ktree}), and thus changes the
prefactor of the term arising from (\ref{s2stringframe}).\footnote{More concretely,
it would cancel the contribution to the correlator (\ref{torustrick})
coming from the fluctuations and only leave those from the zero
modes. We will make this more precise for the case of the modulus $U$ in
appendix \ref{other}, cf.\ equation (\ref{uweyl}). \label{fluctfn}} Another possible source of
modification of the overall factor 
is if the variable $S'$ is corrected again at one-loop,
i.e.\ in addition to the disk level correction present in
(\ref{sstrich}). A similar effect for $S$ was noted in
\cite{ABFPT}. In analogy to that case one might expect a
correction
\be
{\rm Im}(S') ~\rightarrow~  {\rm Im}(S') + \hat c\, \ce_2(A_i,U) {\rm Im}(S)^{-1}
\ee
that would also modify the coefficient of the term proportional to
$\ce_2(A_i,U)$ in the kinetic term for $S'$ in the Einstein frame.
Given that we do not know the constant $\hat c$ exactly, we leave
the overall factor in the one-loop correction to the kinetic term
of $S'$ open for the moment and come back to it in the next
section. Making the replacements\footnote{The power of $\alpha'$
in the second replacement is chosen to give the right dimension,
given that $S'$ is defined to be dimensionless and $g_c \sim
\alpha'^{2}$.}
\be
V_4 ~\rightarrow~d^4 x \sqrt{-g} \quad , \quad (p_1 \cdot p_2) g_c^2 \alpha'^{-4} ~\rightarrow~ \partial_\mu
S_2' \partial^\mu S_2'\ ,
\ee
where we skipped some numerical factors, and using the fact that
\be
K_{S' \bar S'} \partial_\mu S' \partial^\mu \bar S'  = K_{S' \bar S'}
(\partial_\mu S_1' \partial^\mu S_1'+\partial_\mu S_2' \partial^\mu
S_2')\ ,
\ee
we finally read off
\be \label{s2einstein}
K_{S' \bar S'} ~\sim~ \frac{\ce_2(A_i,U)}{(S_0'-\bar S_0')^3(S-\bar S)}  \ .
\ee
%


\subsection{One-loop K\"ahler potential and prepotential}
\label{1loopkpot}

To find the K\"ahler potential that reproduces our one-loop
correction to the K\"ahler metric (\ref{s2einstein}), we can make
use of the fact that for ${\cal N}=2$ supersymmetry the K\"ahler
potential is given by a prepotential according to the  relation
(\ref{n=2kaehler}). In particular, it must be possible to express
the correction to the K\"ahler
potential as a correction to the argument of
the logarithm of 
the ``classical'' K\"ahler potential 
(\ref{kaeN2}). An obvious candidate that
reproduces (\ref{s2einstein}) up to higher order terms is
\beqn \label{tollesK}
K &=&
- \ln\Big[  (S-\bar S)(S'-\bar S')(U-\bar U) \Big]
\non
&&
- \ln\Big[ 1 - \frac{1}{8 \pi} \sum_i \frac{N_i (A_i - \bar A_i)^2}
                                           {(S'-\bar S')(U-\bar U)}
 - \sum_i \frac{c\, \ce_2(A_i,U)}{(S-\bar S)(S'-\bar S')} \Big]\ ,
\eeqn
where $\ce_2(A_i,U)$ is given in
(\ref{e2au_def}) and (\ref{eisen}).
The expression (\ref{tollesK}) 
contains terms of all orders in the string coupling,
when the logarithm is expanded. 
In appendix \ref{other} we 
verify that this K\"ahler potential also correctly 
reproduces all the other
one-loop corrections to the metrics of $\{U,S',A_i\}$ at
the relevant order in the string coupling. From the point of view
of finding the effective Lagrangian that reproduces our string
amplitudes, this justifies the use
of (\ref{tollesK}) as the one-loop corrected K\"ahler potential of
the $\mathbb{Z}_2$ orientifold. Thus, we consider (\ref{tollesK})
one of the main results of this paper.

The constant $c$ is not easily determined directly from the
present calculation of the one-loop correction to the K\"ahler
metric. In addition to the issues we explained in the
previous section, one would have to know the relative
normalization between the tree level and one loop amplitudes.
Instead, we will use the relation
(\ref{km-gc}) to fix it to the value $c=1/(128 \pi^6)$.
This is summarized in appendix \ref{findc}.

To make sure that (\ref{tollesK}) is consistent with $\cn
=2$ supersymmetry, we now express the argument of the logarithm in
terms of a prepotential as in (\ref{n=2kaehler}). Expanding the
prepotential perturbatively into
\beqn \label{prep}
\cf(S,S',U,A_i) = \cf^{(0)}(S,S',U,A_i) + \cf^{(1)}(U,A_i) \ ,
\eeqn
the classical term is given in (\ref{precl}). To find
$\cf^{(1)}(U,A_i)$ we have to convert the correction of the
argument of the logarithm in (\ref{tollesK}) into a prepotential.
This means we must recast 
$U_2 \ce_2(A_i,U)$ in the form of the
argument of equation (\ref{n=2kaehler}). To do so, note that\footnote{See
appendix \ref{appe2} for 
the derivation of this formula and
more details on the function $E_2(A,U)$.
Also, in order to avoid cluttering the formulas with too many
indices we give them only for a single $A=A_1+iA_2$. The
generalization to several $A_i$ is straightforward.}
\beqn \label{UEf}
(U - \bar U) E_2(A,U) & = & \non
&& \hspace{-3cm}
4 i \pi^4 \Big( \frac{1}{90} U_2^3  - \frac13 U_2 A_2^2 + \frac23 A_2^3
 - \frac13 \frac{A_2^4}{U_2} \Big)
+ i \pi \Big[  Li_3(e^{2\pi i A}) + 2 \pi A_2 Li_2(e^{2\pi i A}) + {\rm c.c.} \Big] \non
&& \hspace{-3cm}
+ 2 i \pi^2 \sum_{m>0} \Big[ \left(mU_2 - A_2 \right) Li_2(e^{2 \pi i (m U- A)}) +
                         \left(mU_2 + A_2 \right) Li_2(e^{2 \pi i (m U+ A)}) + {\rm c.c.} \Big]  \non
&& \hspace{-3cm}
+ i \pi \sum_{m>0} \Big[ Li_3(e^{2 \pi i (m U - A)}) + Li_3(e^{2 \pi i (m U + A)}) + {\rm c.c.} \Big]
\ ,
\eeqn
where the polylogarithms are defined in (\ref{polylog}). With the
help of
\be
\frac{d}{dx} Li_n(x) = \frac{1}{x} Li_{n-1}(x)\ ,
\ee
it follows that
\be \label{UEh}
(U - \bar U) E_2(A,U) = - \frac{4 i \pi^4}{3} \frac{A_2^4}{U_2} + 2 h - 2\bar h
 - ( U - \bar U) (\partial_U h + \partial_{\bar U} \bar h)
 - ( A - \bar A) (\partial_A h + \partial_{\bar A} \bar h)
\ee
with
\beqn
h(A,U) &=&
\frac{\pi^4}{2} \Big[ \frac{1}{90} U^3  - \frac13 U A^2 + \frac23 A^3 \Big]
+ \frac{i \pi}{2} Li_3(e^{2\pi i A})
\non
&&
~~~~~~~
+ \frac{i \pi}{2} \sum_{m>0} \Big[ Li_3(e^{2 \pi i (m U - A)}) + Li_3(e^{2 \pi i (m U + A)}) \Big]\ .
\eeqn
Note that the extra term of the form
$A_2^4/U_2$ in (\ref{UEh}) drops out when summing over diagrams
in (\ref{sumE}) and using the anomaly constraint (\ref{anomal}).
We can write
\be \label{uepsilonf}
(U - \bar U) \ce_2(A_i,U) = 2 f - 2\bar f
 - ( U - \bar U) (\partial_U f + \partial_{\bar U} \bar f)
 - \sum_i ( A_i - \bar A_i) (\partial_{A_i} f + \partial_{\bar A_i} \bar f)
\ee
with
\beqn \label{tollesF}
f(A_i,U) &=& \non
&& \hspace{-2cm}
- 4 \sum_{i,j} N_iN_j \big[  h(A_i - A_j,U) +
     h(-A_i + A_j,U)
    - h(A_i + A_j,U)
    - h(-A_i - A_j,U) \big] \nonumber \\
&&\hspace{-2cm}
+ 64 \sum_i N_i \big[  h(A_i,U) + h(-A_i,U)\big]
- 4\sum_i N_i \big[  h(2A_i,U) + h(-2A_i,U)\big] \ .
\eeqn
Thus, the one-loop correction to the prepotential is given by
\beqn \label{prepot}
\cf^{(1)}(A_i,U) = c\, f(A_i,U)\ ,
\eeqn
where the constant $c$ is given in (\ref{tollesc}).
This form is consistent with the prepotentials calculated in the case
of the heterotic string. The duality to type I was discussed in
\cite{Antoniadis:1997gu}. In the case of the heterotic string,
the prepotential generally has the form of a sum of cubic
monomials in the fields with polylogarithms, plus a universal term
involving the coefficient $\zeta(3)$, which is recovered here at
$A=0$ via $Li_k(1)=\zeta(k)$.\footnote{Compare e.g.\ equation
(4.25) of \cite{Harvey:1995fq}.}  We view the existence of the
prepotential (\ref{prepot}) as strong support for the validity of
(\ref{tollesK}). Moreover, as the prepotential (\ref{prep}) does
not receive any further perturbative corrections, the result
(\ref{tollesK}) holds
to all orders of perturbation theory.


\section{The $\cn=1$ orientifold $\mbb T^6/(\mathbb{Z}_2\times\mathbb{Z}_2)$}
\label{z2z2}

We now extend the previous analysis to orientifold
compactifications with $\cn=1$ supersymmetry, treating first the
case $\mathbb T^6 /(\mathbb Z_2\times\mathbb Z_2)$
\cite{Berkooz:1996dw} (see also \cite{Klein:2000qw}). Its orbifold group is generated by two reflection
operators, each non-trivial element leaving a 2-torus in $\mbb T^6
=\mbb T^2_1\times \mbb T^2_2\times\mbb T^2_3$ invariant. We write
the orbifold group $\mbb Z_2\times \mbb Z_2=\{1,\Theta_1,
\Theta_2,\Theta_3=\Theta_1\Theta_2\}$, the eigenvalues $\exp(2\pi i\vec v_I)$ of the
elements being characterized via the three vectors $\vec v_I$
\beqn \label{vs}
\vec v_1=\Big(0,\frac12,-\frac12\Big)\ ,\quad
\vec v_2=\Big(-\frac12,0,\frac12\Big)\ ,\quad
\vec v_3=\vec v_1+\vec v_2=\Big(-\frac12,\frac12,0\Big) \; .
\eeqn
We label
the two-torus $\mbb T^2_I$ by the same index as the
element $\Theta_I$ that leaves it invariant. Similarly $\mbb
T^4_I$ is the four-torus reflected by $\Theta_I$. The model includes
three sets of O5-planes and D5-branes, each wrapped along one
torus and labelled O5$_I$, 
D5$_I$, and O9-planes and  D9-branes wrapping
the entire internal space.

\subsection{The classical Lagrangian}

The closed string moduli space of the $\cn=1$ orientifold on
$\mathbb T^6 /(\mathbb Z_2\times\mathbb Z_2)$ consists of three
copies of the moduli space of a two-torus, plus the
universal axio-dilaton multiplet and the blow-up modes from the 48
twisted sectors. In all, the Hodge numbers of the space are
$(h^{(1,1)},h^{(2,1)}) = (51,3)$. We are only interested in
untwisted moduli and ignore the 48 twisted scalars in the
following. Besides the neutral closed string fields, the moduli
space also includes the D9-brane Wilson lines along the three tori
$\mbb T^2_I$ and the position scalars and Wilson lines of the
D5-branes. The latter we also set to zero. We again
allow for a number of D9-brane stacks with three complex Wilson
lines $A_i^I$ each, and define the scalars\footnote{Note that the
normalization used here for the tree level part of $T^I$ differs
from \cite{Berg:2004ek}. Here, we choose the normalization of
$\sqrt{G^I}$ such that the factors are identical to the $\cn = 2$
case. In addition,
so as not to overload the notation,
we always use ``4'' and ``5'' for the internal 
directions, even though for $I=2$ ($I=3$),
they are of course ``6'' and ``7''
(``8'' and ``9'').}
\beqn\label{scalarsz2}
T^I &=&
\frac{1}{\sqrt{8\pi^2}} ( C^I_{45} + i e^{-\Phi} \sqrt{G^I}) +
\frac{1}{8\pi} \sum_i N_i
( U^I (a^I_i)^2_4 - (a^I_i)_4 (a^I_i)_5 )
\non
&=& T_0^I + \frac{1}{8\pi} \sum_i N_i A_i^I \frac{A_i^I-\bar A_i^{I}}{U^I-\bar U^{I}} \ ,
\non
U^I &=& \frac{1}{G^I_{44}} ( G^I_{45} + i \sqrt{G^I} ) \ , \quad
A_i^I ~=~ U^I (a^I_i)_{4} - (a^I_i)_{5} \; ,
\eeqn
with $G_{mn}^I$ being the metric on each $\mathbb T_I^2$,
$C_{45}^I$ the RR 2-form, and the $(a^I_i)_m$ the components of the Wilson lines of the
D9-brane stack labelled by $i$, $I\in\{1,2,3\}$, see also \cite{Lust:2005dy}.
The dilaton $S$ is defined as in the 
$\cn=2$ case above. The classical K\"ahler potential is given
by
\beqn \label{kaepotN=1tree}
K^{(0)}_{\mathbb{Z}_2^2} &=& -\ln ( S-\bar S) - \sum_{I=1}^3 \ln \big[ ( T_0^I-\bar
T_0^I) (U^I-\bar U^I)\big]
\\
&=& -\ln ( S-\bar S) - \sum_{I=1}^3 \ln \Big[ ( T^I-\bar T^{I}) (U^I-\bar U^{I})
 - \frac{1}{8\pi} \sum_i N_i (A_i^I-\bar A_i^{I})^2 \Big]\ .
\nonumber
\eeqn
The classical gauge kinetic functions are
\beqn
f^{(0)}_{\rm D9} = - iS\ , \quad
f^{(0)}_{{\rm D5}_I} = -iT_0^I\ .
\eeqn
All these expressions can be derived completely analogously
to section \ref{sugra}, via dimensional reduction of the
type I plus Born-Infeld supergravity Lagrangian from ten
dimensions on a product of three tori. 
For the most part, this model can be thought of
as the direct sum of three copies of the $\cn=2$
compactification on K3$\times \mbb T^2$: three non-trivial
elements of the orbifold group act on three $\mbb T^2_I$ with
moduli $\{ T^I, U^I, A_i^I\}$, and there are three sets of
D5$_I$-branes instead of a single one. Only the multiplet $S$ and
the D9-branes are universal. The effective Lagrangian is then very
similar to (\ref{dimred2}) where one only needs to sum over the
three tori in the form
\beqn
\kappa_4^2 \cl_{4d} &=&
\frac{1}{2} R
+ \frac{\partial_\mu S\partial^\mu \bar S}{(S-\bar S)^2}
- \frac14 \kappa_4^2 {\rm Im}(S)\, {\rm tr}\, \cf_{\rm D9}^2 \non
&&
+ \sum_{I=1}^3 \Bigg[
\frac{\partial_\mu U^I\partial^\mu \bar U^I}{(U^I-\bar U^I)^2}
+ \frac{|\partial_\mu T^I_0+\frac{1}{8 \pi} \sum_i N_i ( (a^I_i)_4\partial_\mu (a^I_i)_5-
(a^I_i)_5\partial_\mu (a^I_i)_4)|^2}{(T^I_0-\bar T^I_0)^2}
\non
&&
\hspace{2cm}
+ \frac{\sum_i N_i |U^I\partial_\mu (a^I_i)_4 - \partial_\mu
(a^I_i)_5|^2}{4 \pi (U^I-\bar U^I)(T^I_0-\bar T^I_0)}
- \frac14 \kappa_4^2 {\rm Im}\, (T^I_0)\, {\rm tr}\,
\cf_{{\rm D5}_I}^2 \Bigg] \ .
\nonumber
\eeqn
The classical K\"ahler metric can also be read off from
(\ref{ktree}) since it factorizes into the three tori.

However, there are important differences compared to the $\cn=2$
case. In particular, one cannot deduce the exact form of the
K\"ahler potential as a logarithm of a corrected argument such as
in (\ref{n=2kaehler}), but only its perturbative expansion
\beqn
K = K^{(0)} + \sum_{n=1}^\infty K^{(n)}
\eeqn
where for $n \geq 1$, $n+1$ denotes the power of $e^\Phi$ in
$K^{(n)}$, i.e. $K^{(n)}\propto e^{(n+1)\Phi}$. The sum starts at
$n+1=2$ because the classical piece already includes the disk
diagrams. 
Explicitly, we will find three terms
that behave like $(S-\bar S)^{-1}(T^I-\bar T^I)^{-1}$ at the level
$n=2$, just as for the $\cn=2$ model, plus three terms of the form
$(T^I-\bar T^I)^{-1}(T^J-\bar T^J)^{-1}$.


\subsection{One-loop amplitudes}

The vertex operators for the untwisted moduli of the three $\mbb
T^2_I$ are identical to those derived for the torus in the $\cn=2$
model in section \ref{vertexsec}, and can be 
read off from (\ref{vops}) by 
simply adding the label $I$ for the three tori. This
follows immediately from the fact that the orbifold projection
does not constrain the metric or the Wilson lines, except for
imposing conditions on the CP matrices $\lambda$. Therefore, the
form (\ref{wsaction}) of the world-sheet Lagrangian is formally
unmodified here, only using a different $\lambda$.

We can now follow the steps of section \ref{1loopsection} to
compute the correlators $\langle V_{T_2^I}V_{T_2^I} \rangle$, i.e.\ the equivalent of
(\ref{2pts2}). It will give us the K\"ahler potential upon integration just
as in the case of the $\cn=2$ model.

To evaluate the one-loop amplitudes, it is most useful to split
all contributions into those which are repeated copies of the ones
that already appeared in the computation of section
\ref{1loopsection}, and extra pieces.
The Klein bottle, as all other diagrams,  now includes three non-trivial insertions
$\Theta_I$ in the trace from expanding the orbifold projector
\beqn\label{projz22}
\cp = \frac12 (1+\Theta_1)  \frac12 (1+\Theta_2) = \frac14
(1+\Theta_1+\Theta_2+\Theta_3)\ ,
\eeqn
and a sum over $3\times 16=48$ twisted sectors
at the fixed points of the three $\Theta_I$.
The M\"obius strip diagrams now have boundaries on the D9- and the
D5$_I$-branes and (\ref{projz22}) inserted. For the annulus
diagrams, there are 99, diagrams, 
three sets of $5_I 5_I$ diagrams, and three
sets of $95_I$ diagrams. In addition there are three $5_I
5_J$ diagrams, $I\neq J$, that do not have any analog in the $\cn = 2$ case.
In all, we have the one-loop diagrams
\beqn\label{allN=1}
&& \sum_{\hat I=0}^3 \Bigg[ \ck^{(\hat I)}_1
 + \sum_{J=1}^3 \ck^{(\hat I)}_{\Theta_J}
+ \cm_9^{(\hat I)} + \sum_{J=1}^3 \cm_{5_J}^{(\hat I)}
\non && \hspace{3cm}
+ \ca_{99}^{(\hat I)} +
\sum_{J=1}^3 \Big[ \ca_{5_J 5_J}^{(\hat I)} + \ca_{95_J}^{(\hat I)} + \ca_{5_J9}^{(\hat I)}
+ \sum_{K\neq J} \ca_{5_J 5_K}^{(\hat I)}
\Big]\Bigg] \ ,
\eeqn
where we have defined a label $\hat I\in\{0,1,2,3\}$, which refers
to the insertion of either $\Theta_I$ or the identity into the
trace.\footnote{Once we introduce Wilson lines, we would actually
need to use a notation as in (\ref{actually}). To
keep the notation reasonably compact, we refrain from making that explicit.} The computation of these
amplitudes is now in principle identical to that of the $\cn=2$ case in section \ref{1loopsection}.
The internal partition functions of the annulus and M\"obius
diagrams read
\beqn\label{partN=1}
\cz_{\sigma,\hat I}^{\rm int} \zba{\alpha}{\beta} ~=~
\eta_{\alpha\beta}
\prod_{I=1}^3
\frac{\thbw{\alpha+h_I}{\beta+g_I}(0,\tau)}
     {\thbw{1/2+h_I}{1/2+g_I}(0,\tau)}
\times \left\{
\begin{array}{cl}
( 2\sin(\pi g_I) ) & {\rm for}\ h_I=0 \\
1 & {\rm for}\ h_I=\frac12
\end{array} \right.
\ .
\eeqn
The values of $g_I$ are $0$ or $\pm\frac12$ for a trace with
either the identity or a reflection of $\mbb T^2_I$ inserted, the
sign determined by using (\ref{vs}). The assignment is partly
reversed for the 5-brane M\"obius strip, where world-sheet
parity $\Omega$ acts on a field with Dirichlet (D) boundary
conditions with an extra reflection compared to its operation with
Neumann (N) boundary conditions. The $h_I$ are all 0 in the
M\"obius diagrams, and in the annulus
they are 0 for an open
string sector with NN or DD boundary conditions, or
$\pm\frac12$ for ND and DN boundary conditions.
The partition function of the Klein bottle is
\beqn\label{partN=1b}
\cz_{\sigma,\hat I}^{\rm int} \zba{\alpha}{\beta} ~=~
\eta_{\alpha\beta}
\prod_{I=1}^3
\frac{\thbw{\alpha+h_I}{\beta+2g_I}(0,\tau)}{\thbw{1/2+h_I}{1/2+2g_I}(0,\tau)}
\times \left\{
\begin{array}{cl}
( 2\sin(2\pi g_I) ) & {\rm for}\ h_I=0 \\
16 & {\rm for}\ h_I=\frac12
\end{array} \right.
\ .
\eeqn
Again, $h_I$ and $g_I$ are 0 or $\pm\frac12$ for untwisted and
twisted sectors and insertions of reflections on the three $\mbb
T^2_I$, respectively. For $g=h=0$, one
again has to interpret these formulas
in the sense of the limit (\ref{lim}). 
Note that for the Klein bottle, the
presence of a reflection does not affect the partition function of
the internal string oscillators, since only $2g_I$ appears. The
reflection does, however, determine whether the spectrum of KK
states invariant under the insertion involves winding or momentum
modes (in the open channel).

Using the identity (\ref{simple}) one can evaluate the sum over
spin structures, and obtains the numerical coefficients
$\cq_{\sigma,\hat I}$ as defined in (\ref{spinstructuresums}). The
supersymmetric solution for the orientifold action
on the CP matrices was found in
\cite{Berkooz:1996dw}. It gives a maximal gauge group $Sp(8)^4$
of rank 32, each factor referring to one of the four types of D9-
or D5$_I$-branes. For the moment we just need the
properties\footnote{The matrices are called $M_i$ or $N_1$ in
\cite{Berkooz:1996dw} where they are defined in a table on page
16. See also equations (4.5) and (4.6).}
\beqn \label{gammaN=1}
\gamma_{\Omega \Theta_I 9} = - \gamma_{\Omega \Theta_I 9}^T \ ,
\quad
\gamma_{\Omega 5_I} = - \gamma_{\Omega 5_I}^T \ ,
\quad
\gamma_{\Omega \Theta_I 5_J} = - \gamma_{\Omega \Theta_I 5_J}^T ~~~{\rm for}~~~ {I\neq J}\ ,
\eeqn
and that $\gamma_{\Theta_I 5_J}$ is traceless. Thus, the traces in the M\"obius strip diagrams
behave like those for the matrices $\gamma_{\Omega\Theta i}$ or $\gamma_{\Omega a}$ in
(\ref{gammas}).

We now discuss the evaluation of all the diagrams of
(\ref{allN=1}) piece by piece. First note that the $\cn=4$
subsectors do not contribute because, just as in the $\cn=2$ case,
the relevant factors $\cq_{\sigma,\hat I}$ vanish as in table
\ref{tolletabelle}. In other words, there are four sets of vanishing amplitudes
\beqn
\ck^{(0)}_1 =\cm_9^{(0)} =\ca_{99}^{(0)} = 0 \ , \quad
\ck^{(I)}_1 =\cm_{5_I}^{(I)} =\ca_{5_I 5_I}^{(0)} = 0 \
.
\eeqn
These are just the diagrams of type I string theory
compactified on a torus, respectively its T-dual versions (with 4
T-dualities along $\mbb T^4_I$). Next there are three copies of
the $\cn=2$ diagrams computed in section
\ref{1loopsection}, one for each value of $I$,
\beqn \label{N=2inN=1}
\ck^{(0)}_{\Theta_I} + \ck^{(I)}_{\Theta_I} + \cm_9^{(I)} + \cm_{5_I}^{(0)}
 + \ca_{99}^{(I)} + \ca_{5_I 5_I}^{(I)} + \ca_{95_I}^{(0)} +
\ca_{5_I9}^{(0)} +\ca_{95_I}^{(I)} +
\ca_{5_I9}^{(I)} \ .
\eeqn
The amplitudes $\ca_{5_I 5_I}^{(I)}$ and $\ca_{95_I}^{(I)}$ actually
do not contribute due to the matrices $\gamma$
being traceless. Further, (\ref{gammaN=1}) implies
the cancellation of the Klein bottle diagrams with the 5-brane
M\"obius strip diagrams as in (\ref{sumE}) such that the final set
of amplitudes, that are the analogs of the $\cn=2$ model of the
last section, is
\beqn \label{finalampl}
\sum_{I=1}^3\Bigg[ \cm_9^{(I)} + \ca_{99}^{(I)} + \ca_{95_I}^{(0)}+ \ca_{5_I9}^{(0)}
\Bigg] \ .
\eeqn
They involve the dependence on the D9-brane Wilson line scalars
$A_i^I$, and are each formally identical to (\ref{contribu}) (after
cancellation of the Klein bottle and 5-brane M\"obius diagrams).
The relevant values for the coefficients $\cq_{\sigma,\hat I}$ can
be taken from table \ref{tolletabelle} and the 
matrices needed to evaluate the diagrams can be chosen as in
(\ref{gammas}).  The only difference compared to the case of
$\cn=2$ consists of the choice of the matrices $W_i$ that define
the Wilson lines. We will come to the explicit calculation of
(\ref{finalampl}) in the next section.

The extra diagrams in the $\cn = 1$ case are
\beqn \label{N=1dia}
\sum_{I=1}^3 \sum_{J\neq I} \Bigg[ \ck^{(I)}_{\Theta_J} + \cm_{5_J}^{(I)} + \ca_{5_J 5_J}^{(I)} +
\ca_{95_J}^{(I)} +
\ca_{5_J9}^{(I)} \Bigg] + \sum_{\hat I=0}^3 \sum_{J=1}^3 \sum_{K\neq
J} \ca_{5_J 5_K}^{(\hat I)}\  .
\eeqn
The annulus diagrams with non-trivial insertions among
(\ref{N=1dia}) immediately vanish due to the properties of the CP
matrices. They are proportional to the trace of a traceless
matrix $\gamma_{\Theta_I5_J}$.

This leaves us with the diagrams $\ca_{5_I5_J}^{(0)}$ and
the sum over terms $\ck^{(I)}_{\Theta_J} + \cm_{5_J}^{(I)}$
in (\ref{N=1dia}), which are independent of the Wilson line
moduli. Let us first deal with the latter two.
From the internal partition function (\ref{partN=1}) and (\ref{partN=1b}) it
follows that the numerical coefficients that appear are just equal to the
coefficients of the amplitudes
$\ck^{(0)}_{\Theta_I}+\ck^{(I)}_{\Theta_I}$ and $\cm_{5_I}^{(0)}$, explicitly
\beqn
\cq_{(\Theta_I),J} = \cq_{(\Theta_I),I} = \cq_{(\Theta_I),0} = 16\ , \quad
\cq_{(5_I),J} = \cq_{(5_I),0} = -4\ .
\eeqn
Also, the properties of the orientifold
matrices relevant for the
M\"obius diagrams are identical in both cases, cf.\
(\ref{gammaN=1}). The only real difference lies in the fact that
the spectrum of KK states that contribute in the traces in
$\ck^{(I)}_{\Theta_J} + \cm_{5_K}^{(I)}$, $I\neq J \neq K$,
involves winding states along the untwisted torus $\mbb T^2_J$ (in
the open string channel), as opposed to the momentum states that
contribute in
$\ck^{(0)}_{\Theta_J}+\ck^{(J)}_{\Theta_J}+\cm_{5_J}^{(0)}$. This
appears due to the fact that $\ck^{(I)}_{\Theta_J} +
\cm_{5_K}^{(I)}$, $I\neq J \neq K$, is just mapped to
$\ck^{(0)}_{\Theta_J} + \cm_{5_J}^{(0)}$ upon four T-dualities
along
 $\mbb T^4_I$.

To be more precise, a T-duality transformation along $\mbb T^4_I$,
i.e.\  the simultaneous inversion of all four radii of $\mbb
T^4_I$, maps, for example, $\Omega$ to $\Omega\Theta_I$, thus
permuting the insertions in the Klein bottle. It maps D5$_I$-branes
to D9-branes and permutes the other two types of D5-branes,
it maps
the 9-brane M\"obius strip with $\Omega$ insertion to the
5$_I$-brane M\"obius with $\Omega\Theta_I$ insertion, and so on.

As observed in (\ref{sumE}), contributions of the Klein bottle and
the 5-brane M\"obius strip cancel out according to (both equations
hold for each fixed value of $I$)
\beqn
\ck^{(0)}_{\Theta_I} + \ck^{(I)}_{\Theta_I} + \cm_{5_I}^{(0)} &=& 0\ , \\
\cm_{5_K}^{(L)}|_{L\neq I\neq K} + \sum_{J\neq I} \ck^{(J)}_{\Theta_I} &=& 0 \ . \nonumber
\eeqn
But note that this leaves us with one of the two diagrams
$\cm_{5_K}^{(L)}|_{L\neq I\neq K}$ which both have winding modes
along $\mbb T^2_I$. Together, this now means that the surviving
extra contributions from (\ref{N=1dia}) in $\mbb T^6/(\mbb
Z_2\times \mbb Z_2)$ are given by the following three sets of
diagrams,\footnote{Note that the diagrams (\ref{N=1contr}) do not
depend on the 9-brane scalars since they do not involve D9-branes
at all.}
\beqn \label{N=1contr}
\sum_{J=2}^3 \sum_{I<J} \Big[ \big[ \ca_{5_I5_J}^{(0)} +
\ca_{5_J5_I}^{(0)} \big] + \cm_{5_J}^{(I)} \Big] \ .
\eeqn
Any set of three is T-dual to $\ca_{95_I}^{(0)} + \ca_{5_I9}^{(0)}
+ \cm_{9}^{(I)}$ upon four T-dualities along $\mbb T^4_J$. This is
precisely the contribution that the $\ca_{95}$ and $\cm_9$
diagrams give in the $\cn=2$ model of section \ref{1loopsection}
(after setting $A_i^I$ to zero), and the result can be obtained
from the final result of that section via T-duality. It will be
given in section \ref{1loopkp}.

In order to evaluate (\ref{finalampl}) explicitly one needs to use
more details of the representation of the CP algebra given in
\cite{Berkooz:1996dw}, because these amplitudes depend on the
Wilson line moduli. We now return to a notation where the
D9-branes are broken up into stacks labelled by $i$, each
characterized by its individual value for the three Wilson lines
$A^I_i$. In conventions where the total number of branes is 32, a
minimum of 4 branes is required to make up an independent
stack that can be associated with a modulus $A^I_i$. This
corresponds to the fact that the maximal D9-brane gauge group of
the model is $Sp(8)$ with rank 8, i.e. $4N$ ``elementary''
D9-branes get identified under the orbifold and orientifold
projections to form $Sp(N)$. The D5-branes behave similarly.


\subsection{Wilson lines in $\mbb T^6/(\mbb Z_2\times \mbb Z_2)$}

The matrix $W_i^I$ for the Wilson line along either one of the two
elementary circles of the $\mbb T^2_I$, denoted by the basis
vectors $e^m_{\, M}$, $M=1,2$, is used to define the CP matrix
\be \label{gammawii}
\gamma_{W_i^I}^{(M)} =
\exp (2\pi i(\vec a^{\, I}_i\vec e_{\, M}) W_i^I)\ .
\ee
Its form is determined by solving a number of constraints
\cite{Cvetic:2000aq,Cvetic:2000st,Berg:2004ek}. It has to
satisfy tadpole cancellation conditions, obey
unitarity and be
compatible with the orbifold projection. Solutions to these
conditions are known explicitly only in a few cases, and we will
not go through the procedure in exhaustive detail. We use the
definitions of the matrices $\gamma_{\Theta_I i}$ in
\cite{Berkooz:1996dw} written in terms of $(4N_i) \times (4N_i)$
matrices,
\beqn
\gamma_{\Theta_1 i} =
\left[ \begin{array}{cc} 0 & - {\bf 1}_{2N_i} \\ {\bf 1}_{2N_i} & 0 \end{array}
\right] \oplus {\bf 0}_{32-4N_i} \ , \quad
\gamma_{\Theta_2 i} =
\left[ \begin{array}{cc} i\sigma_{2N_i} & 0 \\ 0 & -i\sigma_{2N_i} \end{array}
\right] \oplus {\bf 0}_{32-4N_i}
\ ,
\eeqn
and $\gamma_{\Theta_3 i} =  \gamma_{\Theta_2 i} \gamma_{\Theta_1
i}$,  using $\sigma_{2N_i}$ from (\ref{sigmas}). For concreteness
we now discuss the conditions and solutions for a Wilson line
along the second torus with $I=2$, but the other cases ($I=1,3$)
can be dealt with analogously. Now (skipping the index $M$ for the
moment) $\gamma_{W^2_i}$ has to satisfy the tadpole constraints
\beqn
{\rm tr}( \gamma_{\Theta_I i} \gamma_{W^2_i} ) = 0 \ , \quad I=1,2,3\ ,
\eeqn
and the compatibility relations
\beqn
( \gamma_{\Theta_1 i} \gamma_{W^2_i})^2 = ( \gamma_{\Theta_3 i}
\gamma_{W^2_i})^2  = -{\bf 1}_{4N_i} \oplus {\bf 0}_{32-4N_i} \ .
\eeqn
There is no such relation for $\gamma_{\Theta_2 i}$ which is
trivial on the second torus by definition. One can now easily
convince oneself that any matrix
\beqn \label{Z2wl}
\gamma_{W^2_i} =
{\rm diag}( e^{-i\varphi} {\bf 1}_{2N_i}, e^{i\varphi}{\bf 1}_{2N_i}) \oplus {\bf 0}_{32-4N_i}
\eeqn
satisfies these relations and is obviously also unitary on the
$(4N_i)\times (4N_i)$ block. In order to make contact to the
calculations of the model with $\cn=2$ supersymmetry, where we
used the diagonal CP matrices (\ref{gammas}) for $\gamma_{\Theta
i}$, we also diagonalize $\gamma_{\Theta_2 i}$ via the unitary
transformation
\beqn
P_4 = \frac{1}{\sqrt 2} \left[ \begin{array}{cccc}
0 & i & 0 & -i \\
0 & 1 & 0 & 1 \\
-i & 0 & i & 0 \\
1 & 0 & 1 & 0
\end{array}\right]
\eeqn
such that
\beqn\label{diagWL}
P_4^\dag \gamma_{\Theta_2 i} P_4 &=& {\rm diag} ( i {\bf
1}_{2N_i}, -i {\bf 1}_{2N_i}) \oplus {\bf 0}_{32-4N_i} \ ,
\non
P_4^\dag \gamma_{W^2_i}^{(M)} P_4 &=&  {\rm diag} ( e^{i\varphi_M}{\bf
1}_{N_i}, e^{-i\varphi_M}{\bf 1}_{N_i}, e^{i\varphi_M}{\bf
1}_{N_i}, e^{-i\varphi_M}{\bf 1}_{N_i} ) \oplus
{\bf 0}_{32-4N_i} \ .
\eeqn
These two matrices are all that is needed to evaluate the traces
that occur in the amplitudes with $I=2$ in (\ref{finalampl}). Note
that $P_4^\dag \gamma_{\Theta_2 i} P_4$ is now identical to
$\gamma_{\Theta i}$ in (\ref{gammas}). The continuous real
parameters $\varphi_M$ are interpreted geometrically as the Wilson
line degrees of freedom, by identifying them with the projection
of the vector $\vec a_i^{\, 2}$ that appears as the shift in the
open string KK momenta onto the two elementary lattice vectors of
the second torus $\mbb T^2_2$,
\beqn
\varphi_M = 2\pi\, \vec e_{\, M} \vec a^{\, 2}_i\ .
\eeqn
Using (\ref{gammawii}) this leads to
\beqn \label{WL2}
W^2_i = {\rm diag}( {\bf 1}_{N_i} , - {\bf 1}_{N_i} ,{\bf
1}_{N_i}, -{\bf 1}_{N_i}) \oplus {\bf 0}_{32-4N_i}\ .
\eeqn
By a unitary change of basis the other Wilson line matrices for
$I=1,3$ can also be brought to this form, so that we use the form
of (\ref{WL2}) for all the Wilson lines when evaluating the traces
in the next section. The two Wilson lines along the two elementary
cycles of the torus are independent, since the orbifold action
does not identify the two, and captured by the two independent
components of the vector $\vec a^{\, 2}_i$ in the dual lattice of
the torus. The special values $\vec e_{\, M} \vec a_i^{\, 2} =
\frac12\ {\rm mod}\ \mbb Z$ correspond to the positions of the
fixed points of the orbifold reflection in the dual lattice. For
these values the Wilson lines are of order 2, i.e.
$\gamma_{W_i^2}^2={\bf 1}_{4N_i}\oplus {\bf 0}_{32-4N_i}$, and
they commute with the projection of the orbifold group since then
$[\gamma_{W_i^2},\gamma_{\Theta_1
i}]=[\gamma_{W_i^2},\gamma_{\Theta_3 i}]=0$. These discrete Wilson
lines lead to points of enhanced symmetry.


\subsection{One-loop K\"ahler potential}
\label{1loopkp}

The full set of corrections is now obtained by adding the
diagrams (\ref{N=1contr}) and (\ref{finalampl}) which we still
have to evaluate.

In order to obtain the final expressions for the analogs of the
$\cn=2$ model (\ref{finalampl}) we use (three copies of) the
corrections calculated in section \ref{1loopsection} and only keep
track of the effect of the modified matrices $W_i^I$ (\ref{WL2})
in place of (\ref{Wmatrix}) whenever Wilson lines appear. These
matrices are not equivalent to the $\cn=2$ matrix in
(\ref{Wmatrix}) since they differ by relative signs within the
$(2N_i)\times (2N_i)$ blocks. One now has to evaluate the
expressions for the 2-point one-loop correlators $\langle
V_{T_2^I} V_{T_2^I} \rangle$, which are the analogs of
(\ref{2pts2}). Following the same steps as in section
\ref{1loopsection}, the correlators are given by a formula as in
(\ref{final}) corrected by an additional factor $\frac12$ for the normalization of the
orbifold projection (\ref{projz22}).

The final result is very similar, except that the trace over CP labels
vanishes in the case of the $99$ annulus $\ca_{ij}$, as follows from the form of
(\ref{WL2}). The evaluation of the
traces produces the result
\beqn\label{sumEZ2}
\sum_{\sigma} c^2_\sigma \sum_{k=0,1} \tr \Big[
E_2({\bf A}^I_\sigma,U^I) \gamma_{\sigma,\hat I} \cq_{\sigma,\hat I} \Big]
&=&
4 \cdot 32 \sum_i N_i \big[  E_2(A^I_i,U^I) + E_2(-A^I_i,U^I)\big] \non
&&\hspace{-2cm}
-~ 8 \sum_i N_i \big[  E_2(2A^I_i,U^I) + E_2(-2A^I_i,U^I)\big] \ ,
\eeqn
where the factors changed as compared to (\ref{sumE}) because now $\sum_i N_i =8$.
We define the quantity
\beqn
\ce^{\mbb Z_2^2}_2(A^I_i,U^I) = \sum_{\sigma} c_\sigma^2
 \sum_{k=0,1} \tr \Big[
E_2({\bf A}^I_\sigma,U^I) \gamma_{\sigma,\hat I} \cq_{\sigma,\hat I} \Big] \ ,
\eeqn
to abbreviate the sum above.

Finally, we have to add the diagrams from (\ref{N=1contr}) which
we already identified as being T-dual to the contributions above
at $A_i^I=0$. Thus, we only have to perform this T-duality on the
final result. To do so, note that the function $E_2(A,U)$ that
appears in (\ref{sumE}) and (\ref{sumEZ2}) is invariant under
T-duality at $A=0$, as we show in appendix \ref{appe2}. Therefore,
the moduli dependence with respect to the $U^I$ appears only
through the function $\ce^{\mbb Z_2^2}_2(0,U^I)$. The denominator
$((S-\bar S)(T^I_0-\bar T_0^I))^{-1}$ from (\ref{tollesK})
transforms under a T-duality along a $\mbb T^4_J$, $J\neq I$, as
\beqn
\frac{1}{(S-\bar S)(T^I_0-\bar T_0^I)} ~\longrightarrow~
 \frac{1}{(T^J_0-\bar T_0^J)(T^K_0-\bar T_0^K)}\Big|_{K\neq I\neq J}\
.
\eeqn
Putting the pieces together, the total one-loop correction to the
K\"ahler potential of $\mbb T^6/(\mbb Z_2\times \mbb Z_2)$ can be
written as (valid up to order ${\co}(e^{2 \Phi})$)
\beqn \label{tollesK2}
K^{(1)}_{\mathbb{Z}_2^2} = \frac12 \sum_{I=1}^3 \frac{c\, \ce^{\mbb Z_2^2}_2(A_i^I,U^I)}{(S-\bar S)(T^I-\bar T^I)}
+ \frac12 \sum_{I=1}^3 \frac{c\, \ce^{\mbb
Z_2^2}_2(0,U^I)}{(T^J-\bar T^J)(T^K-\bar T^K)}\Big|_{K\neq
I\neq J}\ .
\eeqn
The constant of proportionality $c$ is the same as for the corrections
that appeared in the $\cn=2$ model in (\ref{tollesK}) and
given in (\ref{tollesc}).

As explained earlier, there is now no reason to expect that one can
absorb this term into the logarithm of $K^{(0)}$ as in the case of $\cn=2$
supersymmetry, since all higher terms in its expansion would
be subject to higher order perturbative corrections.

We will be content with verifying only the component
$K_{T^I\bar T^I}$ of the K\"ahler metric and not all other
correlators. We expect that everything will go through 
analogously to the case of $\cn=2$ and one can confirm
that all 2-point functions are reproduced by the K\"ahler
potential that combines out of (\ref{kaepotN=1tree}) and
(\ref{tollesK2}) up to the relevant order in the string coupling.


\section{The $\cn =1$ orientifold $\mbb T^6/\mathbb{Z}_6'$}
\label{z6}

The orbifold generator of the orbifold $\mbb T^6/\mathbb{Z}_6'$
model is defined via the vector
\be
\vec{v}= \Big( {1 \over 6}, -{1 \over 2}, {1\over 3}
\Big) \; .
\ee
Since $\Theta^3$ is of order 2, we have D5-branes wrapping the
third torus at its fixed points. The maximal gauge group is
$(U(4)^2\times U(8))_{\rm D9}\times (U(4)^2 \times U(8))_{\rm
D5}$. The torus lattices of the first and third $\mathbb T^2$ must
be invariant under $\Theta$, which fixes the complex structures
$U^1$ and $U^3$, but there are a number of possible choices. This
leaves only $U^2$ as a modulus. In addition, there are three
generic untwisted K\"ahler moduli $T^I$ and $S$, plus twisted
scalars that we do not consider. As for the open string moduli, we
consider the Wilson lines $A_i^I$ of the D9-branes, and set the
coordinates of the D5-branes and their Wilson lines to zero. We
will only present the one-loop correction to the K\"ahler
potential and not discuss the model in full detail, referring 
to \cite{Lust:2005dy} for a thorough discussion of the tree level
K\"ahler potential and coordinates. The form of the untwisted
K\"ahler moduli at leading order is the same as in the case of
$\mbb T^6/(\mathbb{Z}_2\times\mathbb{Z}_2)$ (up to numerical
factors), i.e.\
\be
{\rm Im}(T^I_0) ~\sim~ e^{-\Phi} \sqrt{G^I}\ .
\ee
This is all we need to know to write down a vertex
operator for $T_2^I$ in analogy with the first line of (\ref{vops}).
The complexified Wilson line moduli are taken to be
the same as above (cf.\ equation
(\ref{scalarsz2})) even though 
only $U^2$ is still a modulus. (As it turns out, 
the following calculation does not make use of the exact 
definition of the Wilson line moduli in this case).

Since we discussed K\"ahler potential loop corrections at length
in previous sections, we can already anticipate the form of the
expected correction in this model. This model now contains
also $\cn=1$ sectors, whose only moduli dependence comes through 
the vertex operators (\ref{vops}) and the Weyl rescaling (\ref{Weyl}).
We leave them to future work \cite{gg3} and here we
only focus on the 
contributions from $\cn = 2$ sectors.
For these contributions, we  expect the one-loop
correction to the K\"ahler potential to be of the form
\be
K^{(1)}_{{\mathbb Z}_6'} ={1 \over 3} \,
{c_1 \ce_2^{(2,P)}(A^2_i,U^2) \over (S-\bar{S})(T^2_0-\bar{T}_0^2) } + 
{1 \over 3} \,
{c_2 \ce_2^{(2,W)}(0,U^2) \over (T_0^1-\bar{T}_0^1)(T^3_0-\bar{T}_0^3) } +
{1 \over 3} \, {c_3 \ce_2^{(3)}(A^3_i,U^3) \over
(S-\bar{S})(T^3_0-\bar{T}_0^3)} + \ldots \ ,
\label{K1z6}
\ee
where the ellipsis stand for the $\cn = 1$ sectors 
(we will make a comment about them at the end of this section) 
and the factors of $\frac13$ come from the orbifold projection. 
The superscripts $P$ and $W$ indicate that the corresponding 
zero mode sums are from momentum and winding states (in the open string channel), 
respectively, as will become clearer at the end of the section. As before, in (\ref{K1z6}) 
we already set to zero the 5-brane scalars, that would otherwise have appeared in the 
first argument of $\ce_2^{(2,W)}(0,U^2)$. Let
us now further restrict attention to Wilson lines $A_i^3$ along the third
two-torus, that we will denote simply as $A_i$ in this section.
Here we  determine the form of $\ce_2^{(2,P)}(0,U^2)$, $\ce_2^{(2,W)}(0,U^2)$ and
$\ce_2^{(3)}(A_i,U^3)$, but leave the constants
$c_i$ and the $\cn=1$ sectors for the future \cite{gg3}.

We will only  explain
a few ingredients of this computation, following the steps of the
previous sections.
The mother of all inventions is indolence, so
we would like to find a means of reducing
calculations in this model
to the K3$ \times {\mathbb T^2}$ case as much as possible.To do so 
we first have to introduce some new notation. Define an "untwisted
torus twist-vector component" $v_{\rm untw}=v_2$ for $k=2,4$,
$v_{\rm untw}=v_3$ for $k=3$, and similarly
$v_{\rm tw}=v_3$ for $k=2,4$,
$v_{\rm tw}=v_2$ for $k=3$. Then,
since $kv_{\rm untw}$ is integer,
\beqn
{\thbw{\alpha}{\beta+kv_{\rm untw}} \over \eta^3}
{\thbw{\alpha+h}{\beta+kv_1}
\thbw{\alpha+h}{\beta+kv_{\rm tw}} \over
\thbw{1/2+h}{1/2+kv_1}
\thbw{1/2+h}{1/2+kv_{\rm tw}} }
&=& (-1)^{kv_{\rm untw}}
{\thbw{\alpha}{\beta} \over \eta^3}
{\thbw{\alpha+h}{\beta+kv_1}
\thbw{\alpha-h}{\beta-kv_1} \over
\thbw{1/2+h}{1/2+kv_1}
\thbw{1/2-h}{1/2-kv_1} }
\eeqn
where we used $kv_{\rm tw} = -k(v_1+v_{\rm untw})$.
The phase on the right hand side
is independent of the spin structure.
At most it gives an overall sign
\beqn
(-1)^{k v_{\rm untw}} = \left\{
\begin{array}{rl} -1 \quad& k=2,3 \\ 1 \quad & k=4 \end{array}
\right.\ .
\eeqn
The internal partition function
in $\cn=2$ sectors is then
reduced to K3$ \times {\mathbb T}^2$
form up to overall factors
\beqn
\label{partz6}
\cz_{\sigma,k}^{\rm int} \zba{\alpha}{\beta} &=&
 \eta_{\alpha\beta} \frac{\thba{\alpha}{\beta}(0,\tau)}{\eta^3(\tau)}
\frac{\thba{\alpha+h}{\beta+g}(0,\tau)\thba{\alpha-h}{\beta-g}(0,\tau)}{\thba{1/2+h}{1/2+g}(0,\tau)\thba{1/2-h}{1/2-g}(0,\tau)}
\non
&&
\hspace{4cm}
\times (-1)^{k v_{\rm untw}} \times \left\{
\begin{array}{ll}
 a_k \quad & {\rm for}\ h=0 \\
   1  \quad &{\rm for}\ h=\frac12  \end{array} \right.\ ,
\eeqn
where $h$ is as before, while $g=kv_1$ whenever there is a
$\Theta$ acting on the world-sheet oscillators in the trace. The
trigonometric factors $a_k$ can be read off from appendix
\ref{app:partition} and have been evaluated to
\beqn
a_k = \left\{ \begin{array}{ll}
              3  & k=2 \\  4 & k=3 \\ -3 & k=4
               \end{array} \right. \; .
\eeqn
To perform the sum over spin structures we deal with the first two 
and the third terms in the K\"ahler potential (\ref{K1z6}) separately.
Let us confront the more involved third term first.

The vertex operators in $\langle V_{T_2^3} V_{T_2^3}
\rangle$ are polarized along the third torus. The moduli-dependent
part of the third term in (\ref{K1z6}) then only receives
contributions from the insertions $k=0,3$. For example, using
(\ref{corrperiodicity}) we see that for $\ca_{99}^{(3)}$,
$\cq_{\sigma,k}=a_k(-1)^{kv_{\rm untw}}=-4$. All results we need
are listed in table \ref{tolletabelle2}, in particular all the $\cn=2$, 
$k=2,4$ sector amplitudes vanish (for the 2-point function of $V_{T_2^3}$; this will be 
different for the 2-point function of $V_{T_2^2}$ discussed in a moment).

For the following details on the Chan-Paton factors we refer to
\cite{Aldazabal:1998mr,Berg:2004ek}. The action of $\Theta$ is
\beqn
\gamma_{\Theta i} =
{\rm diag}( \beta {\bf 1}_{4-N_i},\beta^5 {\bf 1}_{4-N_i}, \beta^9
{\bf 1}_{8-N_i},
\bar \beta {\bf 1}_{4-N_i},\bar \beta^5 {\bf 1}_{4-N_i}, \bar \beta^9
{\bf 1}_{8-N_i}, \gamma^{[6N_i]}_{\Theta i})
\eeqn
with
\beqn
\gamma^{[6N_i]}_{\Theta i} =
{\rm diag}( \beta ,\beta^5 , \beta^9 ,
\bar \beta ,\bar \beta^5 , \bar \beta^9 ) \otimes {\bf 1}_{N_i} =
\gamma^{[6]}_{\Theta i} \otimes {\bf 1}_{N_i}
\eeqn
and $\beta=e^{i\pi /6}$.
The remaining gauge group is
\beqn
U(4-N_i)^2 \times U(8-N_i) \times U(N_i)
 \; ,
\eeqn
and tadpole cancellation requires $\sum_i N_i = 4$. In the $k=3$
sector, the action of the orbifold twist on the Chan-Paton factors
is particularly simple
\beqn
(\gamma^{[6]}_{\Theta i})^3={\rm diag}(i {\bf 1}_{3},-i {\bf 1}_{3}) \; .
\eeqn
There is a convenient basis where the Wilson line on the D9-branes
is diagonal,
\beqn \label{w6}
 \gamma_{W_i}  & = &
{\rm diag}( {\bf 1}_{32-6N_i} , \gamma^{[6N_i]}_{W_i})\ , \non
\gamma^{[6N_i]}_{W_i} &=& {\rm diag}
(e^{i \vec e\, \vec a},e^{i \vec e\, \vec a^\Theta},
e^{i \vec e\, \vec a^{\Theta^2}}, e^{-i \vec e\, \vec a},
e^{-i \vec e\, \vec a^\Theta},
e^{-i \vec e\, \vec a^{\Theta^2}}) \otimes {\bf 1}_{N_i}\ .
\eeqn
This is very similar to the matrix (\ref{Wmatrix}) but one has to
sum over images of $\Theta$ in addition. Just like in the K3$
\times {\mathbb T}^2$ case, all other matrices with $\cq_{\sigma,k}\neq
0$ are proportional to identity matrices.
\begin{table}
\beqn
\begin{array}{|c|l|c|c|c|}
\hline
 & {\rm Sector} & k & \cq_{\sigma,k} &
 \gamma_{\sigma,k} \\
\hline
\hline
\ck      & \sigma = (1) &  0,1,2,3,4,5 &  0 &\\
         & \sigma = (\Theta^3) &  0,3^{\spadesuit}
         &  16 & \\
\hline
\ca_{99} & \sigma = (ij) & 0 &  0
         &  \gamma_{i} \otimes \gamma^{-1}_{j} \\
         & \sigma = (ij) & 2,3,4
         &  0,-4,0
         &  \gamma_{\Theta i}^k \otimes \gamma^{-k}_{\Theta j} \\
\ca_{55} & \sigma = (ab) & 0 &  0
         &  \gamma_{a} \otimes \gamma^{-1}_{b} \\
         & \sigma = (ab) & 2,3,4 & 0,-4,0
         & \gamma_{\Theta a}^k \otimes \gamma^{-k}_{\Theta b} \\
\ca_{95} & \sigma = (ia) & 0 &  1
         &  \gamma_{i} \otimes \gamma^{-1}_{a} \\
         & \sigma = (ia) & \;\, 3^{\spadesuit} &  -1
         & \gamma_{\Theta i}^k \otimes \gamma^{-k}_{\Theta a} \\
\hline
\cm_9    & \sigma = (i) & 0 &  0
         &  -\gamma^{T}_{\Omega i}\gamma^{-1}_{\Omega i} \\
         & \sigma = (i) & 2,3,4 & 0,-4,0
         &  -\gamma^{Tk}_{\Omega\Theta i}\gamma^{-k}_{\Omega\Theta i} \\
\cm_5    & \sigma = (a) & 0 &  -4
         &  -\gamma^{T}_{\Omega a}\gamma^{-1}_{\Omega a} \\
         & \sigma = (a) & 2,3,4 &  0 ,0,0
         &  -\gamma^{Tk}_{\Omega\Theta  a}\gamma^{-k}_{\Omega\Theta  a} \\
\hline
\end{array}
\nonumber
\eeqn
\caption{$\cn=2$ sector amplitudes in ${\mathbb Z}_6'$
for $\langle V_{T_2^3} V_{T_2^3} \rangle $. The $\spadesuit$
remind us that $\ca_{95}^{(k=2,4)}, \ck_{\Theta^3}^{(k=2,4)}$ are
$\cn=1$, so they are not included.} \label{tolletabelle2}
\end{table}
Then the only non-vanishing moduli-dependent contributions come
from diagrams
\beqn
\ck_{\Theta^3}^{(0)} + \ck_{\Theta^3}^{(3)} + \ca_{99}^{(3)} +
\ca_{95}^{(0)} + \ca_{59}^{(0)} + \cm_9^{(3)} + \cm_5^{(0)}\ .
\eeqn
The only dependence on the Wilson lines along the third two-torus
comes from $\cn=2$ sectors
\be
\ca_{99}^{(3)}+\ca_{95}^{(0)}+\ca_{59}^{(0)}+
\cm_{9}^{(3)} \; .
\ee
Now we can follow the steps
in section \ref{z2}, in particular
going from (\ref{contribu})
to (\ref{sumE}). This leads to
\beqn \label{sumEz61}
\sum_{\sigma} c_\sigma^2
 \sum_{k=0,3} \tr \Big[
E_2({\bf A}_\sigma,U^3) \gamma_{\sigma,k} \cq_{\sigma,k} \Big]
&=&
4 E_2(0,U^3) \Big[ \cq_{(\Theta^3),0} + \cq_{(\Theta^3),3} \Big] \non
&& \hspace{-5cm}
+
\sum_{i,j} \tr \Big[ E_2({\bf A}_{(ij)},U^3) \gamma_{(ij),3} \cq_{(ij),3} \Big] +
2 \sum_{i,a} \tr \Big[ E_2({\bf A}_{(i)},U^3) \gamma_{(ia),0} \cq_{(ia),0} \Big] \non
&& \hspace{-5cm} +
\sum_{i} \tr \Big[ E_2({\bf A}_{(i)},U^3) \gamma_{(i),3} \cq_{(i),3} \Big] +
\sum_{a} \tr \Big[ E_2(0,U^3) \gamma_{(a),0} \cq_{(a),0} \Big]
\non
&& \hspace{-5.5cm} =
 \sum_{m=0}^2\Bigg\{
\sum_{n=0}^2\sum_{i,j} \Big[-4 N_iN_j \Big[  E_2(A_i^{\Theta_m} - A_j^{\Theta_n},U^3) +
     E_2(-A_i^{\Theta_m} + A_j^{\Theta_n},U^3) \nonumber \\[-.3cm]
&& \hspace{-2cm}
    -  E_2(A_i^{\Theta_m} + A_j^{\Theta_n},U^3)
    - E_2(-A_i^{\Theta_m} - A_j^{\Theta_n},U^3) \Big] \Big] \nonumber \\[.3cm]
&& \hspace{-2.6cm}
+~ 2\cdot 32 \sum_i N_i \big[  E_2(A_i^{\Theta_m},U^3) + E_2(-A_i^{\Theta_m},U^3)\big] \non
&& \hspace{-2.6cm}
-~ 4\sum_i N_i \big[  E_2(2A_i^{\Theta_m},U^3) +
E_2(-2A_i^{\Theta_m},U^3)\big]  \Bigg\}
\ .
\eeqn
We denote this quantity as $\ce_2^{(3)}(A_i,U^3)$ as introduced in
(\ref{K1z6}). As emphasized above, this complex structure $U^3$ is
no longer a modulus, since it is fixed by the action of the
orbifold.

The functions $\ce_2^{(2, P)}(0,U^2)$ and $\ce_2^{(2, W)}(0,U^2)$ in (\ref{K1z6}) 
can be determined by 
considering vertex operators polarized along the second torus in
 $\langle V_{T_2^2} V_{T_2^2}  \rangle $.
Part of the calculation was already done in
\cite{Bain:2000fb} since we have not turned on Wilson lines
along this two-torus. The function $\ce_2^{(2, P)}(0,U^2)$ gets contributions 
from  
\beqn \label{moment}
\ck_{1}^{(2,4)} + \ca_{99}^{(2,4)} + \cm_9^{(2,4)} \ ,
\eeqn
whereas $\ce_2^{(2, W)}(0,U^2)$ gets contributions from
\beqn \label{wind}
\ck_{1}^{(1,5)} + \ca_{55}^{(2,4)} + \cm_5^{(2,4)} \ .
\eeqn
Note that the diagrams in (\ref{moment}) involve a momentum sum 
along the second torus, while the diagrams in (\ref{wind}) involve a sum 
over winding states along the second torus (both in the open channel).
For the Klein bottle this is clear from the appendix of \cite{Aldazabal:1998mr}.
We do not go through the details of the calculation again. As in section 
(\ref{1loopsection}), the diagrams with the momentum sum lead to a term in the
K\"ahler potential proportional to
\be
\frac{E_2(0,U^2)}{(S-\bar S) (T^2 - \bar T^2)}\ ,
\ee
whereas the different volume dependence in the amplitudes with 
winding sums change the volume dependent factor to\footnote{The dependence 
on $U^2$ does not change as can be seen again from the $SL(2,\mbb Z)$-invariance
of $E_2(0,U^2)$, cf.\ appendix \ref{appe2}.} 
\be
\frac{E_2(0,U^2)}{(T^1 - \bar T^1) (T^3 - \bar T^3)}\ .
\ee
This is analogous to the second term in (\ref{tollesK2}), which also comes from 
terms involving winding sums as opposed to momentum sums. Thus, both 
$\ce_2^{(2, P)}(0,U^2)$ and $\ce_2^{(2, W)}(0,U^2)$ are proportional to 
$E_2(0,U^2)$ and the factor of proportionality is given by a trace similar to 
(\ref{sumEz61}). Since we did not determine the constants $c_1,c_2,c_3$ in (\ref{K1z6}), 
we do not give the proportionality factor here, but we hope to come back to a more 
complete study of this model in the future \cite{gg3}. 

Let us just make one final comment on the $\cn=1$ sectors. One can infer from the 
2-point functions of the $T^I$, for instance, that there may be contributions to the K\"ahler potential 
proportional to
\be \label{n1corr}
\frac{C}{\sqrt{(S-\bar S) (T^1 - \bar T^1) (T^2 - \bar T^2) (T^3 - \bar T^3)}}\ ,
\ee
for some constant $C$. 
This can be seen as follows. Let us choose the 2-point function of $T^1$ to be concrete 
(the other cases $I=2,3$ are analogous). Its moduli dependence comes only from the 
vertex operators (giving a factor $(T^1 - \bar T^1)^{-2}$) and the Weyl rescaling (cf.\ (\ref{Weyl})), 
together giving a factor 
\be
\frac{1}{(T^1 - \bar T^1)^2}\frac{e^\Phi}{S-\bar S} \sim \frac{1}{\sqrt{(S-\bar S) (T^1 - \bar T^1)^5 (T^2 - \bar T^2) (T^3 - \bar T^3)}}\ .
\ee
Integrating this twice with respect to $T^1$ and $\bar T^1$ would 
lead to a scaling behavior as in (\ref{n1corr}). However, if the $\cn=1$ sectors 
only produce IR-divergent 
terms (cf.\ \cite{Bain:2000fb}), 
these would not be included in the K\"ahler potential. 


\section{Conclusions}
\label{conclusions}

We would like to conclude with two comments. The first concerns the translation
of our results to the language of D3- and D7-branes, which is more useful
for most phenomenological applications.
To do so, one performs six T-dualities along all internal circles.
This maps D9- to D3-branes, and it maps
D5-branes wrapping some $\mbb T^2$ onto D7-branes on
the transverse $\mbb T^4$. Further, world-sheet parity
$\Omega$ maps to $\Omega(-1)^6(-1)^{F_R}$, where
$(-1)^6$ is the reflection
along all six circles and $F_R$ the right-moving space-time
fermion number. This operation is obviously not a symmetry of the
models we studied in this paper, 
but rather maps an orientifold of one formulation
into another orientifold of the other formulation, the two being
physically identical at the orbifold point.

Thus we can just copy the results of this paper, i.e.\
(\ref{tollesK}) for $\mbb T^4/\mbb Z_2\times \mbb T^2$,
(\ref{tollesK2}) for $\mbb T^6/(\mbb Z_2\times
\mbb Z_2)$ and (\ref{K1z6}) for $\mbb T^6/\mbb Z_6'$,
if we take into account that they now depend on the T-dual
variables.\footnote{The form invariance of $E_2(A,U)$ under
$SL(2,\mbb Z)$-transformations is demonstrated in appendix
\ref{appe2}.} For the $\cn=2$ model of section \ref{z2} these are
given by the T-dual complex structure modulus, the positions of
the D3-branes (cf.\ (\ref{trafo}) and (\ref{tdualA}),
respectively, for $A=D=0$ and $B=-C=1$) and
\beqn
S&\mapsto&  \frac{1}{\sqrt{8\pi^2}} ( C_0+ie^{-\Phi} )\ , \\
S'&\mapsto& \frac{1}{\sqrt{8\pi^2}} ( C_4|_{\mbb T^4} + ie^{-\Phi} \cv_{\rm K3}) + \frac{1}{8 \pi}
\sum_i N_i (U (a^5)_i + (a^4)_i (a^5)_i) \ . \nonumber
\eeqn
Thus, whereas $S$ becomes independent of the volume, the leading
term in the imaginary part of $S'$ becomes the volume of a
4-cycle, as measured in the ten-dimensional Einstein frame metric,
$e^{-\Phi} \cv^{({\rm string})}_{\rm K3}=\cv^{({\rm Einstein})}_{\rm
K3}$. The mapping in the case of $\cn =1$ orientifolds is
analogous. This implies, for instance, that the first correction
term in (\ref{tollesK2}) scales with the total Einstein-frame
volume $\cv^{({\rm Einstein})}$ like
\beqn
\frac{1}{(S-\bar S)(T^I_0-\bar T_0^I)} ~\sim~ e^{\Phi} [\cv^{({\rm Einstein})}]^{-\frac23}\; ,
\eeqn
where we assumed that none of the three 4-cycles degenerates. 
Compared to the tree level $\alpha'$ corrections that were determined in
\cite{Becker:2002nn}, this term dominates at large volume.\footnote{A similar 
scaling of the one-loop correction to the K\"ahler potential in models with D3/D7-branes 
was reported in \cite{Berglund:2005dm}.} We will pursue this issue further in our companion
paper \cite{gg2}.

Moreover, in the T-dual coordinates, the corrections depend on the 
scalars parame\-trizing the 3-brane positions and should be relevant 
for brane inflation models based on mobile D3-branes.  

Finally, we would like to remark that a different (closed string)
approach to corrections to K\"ahler potentials in (warped) string
compactifications has been pursued in \cite{Giddings:2005ff}. It
would be interesting to explore if their results are
related to ours, or if they capture a complementary type of
corrections that one would have to take into account in addition
to ours, in applications to more realistic warped
compactifications.

\vspace{1cm}


\begin{center}
{\bf Acknowledgements}
\end{center}
\vspace{-.3cm}

We would like to thank Adebisi Agboola,
Ignatios Antoniadis, Costas Bachas, David
Berenstein, 
Paul Garrett, Steve Giddings, Thomas Grimm, Arthur Hebecker, Chris
Herzog, Hans Jockers, Jan Louis, Thomas Mohaupt, Joe Polchinski,
Henning Samtleben, Stephan Stieberger, and Tom Taylor for 
plentiful and 
helpful advice and inspiring questions, during discussions and
email correspendence. M.B.\ was supported by the Wenner-Gren
Foundations, and M.H.\ by the German Science Foundation (DFG).
Moreover, the research of M.B.\ and M.H.\ was supported in part by
the National Science Foundation under Grant No. PHY99-07949. The
work of B.~K.~was supported by the DFG, the DAAD, and the European
RTN Program MRTN-CT-2004-503369, and in part by funds provided by
the U.S. Department of Energy (D.O.E.) under cooperative research
agreement $\#$DF-FC02-94ER40818.


\clearpage
\begin{appendix}
\section{Amplitude toolbox}
\label{app:toolbox}

\subsection{Partition functions}
\label{app:partition}

The zero modes and oscillators of the external space time
coordinates and ghosts give the same contribution for all
compactifications considered in this paper:
\beqn
\cz_{4}\zba{\alpha}{\beta}(\tau) = \frac{1}{4\pi^4\tau^2}
\frac{\vartheta[{\alpha \atop \beta}](0,\tau) }{\eta(\tau)^3}\ .
\eeqn
For the $\mbb T^6/{\mathbb Z}_6^\prime$ model, the internal D9-D9
annulus partition functions are:
\beqn
\cz^{\rm int}_{k}\zba{\alpha}{\beta}(\tau) &=&
\eta_{\alpha \beta}\prod_{j=1}^{3} (-2\sin ({\pi k v_j})) \;
\frac{\vartheta[{\alpha \atop {\beta + k v_j}}](0,\tau)}
     {\vartheta[{1/2 \atop {1/2 + k v_j}}](0,\tau) }, \ \ k=1,5 \ , \non
\cz^{\rm int}_{k}\zba{\alpha}{\beta}(\tau) &=&
\eta_{\alpha \beta} \, \frac{\vartheta[{\alpha \atop {\beta + kv_2}}](0,\tau)}
     {\eta(\tau)^3}  \;
\prod_{j=1,3} (2\sin ({\pi k v_j})) \;
\frac{\vartheta[{\alpha \atop {\beta + k v_j}}](0,\tau)}
     {\vartheta[{1/2 \atop {1/2 + k v_j}}](0,\tau) }, \ \ k=2,4 \ , \non
\cz^{\rm int}_{k}\zba{\alpha}{\beta}(\tau) &=&
\eta_{\alpha \beta} \, \frac{\vartheta[{\alpha \atop {\beta + 3v_3}}](0,\tau)}
     {\eta(\tau)^3}  \;
\prod_{j=1,2} (2\sin ({3\pi v_j})) \;
\frac{\vartheta[{\alpha \atop {\beta + 3v_j}}](0,\tau)}
     {\vartheta[{1/2 \atop {1/2 + 3v_j}}](0,\tau) }, \ \ k=3 \; .
\eeqn
The internal D9-D5 annulus partition functions are:
\beqn
\cz^{\rm int}_{k}\zba{\alpha}{\beta}(\tau) \!&=&\!
\eta_{\alpha \beta} \,  (-2\sin ({\pi k v_3})) \;
\frac{\vartheta[{\alpha \atop {\beta + 3v_3}}](0,\tau)}
{\vartheta[{1/2 \atop {1/2 + 3v_3}}](0,\tau)  }
\prod_{j=1}^{2}\frac{\vartheta[{\alpha+1/2 \atop {\beta + k v_j}}](0,\tau)}
     {\vartheta[{1/2+1/2 \atop {1/2 + k v_j}}](0,\tau) }, \ \ k=1,2,4,5\ ,  \non
\cz^{\rm int}_{k}\zba{\alpha}{\beta}(\tau) &=&
\eta_{\alpha \beta} \, \frac{\vartheta[{\alpha \atop {\beta + kv_3}}](0,\tau)}
     {\eta(\tau)^3}  \;
\prod_{j=1,2}  \;
\frac{\vartheta[{\alpha+1/2 \atop {\beta + kv_j}}](0,\tau)}
     {\vartheta[{1/2+1/2 \atop {1/2 + kv_j}}](0,\tau) }, \ \ k=0,3 \ .
\eeqn
The internal D9 M\"obius strip partition functions are:
\beqn
\cz^{\rm int}_{k}\zba{\alpha}{\beta}(\tau) &=&
\eta_{\alpha \beta} \, \prod_{j=1}^{3} (-2\sin ({\pi k v_j})) \;
\frac{\vartheta[{\alpha \atop {\beta + k v_j}}](0,\tau)}
     {\vartheta[{1/2 \atop {1/2 + k v_j}}](0,\tau) }, \ \ k=1,5\ ,  \non
\cz^{\rm int}_{k}\zba{\alpha}{\beta}(\tau) &=&
\eta_{\alpha \beta} \, \frac{\vartheta[{\alpha \atop {\beta + kv_2}}](0,\tau)}
     {\eta(\tau)^3}  \;
\prod_{j=1,3} (2\sin ({\pi k v_j})) \;
\frac{\vartheta[{\alpha \atop {\beta + k v_j}}](0,\tau)}
     {\vartheta[{1/2 \atop {1/2 + k v_j}}](0,\tau) }, \ \ k=2,4\ ,  \non
\cz^{\rm int}_{k}\zba{\alpha}{\beta}(\tau) &=&
\eta_{\alpha \beta} \, \frac{\vartheta[{\alpha \atop {\beta + 3v_3}}](0,\tau)}
     {\eta(\tau)^3}  \;
\prod_{j=1,2} (2\sin ({3\pi v_j})) \;
\frac{\vartheta[{\alpha \atop {\beta + 3v_j}}](0,\tau)}
     {\vartheta[{1/2 \atop {1/2 + 3v_j}}](0,\tau) }, \ \ k=3\ .
\eeqn
%
%
The arguments are $\tau=1/2+it/2$ for the M\"obius strip and $\tau = it$ for the annulus
(see fig.\ \ref{fig:surfaces}). The above amplitudes can all be
read off from Appendix B of \cite{Antoniadis:1999ge}. For the internal partition functions 
of the D5-D5 annulus and the D5 M\"obius strip we refer the reader to the appendix 
of \cite{Aldazabal:1998mr}, where also the 
Klein bottle partition functions are given as
\beqn
\cz^{{\rm int}}_{(1),k}\zba{\alpha}{\beta}&=&
\prod_{i=1}^3{ \thbw{\alpha}{\beta+2 kv_i}(0,\tau) (-2 \sin (2\pi kv_i))
\over \thbw{1/2}{1/2+2 kv_i}(0,\tau) }\ ,  \\
\cz^{{\rm int}}_{(\Theta^3),k}\zba{\alpha}{\beta} &=&
\tilde{\chi}(\Theta^{3},\Theta^k)
\left(
\prod_{i=1}^{2}{
\thbw{\alpha+1/2}{\beta+2 kv_i}(0,\tau)
\over
\thbw{0}{1/2+2 kv_i}(0,\tau)}
\right)
{\thbw{\alpha}{\beta+2kv_3}(0,\tau)(-2 \sin (2\pi kv_3))
\over \thbw{1/2}{1/2+2 kv_3}(0,\tau) } \; ,
\nonumber
\eeqn
where the second argument of the theta functions is $\tau=2it$, as
usual for the Klein bottle, and the number of simultaneous fixed
points of $\Theta^3$ and $\Theta^k$ is
\be
\tilde{\chi}(\Theta^{3},\Theta^k)
=\left\{
\begin{array}{ll}
12 \quad & {k=1,5}  \\ 12 & k=2,4 \\
16 & k=3
\end{array} \right.\; .
\ee

\subsection{world-sheet correlators}

The bosonic correlation function on the torus ${\cal T}$ in the
untwisted directions is
\beqn
\langle X(\nu_1) X(\nu_2)\rangle _{{\cal T}} =
-\frac{\alpha'}{2} \ln \Big|
   \frac{2 \pi}{\vartheta_1'(0,\tau)} \vartheta_1\Big(\frac{\nu_1-\nu_2}{2 \pi},\tau \Big) \Big|^2
 + \alpha'  \frac{({\rm Im}(\nu_1-\nu_2))^2}{4 \pi \,{\rm Im}(\tau)}\ .
 \eeqn
The correlators on the annulus ${\cal A}$, M\"obius strip ${\cal
M}$ and Klein bottle ${\cal K}$  are obtained by symmetrizing this
function under the involutions
\beqn \label{app:is}
I_{{\cal A}}(\nu) = I_{{\cal M}}(\nu) = 2\pi - {\bar \nu} \ ,
\quad I_{{\cal K}}(\nu) = 2\pi - {\bar \nu} + \pi \tau 
\eeqn
producing (cf.\ the appendix of \cite{ABFPT})
\beqn
\langle X(\nu_1) X(\nu_2)\rangle _{\sigma}
=\langle X(\nu_1) X(\nu_2)\rangle _{\sigma} +\langle X(\nu_1)
X(I(\nu_2))\rangle _{\sigma} \ ,
\eeqn
where $\sigma \in \{ \ca, \cm, \ck\}$. For untwisted world-sheet
fermions in the even spin structures, the correlation functions on
the torus and with DN boundary conditions are, respectively,
\beqn \label{fercor}
P_F(s,\nu_1, \nu_2) \delta^{\mu \nu} \equiv \langle
\psi^\mu(\nu_1)\psi^\nu(\nu_2)\rangle ^{\alpha,\beta}_{{\cal
T}}&=&  \frac{1}{2 \pi}
\frac{\vartheta[{\alpha \atop \beta}](\frac{\nu_1-\nu_2}{2 \pi} ,\tau)
{\vartheta_1^\prime}(0,\tau)}
{\vartheta[{\alpha \atop \beta}](0,\tau)
\vartheta_1(\frac{\nu_1-\nu_2}{2 \pi} ,\tau)}\;
\delta^{\mu \nu} \\ [1mm]
&=&
\overline{\langle \tilde \psi^\mu(\bar \nu_1) \tilde \psi^\nu(\bar \nu_2)\rangle ^{\alpha,\beta}_{{\cal T}}} \ ,
\nonumber \\[2mm]
\langle \psi^m(\nu_1)\psi^n(\nu_2)\rangle ^{\alpha,\beta}_{{\cal A}_{95}} &=&
\frac{1}{2 \pi} \frac{\vartheta[{\alpha + 1/2 \atop {\beta}}](\frac{\nu_1-\nu_2}{2 \pi} ,\tau)
{\vartheta_1^\prime}(0,\tau)  }
     {\vartheta[{\alpha + 1/2  \atop {\beta}}](0,\tau)
\vartheta_1(\frac{\nu_1-\nu_2}{2 \pi} ,\tau)}\;
G^{mn} \non [1mm]
&=&
 \overline{\langle \tilde \psi^m(\bar \nu_1) \tilde \psi^n(\bar \nu_2)\rangle ^{\alpha,\beta}_{{\cal A}_{95}} }\ ,
\eeqn
where the relation between $\alpha,\beta$ and $s$ is listed in
table \ref{tab:reminders}. The propagators for twisted world-sheet
fermions in the even spin structures on the torus and with ND
boundary conditions are, respectively,
\beqn \label{fercortwist}
\langle \psi^m(\nu_1)\psi^n(\nu_2)\rangle ^{\alpha,\beta}_{{\cal T}}
&=&
\frac{1}{2 \pi} \frac{\vartheta[{\alpha \atop {\beta + kv_i}}](\frac{\nu_1-\nu_2}{2 \pi} ,\tau)
{\vartheta_1^\prime}(0,\tau)}
{\vartheta[{\alpha \atop {\beta + kv_i}}](0,\tau)
\vartheta_1(\frac{\nu_1-\nu_2}{2 \pi} ,\tau)}\; G^{mn}\ , \\
\langle \psi^m(\nu_1)\psi^n(\nu_2)\rangle ^{\alpha,\beta}_{{\cal A}_{95}} &=&
\frac{1}{2 \pi} \frac{\vartheta[{\alpha + 1/2 \atop {\beta + kv_i}}](\frac{\nu_1-\nu_2}{2 \pi} ,\tau)
{\vartheta_1^\prime}(0,\tau)  }
     {\vartheta[{\alpha + 1/2  \atop {\beta + kv_i}}](0,\tau)
\vartheta_1(\frac{\nu_1-\nu_2}{2 \pi},\tau)}\;
G^{mn}\ .
\eeqn
Just as for bosons, fermion propagators for the remaining surfaces
can be determined from the torus propagators by the method of
images. The result was listed in the appendix of \cite{ABFPT}, but
as we are using slightly different conventions, we summarize the
derivation in section \ref{app:images}.
\beqn \label{correlators}
\langle \psi(\nu_1)\psi(\nu_2)\rangle^{\alpha, \beta}_{\sigma} &=& P_F(s,\nu_1,\nu_2)\; , \quad
\sigma \in \{ \ca, \cm, \ck\}
\non
\langle \psi(\nu_1) \tilde \psi(\bar \nu_2)\rangle^{\alpha, \beta}_{\sigma} &=& i P_F(s,\nu_1, I_\sigma(\nu_2))\ , \non
\langle \tilde \psi(\bar \nu_1)\tilde \psi(\bar \nu_2)\rangle^{\alpha, \beta}_{\sigma} &=& \overline{P_F(\bar s,\nu_1,\nu_2)}\ ,
\eeqn
where $P_F(s,\nu_1,\nu_2)$ was defined in (\ref{fercor}).


\subsection{Mathematical identities}

The fermionization identity in \cite{Polchinski:rr}
(13.4.20)-(13.4.21) when at least one of the $\phi_i'$ is zero can
be reorganized as
\[
\sum_{{\alpha\beta}\atop{\rm even}} \eta_{\alpha \beta} \prod_{i=1}^4
\thba{\alpha}{\beta}(\phi_i,\tau)=
 \prod_{i=1}^4
\thba{1/2}{1/2}(\phi_i,\tau)\ ,
\]
which greatly simplifies the integrands for even spin structures
(here ``even" means all except $(1/2,1/2)$). Note that the
``angles" $\phi_i$ are equivalent to shifts in the characteristic
$\beta$. Using periodicity properties one can generalize this
formula to allow for shifts also in the $\alpha$ characteristic:
\beqn
\sum_{{\alpha\beta}\atop{\rm even}} \eta_{\alpha \beta} \,
\thba{\alpha}{\beta}(\nu,\tau)\thba{\alpha+h_1}{\beta+g_1}
(\nu,\tau)\prod_{i=2}^3
\thba{\alpha+h_i}{\beta+g_i}(0,\tau)\\
&& \hspace{-4cm}
=\thba{1/2}{1/2}(\nu,\tau)
\thba{1/2+h_1}{1/2+g_1}(\nu,\tau)\prod_{i=2}^3
\thba{1/2+h_i}{1/2+g_i}(0,\tau) \nonumber
\label{app:spinstr}
\eeqn
where $\sum g_i = \sum h_i = 0$.
\begin{table}
\[
\begin{array}{|c|c|r|r|r|}\hline
\rule[-4mm]{0mm}{10mm}\Big[ {\alpha \atop \beta} \Big]
& \; s \;& \eta_{\alpha \beta} & (-1)^{s+1} & \mbox{
$\tau \rightarrow -1/\tau$ }
\\[2mm] \hline\hline
\rule[-4mm]{0mm}{10mm}\Big[ {1/2 \atop 1/2} \Big] & 1 & -1 & 1 & \rightarrow
-\Big[ {1/2 \atop 1/2} \Big] \\[2mm]\hline
\rule[-4mm]{0mm}{10mm}\Big[ {1/2 \atop 0} \Big] & 2 & -1& -1 &  \rightarrow
\Big[ {0 \atop 1/2} \Big] \\[2mm]\hline
\rule[-4mm]{0mm}{10mm}\Big[\; {0\atop 0} \; \Big]  & 3 & 1 & 1 &\rightarrow
\Big[\; {0 \atop 0} \; \Big] \\[2mm] \hline
\rule[-4mm]{0mm}{10mm} \Big[ {0 \atop 1/2} \Big] &4 & -1 & -1 &\rightarrow
\Big[ {1/2 \atop 0} \Big] \\[2mm]\hline
\end{array}
\]
\caption{Some reminders for the summation
over spin structures. The last column denotes the transformation
of the partition function $\cz [{\alpha \atop \beta}]$, so after
modular transformation, $\eta_{\alpha \beta}$ is equivalent to
$(-1)^{1+s}$, which is sometimes used in the literature.}
\label{tab:reminders}
\end{table}
We will also need the periodicity formulas
\be \label{nuperiod}
\thba{\alpha}{\beta} (\nu \pm \tau , \tau) = e^{-i \pi \tau  \mp 2 i \pi (\nu +\beta)} \thba{\alpha}{\beta} (\nu , \tau)\ ,
\ee
that can easily be derived from the sum representations for the
theta functions in \cite{Polchinski:rr}, as is 
\be \label{corrperiodicity}
\thba{\alpha}{\beta+n} (\nu, \tau) = (-1)^{2 \alpha n} \thba{\alpha}{\beta} (\nu, \tau)
\ee
for integer $n$. Finally, we often make use of the Poisson resummation  formula
\beqn
\tht\ba{\vec \alpha}{\vec 0}(0,it G^{-1}) &=& \sqrt{G} \, t^{-N/2}\,  
 \tht\ba{\vec 0}{\vec 0 }(\vec \alpha,it^{-1} G) \ , 
\eeqn
for 
\beqn 
\tht\ba{\vec \alpha}{\vec\beta}(\vec \nu,G) &=& \sum_{\vec n\in \mathbb{Z}^N} 
 e^{i\pi(\vec n+\vec \alpha)^{\rm T} G (\vec n+\vec \alpha)} 
e^{2\pi i(\vec \nu+\vec \beta)^{\rm T}(\vec n+\vec\alpha)}  \ , 
\label{thetamatrix}
\eeqn
where $G$ is an $N\times N$ matrix with  ${\rm Im}(G)>0$.

\subsection{Method of images for fermions}
\label{app:images}

\begin{figure}[h]
\begin{center}
\psfrag{2pitau}[bc][bc][1][0]{$2\pi \tau = i \pi t$}
\psfrag{2pitauM}[bc][bc][1][0]{$2\pi \tau = \pi+i \pi t$}
\psfrag{2pitauK}[bc][bc][1][0]{$2\pi \tau = 4 i \pi t$}
\psfrag{2pi}[bc][bc][1][0]{$2\pi$}
\psfrag{0}[bc][bc][1][0]{$0$}
\psfrag{z}[bc][bc][1][0]{$\nu$}
\psfrag{IzA}[bc][bc][1][0]{$I_{\mathcal A}(\nu)$}
\psfrag{IzM}[bc][bc][1][0]{$I_{\mathcal M}(\nu)$}
\psfrag{IzK}[bc][bc][1][0]{$I_{\mathcal K}(\nu)$}
\psfrag{C}[bc][bc][1][0]{$C$}
\psfrag{Cp}[bc][bc][1][0]{$C'$}
\psfrag{D}[bc][bc][1][0]{$D$}
\includegraphics[width=0.8\textwidth]{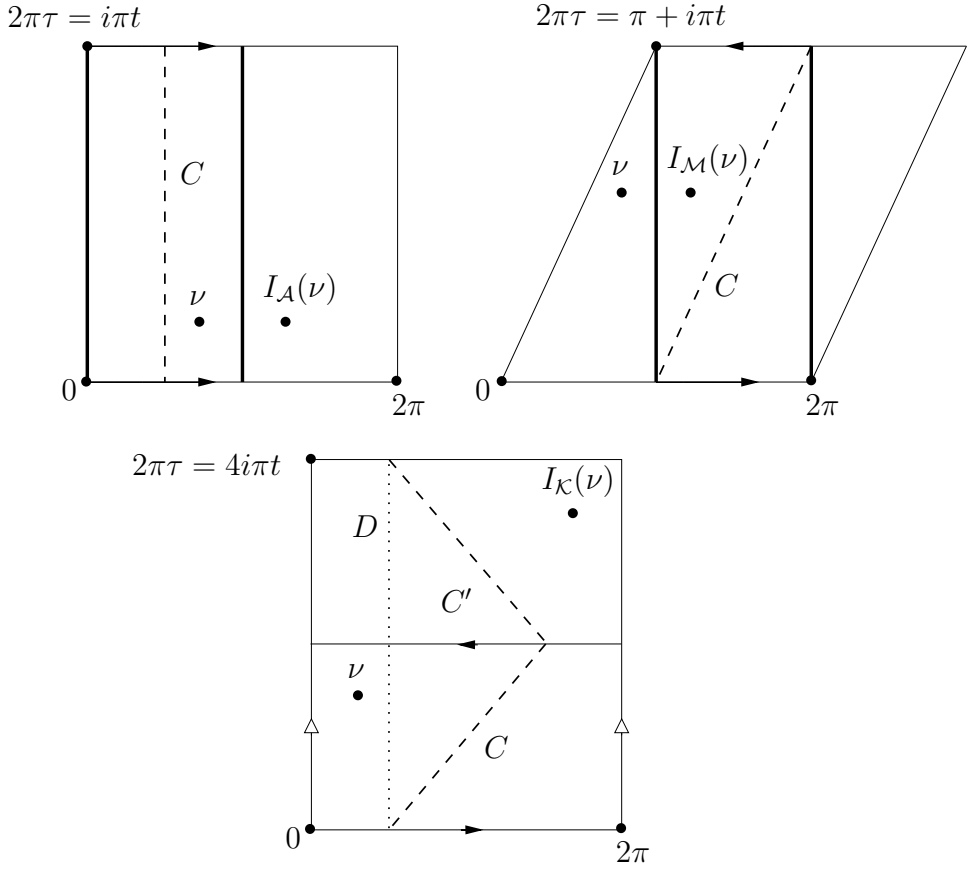}
\caption{Annulus, M\"obius strip and Klein bottle
obtained from covering tori by the involutions (\ref{app:is}). Thick
lines are boundaries (fixed lines under the involution). Each
surface has one 1-cycle $C$. On the Klein bottle, the equality
$CC'=D$ gives rise to the constraint (\ref{constraint}).}
\label{fig:surfaces}
\end{center}
\end{figure}

In this subsection we give some more details about the fermion
correlators in (\ref{correlators}). Throughout, we will use the
conventions of
\cite{Polchinski:rr}, in particular $z=e^{-i \nu}$.
Let us begin by checking that the correlators given in
(\ref{fercor}) and (\ref{fercortwist}) have short-distance
expansions consistent with their OPEs. To see this, we take the
same Laurent expansion in $z$ as \cite{Polchinski:rr}, i.e.\
\be
\psi^\mu (z) = \sum_{r \in \mathbb{Z} + \gamma} \frac{\psi^\mu_r}{z^{r+1/2}} \quad , \quad
\tilde \psi^\mu (\bar z) = \sum_{r \in \mathbb{Z} + \tilde \gamma} \frac{\tilde \psi^\mu_r}{\bar z^{r+1/2}}\ ,
\ee
where $\gamma$ and $\tilde \gamma=0$ $(1/2)$ in the R (NS) sector.
We also take the same OPEs, i.e.\
\be \label{opez}
\psi^\mu (z_1) \psi^\nu (z_2) \sim \frac{\eta^{\mu \nu}}{z_1 -z_2} \quad , \quad
\tilde \psi^\mu (\bar z_1) \tilde \psi^\nu (\bar z_2) \sim \frac{\eta^{\mu \nu}}{\bar z_1 - \bar z_2}\ .
\ee
Using the conformal transformation to torus coordinates
\beqn \label{conftrans}
\psi^\mu (z) & = & (\partial_z \nu)^{1/2} \psi^\mu (\nu) = (-i z)^{-1/2} \psi^\mu (\nu)\ , \non
\tilde \psi^\mu (\bar z) & = & (\partial_{\bar z} \bar \nu)^{1/2} \tilde \psi^\mu (\bar \nu) = (i \bar z)^{-1/2} \tilde \psi^\mu (\bar \nu) \ ,
\eeqn
leads to the following Fourier expansions in torus coordinates
\be \label{psinu}
\psi^\mu (\nu) = (-i)^{1/2} \sum_{r \in \mathbb{Z} + \gamma} \psi^\mu_r e^{-2\pi i r \nu} \quad , \quad
\tilde \psi^\mu (\bar \nu) = i^{1/2} \sum_{r \in \mathbb{Z} + \tilde \gamma} \tilde \psi^\mu_r e^{2\pi i r \bar \nu}\ .
\ee
Using (\ref{opez}) and (\ref{conftrans}), it is also easy to see
that the OPEs in torus coordinates are given by
\be \label{openu}
\psi^\mu (\nu_1) \psi^\nu (\nu_2) \sim \frac{\eta^{\mu \nu}}{\nu_1 -\nu_2} \quad , \quad
\tilde \psi^\mu (\bar \nu_1) \tilde \psi^\nu (\bar \nu_2) \sim \frac{\eta^{\mu \nu}}{\bar \nu_1 - \bar \nu_2}\ .
\ee
Using
\be
\vartheta_1 (\nu,\tau) \;
\stackrel{\nu\rightarrow 0}{\longrightarrow} \;
2 \pi \eta(\tau)^3 \nu
\ee
makes it clear that the correlators (\ref{fercor}) and
(\ref{fercortwist}) have short distance behavior consistent with
the OPEs (\ref{openu}). The differing short distance behavior of
the correlator (A.9) in \cite{ABFPT} can now be understood by the
redefinitions
\be \label{psiabfpt}
\psi^{\rm (ABFPT)} (\nu) = (-2 i)^{-1/2} \psi^{\rm(here)} (\nu) \quad , \quad
\tilde \psi^{\rm (ABFPT)} (\bar \nu) = (2 i)^{-1/2} \tilde
\psi^{\rm(here)} (\bar \nu) \; .
\ee
We proceed to construct the fermionic correlators
(\ref{correlators}) by images under the involutions
(\ref{app:is}). Consider the two Clifford algebras
\be
\{ \gamma_a, \gamma_b \} = \pm 2 \eta_{ab} {\bf 1}\ ,
\label{Cliff}
\ee
where $\eta_{ab}$ is the world-sheet metric. For non-orientable
surfaces, we must a priori consider both signs in (\ref{Cliff}).
With fixed Euclidean signature $\eta_{ab}$, that we take to be
$(++)$, the two algebras generate two different groups Pin$^+(2)$
and Pin$^-(2)$ (see \cite{Berg:2000ne} for an introduction to
these groups). In both cases there is a choice of (2-dimensional,
chiral) representation\footnote{By chiral representation we mean
one for which $\gamma_1 \gamma_2$ is diagonal.} for $\gamma_1$ and
$\gamma_2$. For Pin$^+(2)$ the matrices square to $+{\bf 1}$ by
(\ref{Cliff}), so we can pick $\gamma_1$, $\gamma_2$ to be the
Pauli matrices
\be \label{rep1}
  \left( \begin{array}{cc}
                                 0 & 1 \\
                                 1 & 0
                                 \end{array}
                        \right)\quad , \quad
 \left( \begin{array}{cc}
                                 0 & -i \\
                                 i & 0
                                 \end{array}
                        \right)\; ,
\ee
whereas for Pin$^-(2)$ we can take
\be \label{rep2}
  \left( \begin{array}{cc}
                                 0 & -1 \\
                                 1 & 0
                                 \end{array}
                        \right)\quad , \quad
 \left( \begin{array}{cc}
                                 0 & i \\
                                 i & 0
                                 \end{array}
                        \right)\; .
\ee
The involutions (\ref{app:is}) precisely correspond to the parity
transformation on the respective surfaces (i.e.\ one returns to
the original point and changes the orientation, see
fig.~\ref{fig:surfaces}).
The corresponding reflection matrix is $P^a_b = {\rm diag}(-1,1)$,
because in $\nu = \sigma_1 + i \sigma_2$ the world-sheet space
coordinate is given by $\sigma_1$. As is familiar from four
dimensions, the Lorentz matrix ${\Lambda^a}_b =P^a_b$ acting on
the coordinates induces an action on the (s)pinors by
\be  \label{psimap}
\left( \begin{array}{c}
         \psi(\nu) \\
         \tilde \psi(\bar \nu)
         \end{array}
\right)
\; \stackrel{P}{\longrightarrow} \;
S_P
\left( \begin{array}{c}
         \psi(I_{\sigma}(\nu)) \\
         \tilde \psi(I_{\sigma}(\bar \nu))
         \end{array}
\right)
\ee
for $\sigma \in \{ \ca,\cm,\ck \}$, where the matrix $S_P$ is
given by
\be
\label{homomorphism}
S_P^{-1} \gamma^a S_P = P^a_b \gamma^b\ .
\ee
Using the algebra (\ref{Cliff}), equation (\ref{homomorphism})
yields $S_P = \pm \gamma_2$ and we take\footnote{The overall sign
is a matter of convention; changing it merely maps the pin
structures into each other. An overall factor of $i$ changes the
algebra.} $S_P = \gamma_2$. We will now argue that for the annulus
and M\"obius strip, $\gamma_2$ must be the second matrix in
(\ref{rep1}); if we were to pick the first matrix, the propagator
of Majorana fermions would vanish.\footnote{For the annulus and
M\"obius strip, we pick Pin${}^+$ in our conventions, 
whereas for the Klein bottle we need to allow both Pin${}^+$ and Pin${}^-$, as
expressed in eq.\ (\ref{constraint}) below. This choice is in accord with \cite{ABFPT}.
It would be
interesting to work out precisely how the topological obstructions
in
\cite{Carlip:1988gw,Distler:1992rr}
 manifest themselves here.}
Define the Majorana pinor
\be
{\Psi}(\nu, \bar \nu) = \left(
         \begin{array}{c}
         \psi(\nu) \\
         \tilde \psi(\bar \nu)
         \end{array}
         \right)
\ee
as in \cite{VanNieuwenhuizen:1985be}, i.e.\ by demanding
\be
\overline{\Psi}_{\rm D} = \Psi^\dagger \gamma_2
\stackrel{!}{=} \Psi^T C = \overline{\Psi}_{\rm M}\ ,
\ee
where $\overline{\Psi}_{\rm D}$ is the Dirac adjoint,
$\overline{\Psi}_{\rm M}$ is the Majorana conjugate and $C$ is the
charge conjugation matrix, whose explicit form we do not need in
the following. Then the Dirac propagator on the torus is given by
\cite{ABFPT}
\beqn \label{proptorus}
\langle \Psi(\nu_1,\bar \nu_1) \overline{\Psi}_{\rm D}(\nu_2,\bar \nu_2)\rangle_{{\cal T}}  &=&
\langle \Psi(\nu_1,\bar \nu_1) \Psi^T(\nu_2,\bar \nu_2)\rangle_{{\cal T}} C \\
& = & \left( \begin{array}{cc}
\langle \psi(\nu_1)\psi(\nu_2)\rangle ^{\alpha,\beta}_{{\cal T}} & 0 \\
0 & \langle \tilde \psi(\bar \nu_1) \tilde \psi(\bar
\nu_2)\rangle ^{\bar \alpha, \bar \beta}_{{\cal T}}
\end{array} \right) C\ .
\nonumber
\eeqn
The propagators on the other surfaces are determined by
symmetrizing (\ref{proptorus}) with respect to the involutions
(\ref{app:is}), producing
\beqn \label{propsigma}
\langle \Psi(\nu_1) \overline{\Psi}_{\rm D}(\nu_2)\rangle_{\sigma} &=&  \frac12 \Big(
\langle \Psi(\nu_1) \Psi^T(\nu_2)\rangle_{{\cal T}} + \gamma_2 \langle \Psi(I_\sigma(\nu_1)) \Psi^T(\nu_2)\rangle_{{\cal T}} \\
&& \hspace{0cm}
+ \langle \Psi(\nu_1) \Psi^T(I_\sigma(\nu_2)) \rangle_{{\cal T}}
\gamma_2^T +  \gamma_2 \langle\Psi(I_\sigma(\nu_1))
\Psi^T(I_\sigma(\nu_2))\rangle_{{\cal T}} \gamma_2^T \Big) C\ ,
\nonumber
\eeqn
with $\sigma \in \{ {\cal A},{\cal M},{\cal K}\}$. To go further,
we have to distinguish between annulus and M\"obius strip on the
one hand and Klein bottle on the other. For $\sigma = {\cal A},
{\cal M}$, using (\ref{proptorus}) and (\ref{rep1}), the
propagator (\ref{propsigma}) becomes
\be
\frac12  \left[ \begin{array}{cc}
P_F(s,\nu_1,\nu_2) - \overline{P_F(\bar
s,I_\sigma(\nu_1),I_\sigma(\nu_2))} & i ( P_F(s, \nu_1,
I_\sigma(\nu_2)) - \overline{P_F(\bar s, I_\sigma(\nu_1),\nu_2)})
\\  i (P_F(s,I_\sigma(\nu_1),\nu_2) - \overline{P_F(\bar s,
\nu_1,I_\sigma(\nu_2))})
 & \overline{P_F(\bar s, \nu_1,\nu_2)} - P_F(s, I_\sigma(\nu_1),I_\sigma(\nu_2))
\end{array}
\right] C\ .
\ee
For the annulus, the spin structures for left- and right-movers
are the same on the covering torus ($s = \bar s$), and the complex
structure modulus of the covering torus is purely imaginary. Then,
from the definition of the theta functions in \cite{Polchinski:rr}
we see that $\overline{P_F(s, \nu_1, \nu_2)} = P_F(s, \bar \nu_1,
\bar \nu_2)= -P_F(s, -\bar \nu_1, -\bar \nu_2)$. On the other
hand, for the M\"obius strip, $(s,\bar s) = (2,2),(3,4),(4,3)$ for
the even spin structures, cf.\ \cite{ABFPT}. The M\"obius complex
structure has real part $1/2$, so that
\beqn
\overline{\left( \frac{\vartheta_{3/4} (\frac{\nu_1 - \nu_2}{2 \pi},\tau)}{\vartheta_{3/4} (0,\tau)}\right)} _{{\cal M}}& =&  \left( \frac{\vartheta_{4/3} (\frac{\bar \nu_1 - \bar \nu_2}{2 \pi} ,\tau)}{\vartheta_{4/3} (0,\tau)} \right)_{{\cal M}}\ , \non
\overline{\left(\frac{\vartheta_{2} (\frac{\nu_1 - \nu_2}{2 \pi} ,\tau)}{\vartheta_{2} (0,\tau)}\right)}  _{{\cal M}}&=& \left( \frac{\vartheta_{2} (\frac{\bar \nu_1 - \bar \nu_2}{2 \pi} ,\tau)}{\vartheta_{2} (0,\tau)} \right)_{{\cal M}}\ , \non
\overline{\left(\frac{\vartheta_{1} (\frac{\nu_1 - \nu_2}{2 \pi} ,\tau)}{\vartheta'_{1} (0,\tau)}\right)} _{{\cal M}} &=&  \left(\frac{\vartheta_{1} (\frac{\bar \nu_1 - \bar \nu_2}{2 \pi} ,\tau)}{\vartheta'_{1} (0,\tau)}\right)_{{\cal M}}\ .
\eeqn
We arrive at
\be \label{diracprop}
\langle \Psi(\nu_1) \overline{\Psi}_{\rm D}(\nu_2)\rangle_{\sigma} = \left( \begin{array}{cc}
P_F(s,\nu_1,\nu_2) & i P_F(s, \nu_1, I_\sigma(\nu_2))\\ i
P_F(s,I_\sigma(\nu_1),\nu_2)
 & \overline{P_F(\bar s, \nu_1,\nu_2)}
\end{array}
\right) C\ .
\ee
(Recall that up to now, this only holds for $\sigma={\cal A},
{\cal M}$; however, we will see in a moment that the same holds
also for $\sigma={\cal K}$.) Note that if we had chosen $\gamma_2$
as the first matrix in (\ref{rep1}), the propagator had come out
to be identically zero. That is why we have to choose $\gamma_2$
as the second matrix in (\ref{rep1}) for our conventions of
fermions.\footnote{For the conventions of \cite{ABFPT}, one has
for instance for the annulus $\overline{P_F(s, \nu_1, \nu_2)} = -
P_F(s, \bar \nu_1, \bar \nu_2)$, as there is an additional $i$ in
the definition of $P_F$, cf.\ their (A.9). Thus, for them the
first matrix in (\ref{rep1}) is the right choice for $\gamma_2$,
which is what they report in (A.10).}

We now proceed to show that the same correlators also hold for the
Klein bottle $\ck$. To see this, first note that there is a
complication that only appears for $\ck$ and not for $ {\cal A},
{\cal M}$. Referring to figure
\ref{fig:surfaces},
we see that going twice around the Klein-bottle cycle $C$ (which
is equivalent to first traversing $C$, then $C'$) has the
same effect as following the path $D$ on the covering torus. This
puts a constraint on the action of the square of parity $S_P^2$ on
the pinors, depending on their spin structure on the covering
torus. More precisely, in (\ref{psimap}) we need to introduce
different $S_{P}$ for different spin structures, satisfying
\be \label{constraint}
(S_{P}\zba{\alpha}{\beta})^2 = (-1)^{2\beta + 1} {\bf 1}\quad\quad
\mbox{(Klein bottle only)}\ .
\ee
Thus,  for $s=4$ and $s=1$ (again we refer to table
\ref{tab:reminders}), the constraint
(\ref{constraint}) leads to Pin$^+(2)$ as in the case of the
annulus and M\"obius strip. For $s=2$ and $s=3$, the constraint
(\ref{constraint}) forces us to choose Pin$^-(2)$. Retracing the
above steps for $s=2$ and $s=3$ leads to the Dirac propagator
\beqn \label{DKlein}
&&\langle \Psi(\nu_1,\bar \nu_1) \overline{\Psi}_{\rm D}(\nu_2,\bar \nu_2)\rangle_{{\cal K}}  = \\
&&\hspace{-.8cm} \frac12  \left[ \begin{array}{cc}
P_F(s,\nu_1,\nu_2) - \overline{P_F(\bar s,I_{\cal
K}(\nu_1),I_{\cal K}(\nu_2))} & i (P_F(s, \nu_1, I_{\cal
K}(\nu_2)) + \overline{P_F(\bar s, I_{\cal K}(\nu_1),\nu_2)})\\ i
(P_F(s,I_{\cal K}(\nu_1),\nu_2) + \overline{P_F(\bar s,
\nu_1,I_{\cal K}(\nu_2))})
 & \overline{P_F(\bar s, \nu_1,\nu_2)} - P_F(s, I_{\cal K}(\nu_1),I_{\cal K}(\nu_2))
\end{array}
\right] C\ .
\nonumber
\eeqn
For the Klein bottle, like for the annulus, the spin structures
for left and right movers are the same on the covering torus ($s =
\bar s$) and the complex structure of the covering torus is purely
imaginary,  so again $\overline{P_F(s, \nu_1, \nu_2)} = -P_F(s,
-\bar \nu_1, -\bar \nu_2)$. For the off-diagonal elements, one can
show that the periodicity (\ref{nuperiod}) implies
\beqn
\overline{P_F(\bar s, I_{\cal K}(\nu_1),\nu_2)} &=& P_F(s, \nu_1, I_{\cal K}(\nu_2))\ , \quad s=2,3 \ , \non
\overline{P_F(\bar s, I_{\cal K}(\nu_1),\nu_2)} &=& - P_F(s, \nu_1, I_{\cal K}(\nu_2))\ , \quad s=4 \; .
\eeqn
Substituting this into (\ref{DKlein}) one ends up with the same
results as in (\ref{diracprop}) above also for $s=2,3$, and hence
for all even spin structures $s=2,3,4$. This concludes the
derivation of the correlators summarized in (\ref{correlators}).

It may be useful to note that had we blindly tried to use the
correlators of \cite{ABFPT} without adjusting to our conventions,
the ``torus trick" in equation (\ref{abfpttrick}) would not work.
Also see \cite{Burgess:1986ah,Burgess:1986wt,Epple:2004nh} 
for some related discussion.


\section{Some properties of $E_2(A,U)$}
\label{appe2}

In this appendix we discuss two issues related to the function
$E_2(A,U)$ which we defined in (\ref{eisen}). We show how to
rewrite it in terms of polylogarithms, which is useful for
expressing the one-loop correction to the K\"ahler potential in
terms of a prepotential. Furthermore, we display its
transformation property under $SL(2,\mbb Z)$ transformations of
its arguments, i.e.\ under T-duality.

\subsection{Writing $E_2(A,U)$ in terms of polylogarithms}

The goal is to rewrite $E_2(A,U)$ such that $U_2E_2(A,U)$ can be
expressed in a form like the argument of the logarithm in
(\ref{n=2kaehler}) via some holomorphic prepotential $\cf(A,U)$.
This can be derived in the same way as (1.4) of
\cite{Obers:1999um}. We start with
$E_2(A,U)$ defined in (\ref{eisen}), split off the term with
$m=0$, in the rest use
\be
y^{-s} = \frac{1}{\Gamma (s)} \int_0^\infty dx\, x^{s-1} e^{-yx}\ .
\ee
Next we perform a Poisson resummation over $n$ and then split off
the term with $n=0$. One also has to use
\be
\int_0^\infty dx\, x^{\nu-1} e^{-\beta/x-\gamma x} = 2 \Big(\frac{\beta}{\gamma} \Big)^{\nu/2} K_\nu (2 \sqrt{\beta \gamma})\ , \quad {\rm Re}(\beta)>0 \ , \ {\rm Re}(\gamma)>0 \ .
\ee
Doing so one derives
\beqn
E_2(A,U) & = & 2 U_2^2 \pi^4 \Big( \frac{1}{90}  - \frac13 \frac{A_2^2}{U_2^2} + \frac23
\frac{A_2^3}{U_2^3} - \frac13 \frac{A_2^4}{U_2^4} \Big) \\
&& +\frac{\pi}{2 U_2} \Big[  Li_3(e^{2\pi i A}) + Li_3(e^{-2 \pi i \bar A})\Big]
+ \pi^2 \frac{A_2}{U_2} \Big[ Li_2(e^{2\pi i A}) + Li_2(e^{-2 \pi i \bar A}) \Big] \non
&& + 2 \pi^2 \sqrt{U_2} \sum_{m \neq 0,n \neq 0} \Big| \frac{m-\frac{A_2}{U_2}}{n}\Big|^{3/2}
K_{3/2} \Big(2 \pi \Big|nm-n\frac{A_2}{U_2}\Big| U_2\Big) e^{2 \pi i n (m U_1 - A_1)}\ , \nonumber
\eeqn
using
\be \label{bessel}
K_{3/2}(z) = \left( \frac{\pi}{2 z} \right)^{1/2} e^{-z} (1 + z^{-1})
\ee
and
\be \label{polylog}
Li_n(z) = \sum_{k=1}^{\infty} \frac{z^k}{k^n} \ .
\ee
Using (\ref{bessel}) again, this can be rewritten as (assuming $a_1=A_2/U_2 < 1$)
\beqn
E_2(A,U) & = & 2 U_2^2 \pi^4 \Big( \frac{1}{90}  - \frac13 \frac{A_2^2}{U_2^2} + \frac23
\frac{A_2^3}{U_2^3} - \frac13 \frac{A_2^4}{U_2^4} \Big) \non
&& \hspace{-1cm}
+\frac{\pi}{2 U_2} \Big[  Li_3(e^{2\pi i A}) + Li_3(e^{-2 \pi i \bar A})\Big]
+ \pi^2 \frac{A_2}{U_2} \Big[ Li_2(e^{2\pi i A}) + Li_2(e^{-2 \pi i \bar A}) \Big] \non
&& \hspace{-1cm}
+ \pi^2 \sum_{m>0} \Big(m-\frac{A_2}{U_2}\Big) \Big[ Li_2(e^{2 \pi i (m U - A)}) +
Li_2(e^{-2 \pi i (m \bar U - \bar A)}) \Big] \non
&& \hspace{-1cm}
+ \pi^2 \sum_{m>0} \Big(m+\frac{A_2}{U_2}\Big) \Big[ Li_2(e^{2 \pi i (m U + A)}) +
Li_2(e^{-2 \pi i (m \bar U + \bar A)}) \Big]  \non
&&\hspace{-1cm}
+\frac{\pi}{2 U_2} \sum_{m>0} \Big[ Li_3(e^{2 \pi i (m U - A)}) +
Li_3(e^{-2 \pi i (m \bar U - \bar A)}) \non
&&\hspace{2cm}
+ Li_3(e^{2 \pi i (m U + A)}) +
Li_3(e^{-2 \pi i (m \bar U + \bar A)}) \Big]\ .
\eeqn
This formula is essential to derive (\ref{UEh}).

\subsection{The $SL(2,\mbb Z)$ transformation of $E_2(A,U)$}

We now show how $E_2(A,U)$ transforms under $SL(2,\mbb Z)$. Define
\be \label{z}
Z(\vec{a}, U) = \sum_{\vec n=(n,m)^T} \!\!\!\!\!\! ' \; \; \;
\frac{e^{2 \pi i \vec{n} \vec{a}}}{|n+m U|^4}\ ,
\ee
where $\vec{a}$ is a real vector. More precisely, it determines a
position $a_1 {e}_M^{1} + a_2 {e}_M^{2}$ on a torus, whose lattice
is defined by the vielbein ${e}^{m}_M$. Now perform an $SL(2,
\mbb{Z})$ transformation
\be \label{trafo}
U ~\rightarrow~ \frac{A U + B}{C U + D}\ , \quad a_m ~\rightarrow a^m~\ ,
\ee
where
\be
\left( \begin{array}{cc}
       A & B \\
       C & D
       \end{array}
\right) \in SL(2, \mbb{Z})
\ee
and $a^m$ defines a position $a^1 {e}^M_1 + a^2 {e}^M_2$ on the
transformed lattice that we will again, by slight abuse of
notation, denote by $\vec{a}$. Now (\ref{z}) transforms under
(\ref{trafo}) according to
\beqn
Z(\vec{a}, U) & \rightarrow &
\sum_{\vec n=(n,m)^T} \!\!\!\!\!\! ' \; \; \; 
\frac{e^{2 \pi i \vec{n} \vec{a}}}{|n + \frac{A U + B}{C U + D} m|^4} \non
&=& \sum_{\vec n=(n,m)^T} \!\!\!\!\!\! ' \; \; \;
\frac{e^{2 \pi i \vec{n} \vec{a}}}{|D n + B m +
(A m + C n) U|^4} |CU+D|^4\ .
\eeqn
Introducing new variables
\be
\tilde n = D n + B m\ , \quad \tilde m = A m + C n
\ee
or
\be
n = A \tilde n - B \tilde m\ , \quad m = D \tilde m - C \tilde n\ ,
\ee
and using
\be
\vec{n} \vec{a} = \tilde n (A a^1 - C a^2) + \tilde m (-a^1 B + a^2 D)\ ,
\ee
we arrive at
\be \label{ztrafo}
Z\Big(\vec{a}, \frac{A U + B}{C U + D}\Big) = |CU+D|^4 Z(\tilde{\vec{a}}, U)\ ,
\ee
with
\be
\tilde{\vec{a}} = \left( \begin{array}{c}
       \tilde a^1 \\
       \tilde a^2
       \end{array} \right) = \left( \begin{array}{cc}
       A & -C \\
       -B & D
       \end{array}
\right) \left( \begin{array}{c}
       a^1 \\
       a^2
       \end{array} \right)\ .
\ee
On the other hand
\be \label{u2trafo}
U_2^2  ~\rightarrow~  U_2^2 |CU+D|^{-4}\ .
\ee
Taking (\ref{ztrafo}) and (\ref{u2trafo}) together, we arrive at
\be
E_2(A,U) ~\rightarrow~ E_2(\tilde{A},U)\ ,
\label{sl2ze2}
\ee
where
\be\label{tdualA}
\tilde{A} = U \tilde{a}^1 - \tilde{a}^2\ .
\ee
Equation (\ref{sl2ze2}) is the desired result.
As a special case for $A=0$, we recover
 $SL(2, \mbb{Z})$ invariance
of the nonholomorphic Eisenstein series $E_2(0,U)$.


\section{The other one-loop 2-point functions}
\label{other}

In this appendix we compute the remaining 2-point functions for
the K\"ahler variables $\{S',U,A_i\}$ of the
$\cn=2$ orientifold discussed in section \ref{z2}, where only the
2-point function $\langle V_{S_2'}V_{S_2'} \rangle$ was considered. Doing
so, we confirm the formula (\ref{tollesK}) for the K\"ahler potential
that was the main result of section \ref{z2}.

Let us start by giving the metric components derived from the
K\"ahler potential (\ref{tollesK}) that we would like to reproduce by
the 2-point functions. Up to  one-loop order they are given by
\beqn
K_{U\bar U} &=& K^{(0)}_{U\bar U} + \frac{c}{(S - \bar S)(S_0'-\bar S_0')} \partial_U \partial_{\bar U} \ce_2(A_k,U) + \co (e^{3 \Phi}) \ , \non
K_{A_i\bar A_j} &=& K^{(0)}_{A_i\bar A_j} + \frac{c}{(S - \bar S)(S_0'-\bar S_0')} \partial_{A_i} \partial_{\bar A_j} \ce_2(A_k,U) + \co (e^{3 \Phi}) \ , \non
K_{U\bar S'} &=& K^{(0)}_{U\bar S'} + \frac{c}{(S - \bar S)(S_0'-\bar S_0')^2} \partial_{U} \ce_2(A_k,U) +\co (e^{4 \Phi}) \ , \non
K_{U\bar A_i} &=& K^{(0)}_{U\bar A_i} + \frac{c}{(S - \bar S)(S_0'-\bar S_0')} \partial_{U} \partial_{\bar A_i} \ce_2(A_k,U) + \co (e^{3 \Phi}) \ , \non
K_{A_i \bar S'} &=& K^{(0)}_{A_i \bar S'} + \frac{c}{(S - \bar S)(S_0'-\bar S_0')^2} \partial_{A_i} \ce_2(A_k,U) +\co (e^{4 \Phi})\ , \label{k1loop}
\eeqn
where the terms with a superscript $(0)$ were already given in
(\ref{ktree}). The 2-point functions that we want to compute are
\beqn \label{2ptann}
\langle V_{U} V_{\bar U} \rangle &=& -\sum_\sigma \frac{4 g_c^2 \alpha'^{-4}}{(U-\bar U)^2} \langle V_{ZZ}^{(0,0)}V_{\bar Z\bar Z}^{(0,0)} \rangle_\sigma \\
&& \hspace{-1cm}
+ \sum_{\sigma} \left[ \sum_B \frac{8 \pi g_c^2 \alpha'^{-4} (A_{s[B]} - \bar A_{s[B]})}{(U-\bar U)^{5/2} (T -\bar T)^{1/2}} \left( \langle V_{ZZ}^{(0,0)} V_{\bar Z}^{(0)B} \rangle_\sigma + \langle V_Z^{(0)B} V_{\bar Z\bar Z}^{(0,0)} \rangle_\sigma \right) \right. \non
 && \hspace{-1cm}
- \left. \sum_{B,C} \frac{16 \pi^2 g_c^2 \alpha'^{-4}
(A_{s[B]}-\bar A_{s[B]}) (A_{s[C]}-\bar A_{s[C]})}{(U-\bar
U)^3(T-\bar T)} \langle V_Z^{(0)B} V_{\bar Z}^{(0)C}
\rangle_\sigma \right] + {\cal O} (e^{\Phi})\ ,
\non
\langle V_{A_i} V_{\bar A_j} \rangle &=& - \sum_{\sigma } \sum_{B,C} \frac{16 g_o^2 \alpha'^{-4} \delta_{i s[B]} \delta_{j s[C]}}{(U-\bar U)(T-\bar T)} \langle V_Z^{(0)B} V_{\bar Z}^{(0)C} \rangle_\sigma + {\cal O} (e^{\Phi})\ , \non
\langle V_{U} V_{S_2'} \rangle &=& - \sum_\sigma \frac{i 2 g_c^2 \alpha'^{-4}}{(U -\bar U)(S_0'-\bar S_0')} \langle V_{ZZ}^{(0,0)}V_{Z\bar Z}^{(0,0)} \rangle_\sigma \non
&& + \sum_{\sigma } \sum_B \frac{i 4 \pi g_c^2 \alpha'^{-4}(A_{s[B]} - \bar A_{s[B]})}{(U - \bar U)^{3/2} (T -\bar T)^{1/2} (S_0'-\bar S_0')} \langle V_Z^{(0)B} V_{Z\bar Z}^{(0,0)} \rangle_\sigma + {\cal O} (e^{2 \Phi})\ , \non
\langle V_{U} V_{\bar A_i} \rangle &=& \sum_{\sigma } \left[ - \sum_B \frac{8 g_c g_o \alpha'^{-4} \delta_{i s[B]}}{(U - \bar U)^{3/2} (T -\bar T)^{1/2}} \langle V_{ZZ}^{(0,0)} V_{\bar Z}^{(0)B} \rangle_\sigma \right. \non
&& \left. \hspace{1cm} + \sum_{B,C} \frac{16 \pi g_c g_o \alpha'^{-4} (A_{s[B]}-\bar A_{s[B]}) \delta_{i s[C]}}{(U-\bar U)^2(T-\bar T)} \langle V_Z^{(0)B} V_{\bar Z}^{(0)C} \rangle_\sigma \right] +  {\cal O} (e^{\Phi}) \ , \non
\langle V_{A_i} V_{S_2'} \rangle &=& - \sum_{\sigma } \sum_B \frac{i 4 g_c g_o \alpha'^{-4} \delta_{i s[B]}}{(U - \bar U)^{1/2} (T -\bar T)^{1/2}(S_0'-\bar S_0')} \langle V_Z^{(0)B} V_{Z\bar Z}^{(0,0)} \rangle_\sigma + {\cal O} (e^{2 \Phi})\ . \nonumber
\eeqn
The summation over $\sigma$ effectively only runs over $\{ (ij),
(ia), (ai)\}$ and $\{(i)\}$ for the correlators involving the open
string vertex operators $V^{(0)B}$, because we are only
considering Wilson line moduli on the 9-branes. We have given only
the leading contribution in an expansion in the dilaton. This is
the only one that we surely have to be able to reproduce with the
K\"ahler potential that we suggested in \ref{1loopkpot}. Higher
order terms (arising from the higher order corrections in the
vertex operators (\ref{vops})) might in general also get
contributions from higher genus world-sheets in string
perturbation theory. We do not attempt to calculate these higher
order terms here. All amplitudes are expressed by the following
basic correlators
\beqn \label{ZZbZbZ}
\langle V_{Z Z}^{(0,0)}V_{\bar Z \bar Z}^{(0,0)} \rangle_\sigma &=&
- V_4 \frac{(p_1 \cdot p_2) \sqrt{G}}{16 (4 \pi^2 \alpha')^2} \int_0^\infty \frac{dt}{t^4}
  \int_{\cf_\sigma} d^2\nu_1 d^2\nu_2 \\
&&  \hspace{-1cm} \times \sum_{k=0,1} \sum_{\vec n=(n,m)^T}  \tr
\Bigg[
 e^{-\pi \vec{n}^{T} G \vec{n} t^{-1}} e^{2 \pi i \vec{\bf
 A}_\sigma \cdot \vec{n}} \sum_{{\alpha\beta}\atop{\rm even}}
\frac{\thba{\alpha}{\beta}(0,\tau)}{\eta^3(\tau)}
 \gamma_{\sigma,k} \cz_{\sigma,k}^{\rm int} \zba{\alpha}{\beta} \non
&& \hspace{-1cm}
 \times
\Big[ \langle \bar \partial Z(\bar \nu_1)
\bar \partial \bar{Z}(\bar \nu_2) \rangle_\sigma \langle \Psi(\nu_1)
\bar \Psi(\nu_2) \rangle_\sigma^{\alpha, \beta} \langle \psi(\nu_1)
\psi(\nu_2) \rangle_\sigma^{\alpha, \beta} \non
&& \hspace{-.5cm}
+ \langle \bar \partial Z(\bar \nu_1)
\partial \bar Z(\nu_2) \rangle_\sigma \langle \Psi(\nu_1)
\bar{\tilde{\Psi}}(\bar \nu_2) \rangle_\sigma^{\alpha, \beta} \langle \psi(\nu_1) \tilde
\psi(\bar \nu_2) \rangle_\sigma^{\alpha, \beta} + {\rm c.c.} \Big]
\Bigg] + {\cal O} (\delta^2)\ , \non
\non
\langle V_{Z}^{(0)B} V_{\bar Z}^{(0)C}\rangle_\sigma&=&
- V_4 \frac{(p_1 \cdot p_2)  \alpha' \sqrt{G}}{8 (4 \pi^2 \alpha')^2} \int_0^\infty \frac{dt}{t^4} \int_{(\partial \Sigma)_B} d\tilde{\nu}_1
\int_{(\partial \Sigma)_C} d\tilde{\nu}_2 \label{ZbZ}  \\
&& \hspace{-1cm} \times \sum_{k=0,1} \sum_{\vec n=(n,m)^T}
 \tr \Bigg[ \lambda_{s[B]} \lambda^\dag_{s[C]}
 e^{-\pi \vec{n}^{T} G \vec{n} t^{-1}}
e^{2 \pi i \vec{\bf A}_\sigma \cdot \vec{n}}
\sum_{{\alpha\beta}\atop{\rm even}}
\frac{\thba{\alpha}{\beta}(0,\tau)}{\eta^3(\tau)}
 \gamma_{\sigma,k} \cz_{\sigma,k}^{\rm int} \zba{\alpha}{\beta}
\non
&& \hspace{1.5cm}
 \times \langle \Psi(\nu_1) \bar \Psi(\nu_2) \rangle_\sigma^{\alpha, \beta}
\langle \psi(\nu_1) \psi(\nu_2) \rangle_\sigma^{\alpha, \beta} \Bigg] + {\cal O} (\delta^2)\ ,
 \non
\langle V_{ZZ}^{(0,0)}V_{Z\bar Z}^{(0,0)} \rangle_\sigma & = & - V_4 \frac{(p_1 \cdot p_2)  \sqrt{G}}{16 (4 \pi^2 \alpha')^2} \int_0^\infty \frac{dt}{t^4}
  \int_{\cf_\sigma} d^2\nu_1 d^2\nu_2 \\
 && \hspace{-1cm}
 \times \sum_{k=0,1} \sum_{\vec n=(n,m)^T}
\tr \Bigg[ e^{-\pi \vec{n}^{T} G \vec{n} t^{-1}}
e^{2 \pi i \vec{\bf A}_\sigma \cdot \vec{n}}
\sum_{{\alpha\beta}\atop{\rm even}}
\frac{\thba{\alpha}{\beta}(0,\tau)}{\eta^3(\tau)}
 \gamma_{\sigma,k} \cz_{\sigma,k}^{\rm int} \zba{\alpha}{\beta} \non
 && \hspace{-1cm}
 \times \Big[
\langle \bar \partial Z(\bar \nu_1)
\bar \partial Z(\bar \nu_2) \rangle_\sigma \langle \bar \Psi(\nu_1)
\Psi(\nu_2) \rangle_\sigma^{\alpha, \beta} \langle \psi(\nu_1) \psi(\nu_2) \rangle_\sigma^{\alpha, \beta} \non
&& \hspace{-.5cm} + \langle \partial Z(\nu_1)
\bar \partial Z(\bar \nu_2) \rangle_\sigma \langle \bar{\tilde \Psi}(\bar \nu_1)
\Psi(\nu_2) \rangle_\sigma^{\alpha, \beta} \langle \tilde \psi(\bar \nu_1)
\psi(\nu_2) \rangle_\sigma^{\alpha, \beta} + {\rm c.c.}
 \Big] \Bigg] + {\cal O} (\delta^2)\ ,\non
\langle V_{ZZ}^{(0,0)} V_{\bar Z}^{(0)B} \rangle_\sigma & = &- i V_4 \frac{(p_1 \cdot p_2)   \sqrt{2 \alpha'} \sqrt{G}}{16 (4 \pi^2 \alpha')^2} \int_0^\infty \frac{dt}{t^4}
  \int_{\cf_\sigma} d^2\nu_1 \int_{(\partial \Sigma)_B} d\tilde{\nu}_2 \label{ZZZbar} \\
&& \hspace{-1cm}\times \sum_{k=0,1} \sum_{\vec n=(n,m)^T} \tr \Bigg[ \lambda^\dag_{s[B]}
e^{-\pi \vec{n}^{T} G \vec{n} t^{-1}} e^{2 \pi i \vec{\bf
A}_\sigma \cdot \vec{n}}
 \sum_{{\alpha\beta}\atop{\rm even}} \frac{\thba{\alpha}{\beta}(0,\tau)}{\eta^3(\tau)}
 \gamma_{\sigma,k} \cz_{\sigma,k}^{\rm int} \zba{\alpha}{\beta} \non
&& \hspace{-1cm} \times \Big[
\langle i \bar \partial Z(\bar \nu_1) \rangle_\sigma \langle \Psi(\nu_1)
\bar \Psi(\nu_2) \rangle_\sigma^{\alpha, \beta} \langle \psi(\nu_1) \psi(\nu_2) \rangle_\sigma^{\alpha, \beta} \non
&& \hspace{-.5cm} + \langle i \partial Z(\nu_1) \rangle_\sigma \langle {\tilde \Psi}(\bar \nu_1)
\bar \Psi(\nu_2) \rangle_\sigma^{\alpha, \beta} \langle \tilde \psi(\bar \nu_1)
\psi(\nu_2) \rangle_\sigma^{\alpha, \beta}
 \Big] \Bigg] + {\cal O} (\delta^2)\ ,\non
\langle V_Z^{(0)B} V_{Z\bar Z}^{(0,0)} \rangle_\sigma & = & - iV_4 \frac{(p_1 \cdot p_2) \sqrt{2 \alpha'} \sqrt{G}}{16 (4 \pi^2 \alpha')^2} \int_0^\infty \frac{dt}{t^4}
  \int_{(\partial \Sigma)_B} d\tilde{\nu}_1 \int_{\cf_\sigma} d^2\nu_2 \label{ZZZb} \\
&& \hspace{-1cm}\times \sum_{k=0,1} \sum_{\vec n=(n,m)^T}  \tr \Bigg[ \lambda_{s[B]}
e^{-\pi \vec{n}^{T} G \vec{n} t^{-1}} e^{2 \pi i \vec{\bf
A}_\sigma \cdot \vec{n}}
\sum_{{\alpha\beta}\atop{\rm even}} \frac{\thba{\alpha}{\beta}(0,\tau)}{\eta^3(\tau)}
 \gamma_{\sigma,k} \cz_{\sigma,k}^{\rm int} \zba{\alpha}{\beta} \non
&& \hspace{-1cm} \times  \Big[
\langle i \bar \partial Z(\bar \nu_2) \rangle_\sigma \langle \Psi(\nu_1)
\bar \Psi(\nu_2) \rangle_\sigma^{\alpha, \beta} \langle \psi(\nu_1) \psi(\nu_2) \rangle_\sigma^{\alpha, \beta} \non
&& \hspace{-.5cm} + \langle i \partial Z(\nu_2) \rangle_\sigma \langle \Psi(\nu_1)
\bar {\tilde \Psi}(\bar \nu_2) \rangle_\sigma^{\alpha, \beta} \langle \psi(\nu_1)
\tilde \psi(\bar \nu_2) \rangle_\sigma^{\alpha, \beta}
 \Big] \Bigg] + {\cal O} (\delta^2)\ . \nonumber
\eeqn
The last three correlation functions obviously only get contributions
from the $Z$-zero modes (\ref{zeromode}), (note that the comment after
(\ref{ZZbZZb}), concerning
the sum over zero modes, also applies for formulas (\ref{ZZbZbZ})-(\ref{ZZZb})).
Another comment is in order here, concerning the minus sign in
(\ref{ZbZ}) and the factors of $i$ in (\ref{ZZZbar}) and
(\ref{ZZZb}). When we change coordinates in the open string amplitudes
via $\theta=e^{-i \nu}$, the boundary that ran along the real axis
before now runs along the imaginary axis, i.e.\ $\nu=i
\tilde{\nu}$. If we then change the integration variable from $\nu$ to
$\tilde{\nu}$ and define the derivative $\dot Z$ in the vertex
operator of the open string according to $\dot Z =
\partial_{\tilde{\nu}}Z$, the open string vertex operators become
\beqn \label{vopnutilde}
V_Z^{(0)B} &=& \frac{1}{\sqrt{2 \alpha'}} \lambda_{s[B]} \int_{(\partial\Sigma)_B}d \tilde{\nu} \big[i \dot Z + i 2\alpha' (p \cdot \psi) \Psi \big] e^{ipX} \ ,
\non
V_{\bar Z}^{(0)B} &=&  \frac{1}{\sqrt{2 \alpha'}} \lambda_{s[B]}^\dagger \int_{(\partial\Sigma)_B}d \tilde{\nu}\ \big[i \dot{\bar Z} + i 2\alpha' (p \cdot \psi) \bar \Psi \big] e^{ipX} \ ,
\eeqn
in particular there is a factor of $i$ now also in front of the
fermionic terms.

Following the same steps as in section \ref{1loopsection}, we see
that the theta functions from the internal partition function and
the fermionic  world-sheet correlators again drop out due to
(\ref{spinstructuresums}). The remaining bosonic world-sheet
correlators can be dealt with analogously as in the main text
using (\ref{torustrick}) and
\beqn
&& \int_{\cf_\sigma} d^2\nu_1 d^2\nu_2
\Big[ \langle \bar \partial Z(\bar \nu_1)
\bar \partial Z(\bar \nu_2) \rangle_\sigma
- \langle \partial Z(\nu_1) \bar \partial Z(\bar \nu_2) \rangle_\sigma + {\rm c.c.} \Big] \\
&& \hspace{1cm} = -2 \pi^4 c_\sigma^2 \frac{T_2}{U_2} (n + m \bar U)^2 \alpha'
= \left\{
\begin{array}{ll}
      -2 \pi^4 \frac{T_2}{U_2} (n + m \bar U)^2 \alpha' & {\rm for}\ \ca , \cm \\[.1cm]
     -8 \pi^4 \frac{T_2}{U_2} (n + m \bar U)^2 \alpha' & {\rm for}\ \ck
\end{array}
\right. \ , \non
&& \int_{\cf_\sigma} d^2\nu_1 \int_{\partial \Sigma}  d \tilde{\nu}_2
\Big[ \langle i \bar \partial Z(\bar \nu_1) \rangle_\sigma
- \langle i \partial Z(\nu_1) \rangle_\sigma \Big] \label{zeromode2} \\
&& \hspace{1cm} = -2 \pi^3 c_\sigma^3 d_\sigma \sqrt{\frac{T_2}{2 U_2}} (n + m \bar U) t \sqrt{\alpha'} = \left\{
\begin{array}{ll}
      -2 \pi^3 \sqrt{\frac{T_2}{2 U_2}} (n + m \bar U) t \sqrt{\alpha'} & {\rm for}\ \ca \\[.1cm]
      -4 \pi^3 \sqrt{\frac{T_2}{2 U_2}} (n + m \bar U) t \sqrt{\alpha'} & {\rm for}\ \cm \\[.1cm]
      0 & {\rm for}\ \ck
\end{array}
\right. \ , \nonumber
\eeqn
which only get contributions from the zero modes. In (\ref{zeromode2})
we had to introduce another constant $d_\sigma$. For the sake of
completeness let us here list all three constants $c_\sigma$, $d_\sigma$
and $e_\sigma$ that we use in various places throughout this paper,
\beqn \label{constants}
c_\sigma= \left\{
\begin{array}{ll}
      1 & {\rm for}\ \ca \\
      1 & {\rm for}\ \cm \\
      2 &  {\rm for}\ \ck
\end{array}
\right. \ , \quad
d_\sigma= \left\{
\begin{array}{ll}
      1 & {\rm for}\ \ca \\
      2 & {\rm for}\ \cm \\
      0 &  {\rm for}\ \ck
\end{array}
\right. \ , \quad
e_\sigma= \left\{
\begin{array}{ll}
      1 & {\rm for}\ \ca \\
      4 & {\rm for}\ \cm \\
      4 &  {\rm for}\ \ck
\end{array}
\right. \ .
\eeqn
The integrals over the world-sheet modulus $t$ can now be
performed similarly to (\ref{intKK}) where we have to regulate the
integrals again with a UV cutoff $\Lambda$ and, where necessary,
with an IR cutoff $\chi$  (UV and IR referring to the open string
channel),
\beqn
&&  \int_0^\infty \frac{dt}{t^4}
\sum_{\vec n=(n,m)^T} e^{-\pi \vec{n}^{T} G \vec{n} t^{-1}}
e^{2 \pi i \vec{\bf A}_\sigma \cdot \vec{n}} \Big(
-2 \pi^4 c_\sigma^2 \alpha' \frac{T_2}{U_2} (n + m \bar U)^2 \Big) \non
&& \hspace{1cm} = -4 \pi c_\sigma^2 \alpha' \frac{U_2^2}{T_2^2}
\sum_{\vec n=(n,m)^T} \!\!\!\!\!\! ' \; \; \; \frac{e^{2 \pi i \vec{\bf A}_\sigma \cdot \vec{n}}}{(n+m U)^3 (n+m \bar U)}\ , \\
&& \non
&& \int_0^\infty \frac{dt}{t^3}
\sum_{\vec n=(n,m)^T} e^{-\pi \vec{n}^{T} G \vec{n} t^{-1}}
e^{2 \pi i \vec{\bf A}_\sigma \cdot \vec{n}} \Big(
-2 \pi^3 c_\sigma^3 d_\sigma \sqrt{\alpha'} \sqrt{\frac{T_2}{2 U_2}} (n + m \bar U) \Big) \non
&& \hspace{1cm} = -\sqrt{2} \pi c_\sigma^3 d_\sigma \sqrt{\alpha'} \frac{U_2^{3/2}}{T_2^{3/2}}
\sum_{\vec n=(n,m)^T} \!\!\!\!\!\! ' \; \; \; \frac{e^{2 \pi i \vec{\bf A}_\sigma \cdot \vec{n}}}{|n+m U|^4} (n+m \bar U)\ , \\
&& \non
&& \int_{1/(e_\sigma \Lambda^2)}^\infty \frac{dt}{t^2}
\sum_{\vec n=(n,m)^T} e^{-\pi \vec{n}^{T} G \vec{n} t^{-1}}
e^{2 \pi i \vec{\bf A}_\sigma \cdot \vec{n}} \Big( \pi^2 d_\sigma^2 e^{-2 \pi \chi t} \Big) \non
&& \hspace{1cm} =  \pi^2 d_\sigma^2 e_\sigma \Lambda^2 + \pi d_\sigma^2 T_2^{-1}
\tilde{E}_1({\bf A}_\sigma,U)+ \ldots \ .
\label{KKe1}
\eeqn
We defined the function $\tilde{E}_1(A,U)$ as
\beqn
\tilde{E}_1(A,U) = E_1(A,U) -
 \pi \ln \Big( 1+ 2 \pi \chi \frac{\sqrt{G}U_2}{|A|^2}
\Big)  \ ,
\eeqn
where $E_1(A,U)$ was defined in (\ref{eisen}) for $s=1$.
Note that $\tilde{E}_1(A,U)$ has a smooth limit for $A=0$. To see this one has to make use of
\be
E_1(A,U) = - \pi \ln\left| \frac{\vartheta_1 (A|U)}{\eta(U)} \right|^2 + 2 \pi^2 U_2 a_4^2\ ,
\ee
and then proceed as in \cite{Berg:2004ek} to show that
\be
\tilde{E}_1(0,U) = - \frac{\pi}{T_2} \ln(8 \pi^3 \chi T_2 U_2 |\eta(U)|^4) \ .
\ee
On the other hand, for $A \neq 0$, it has the expansion
\be
\tilde{E}_1(A,U) = E_1(A,U) + {\cal O}\Big(\frac{\chi \sqrt{G}U_2}{|A|^2}\Big)\ ,
\ee
i.e.\ there is no need for an IR cutoff $\chi$ which can be set
to zero in the corresponding terms. Altogether, the building block
correlators of (\ref{ZZbZbZ})-(\ref{ZZZb}), turn out to be
\beqn
\langle V_{ZZ}^{(0,0)}V_{\bar Z\bar Z}^{(0,0)} \rangle_\sigma &=&(p_1 \cdot p_2) \alpha' \frac{V_4}{(4 \pi^2 \alpha')^2} \frac{3 c_\sigma^2 \pi}{16 T_2}
\sum_{k} \tr \Big[ E_2({\bf A}_\sigma,U) \gamma_{\sigma,k} \cq_{\sigma,k} \Big]\ ,
\label{final2} \\
\langle V_{Z}^{(0)B}V_{\bar Z}^{(0)C} \rangle_\sigma &=& - (p_1 \cdot p_2) \alpha' \frac{V_4}{(4 \pi^2 \alpha')^2}
\frac{d_\sigma^2 \pi}{8}
\sum_{k} \tr \Big[ \lambda_{s[B]} \lambda^\dagger_{s[C]} \tilde{E}_1({\bf A}_\sigma,U)
\gamma_{\sigma,k} \cq_{\sigma,k} \Big] \ , \non
\langle V_{ZZ}^{(0,0)}V_{Z\bar Z}^{(0,0)} \rangle_\sigma &=&(p_1 \cdot p_2) \alpha' \frac{V_4}{(4 \pi^2 \alpha')^2} \frac{c_\sigma^2 \pi U_2^2}{4 T_2} \times \non
&& \hspace{2cm} \times
\sum_{k} \tr \Big[ \sum_{\vec n=(n,m)^T} \!\!\!\!\!\! ' \; \; \; \frac{e^{2 \pi i \vec{\bf A}_\sigma \cdot \vec{n}}}{(n+m U)^3 (n+m \bar U)} \gamma_{\sigma,k} \cq_{\sigma,k} \Big]\ , \non
\langle V_{ZZ}^{(0,0)} V_{\bar Z}^{(0)B} \rangle_\sigma &=& i (p_1 \cdot p_2) \alpha' \frac{V_4}{(4 \pi^2 \alpha')^2}
\frac{c_\sigma^3 d_\sigma \pi U_2^{3/2}}{8 T_2^{1/2}} \times \non
&& \hspace{2cm} \times
\sum_{k} \tr \Big[ \lambda^\dagger_{s[B]} \sum_{\vec n=(n,m)^T} \!\!\!\!\!\! ' \; \; \; \frac{e^{2 \pi i \vec{\bf A}_\sigma \cdot \vec{n}}}{|n+m U|^4}(n+m \bar U) \gamma_{\sigma,k} \cq_{\sigma,k} \Big] \ , \non
\langle V_{Z}^{(0)B} V_{Z\bar Z}^{(0,0)} \rangle_\sigma &=& i (p_1 \cdot p_2) \alpha' \frac{V_4}{(4 \pi^2 \alpha')^2}
\frac{c_\sigma^3 d_\sigma \pi U_2^{3/2}}{8 T_2^{1/2}} \times \non
&& \hspace{2cm} \times
\sum_{k} \tr \Big[ \lambda_{s[B]} \sum_{\vec n=(n,m)^T} \!\!\!\!\!\! ' \; \; \; \frac{e^{2 \pi i \vec{\bf A}_\sigma \cdot \vec{n}}}{|n+m U|^4}(n+m \bar U) \gamma_{\sigma,k} \cq_{\sigma,k} \Big] \ . \nonumber
\eeqn

In the 2-point functions for the open string moduli, the only contributions now come from
\beqn
\sum_{i,j} \ca_{ij}^{(1)} + \sum_{i,a} \big[ \ca_{ai}^{(0)}
 + \ca_{ia}^{(0)} \big] + \sum_i \cm_i^{(1)} \ .
\eeqn
To evaluate them we have to know the form of the Chan-Paton
matrices. They are given by
\beqn
\cm : \quad \lambda_{s[1]} & = & {\rm diag} ({\bf 1}_{N_{s[1]}}, -{\bf 1}_{N_{s[1]}}) \oplus {\bf 0}_{32-2N_{s[1]}} \ , \label{lambdas} \\
\ca : \quad \lambda_{s[1]} & = & \Big({\rm diag} ({\bf 1}_{N_{s[1]}}, -{\bf 1}_{N_{s[1]}}) \oplus {\bf 0}_{32-2N_{s[1]}}\Big) \otimes {\bf 1}_{32}\ , \non
\lambda_{s[2]}   & = & {\bf 1}_{32} \otimes \Big(-{\rm diag} ({\bf 1}_{N_{s[2]}}, -{\bf 1}_{N_{s[2]}}) \oplus {\bf 0}_{32-2N_{s[2]}} \Big)\ , \nonumber
\eeqn
where the extra minus in $\lambda_{s[2]}$ comes from the charge
$q_B$ at the string end point (compare the comment below equations
(\ref{vops})). These matrices then lead to
\beqn
\sum_{\sigma} d_\sigma^2 \sum_{k=0,1} \tr \Big[
\lambda_{s[B]} \lambda_{s[C]}^\dag \tilde E_1({\bf A}_\sigma,U) \gamma_{\sigma,k} \cq_{\sigma,k} \Big] &&  \label{sumE1} \\
&& \hspace{-8cm} ~=~  -4 \sum_{i,j} N_iN_j \Big[ ( \delta_{i s[B]} \delta_{i s[C]} + \delta_{j s[B]} \delta_{j s[C]} )  [  \tilde E_1(A_i - A_j,U) + \tilde E_1(-A_i + A_j,U) \non
&& \hspace{-.3cm} - \tilde E_1(A_i + A_j,U) - \tilde E_1(-A_i - A_j,U) ] \non
&& \hspace{-5cm} - ( \delta_{i s[B]} \delta_{j s[C]} + \delta_{j s[B]} \delta_{i s[C]} )  [  \tilde E_1(A_i - A_j,U) + \tilde E_1(-A_i + A_j,U) \non
&& \hspace{-.3cm} + \tilde E_1(A_i + A_j,U) + \tilde E_1(-A_i - A_j,U) ] \Big] \non
&& \hspace{-7cm}  + 2\cdot 32 \sum_i N_i \delta_{i s[B]} \delta_{i s[C]} [  \tilde E_1(A_i,U) + \tilde E_1(-A_i,U) ] \non
&&  \hspace{-7cm}  - 16 \sum_i N_i \delta_{i s[B]} \delta_{i s[C]} [  \tilde E_1(2A_i,U) + \tilde E_1(-2A_i,U) ]\ .
\nonumber
\eeqn
For future reference, we introduce a closely related quantity, where both lambda matrices are
inserted on the same stack of branes that we choose to be the $i$th stack, i.e.\ $s[B] = s[C] = i$,
\beqn \label{defE1}
\ce_1^{(i)}(A_l,U) &=& \sum_{\sigma} d_\sigma^2 \sum_{k=0,1} \tr \Big[
\lambda_{i} \lambda_{i}^\dag \tilde E_1({\bf A}_\sigma,U)
\gamma_{\sigma,k} \cq_{\sigma,k} \Big] \\
&&\hspace{-1cm} = -~8 N_i \sum_l N_l [\te_1(A_i - A_l,U)+\te_1(-A_i + A_l,U)
\nonumber \\[-.3cm]
&& \hspace{3cm} -\te_1(A_i + A_l,U)-\te_1(-A_i - A_l,U)] \non
&&\hspace{-1cm} +~ 8 N_i^2 [2 \te_1(0,U) + \te_1(2A_i,U)+\te_1(-2A_i,U)] \non
&&\hspace{-1cm} +~64 N_i [\te_1(A_i,U) + \te_1(-A_i,U)] - 16 \delta_{ij} N_i [\te_1(2 A_i,U) + \te_1(-2 A_i,U)] . \nonumber
\eeqn
The UV divergences of the 95 annulus and the M\"obius strip
proportional to $\Lambda^2$ cancel against each other. All other
terms involving the UV cutoff become proportional to $\sum_i N_i
\partial_\mu A_i$ when inserted into the effective Lagrangian,
which is zero due to the anomaly constraint
\be \label{anomal}
\sum_i N_i A_i =0\ ,
\ee
that  ensures a decoupling of the anomalous overall $U(1)$ in
$U(16)$
\cite{Berkooz:1996iz}.

We next evaluate the traces of the other expressions in (\ref{final2})
\beqn
\sum_{\sigma} c_\sigma^2
 \sum_{k=0,1} \tr \Big[
\sum_{\vec n=(n,m)^T} \!\!\!\!\!\! ' \; \; \; \frac{e^{2 \pi i \vec{\bf A}_\sigma \cdot \vec{n}}}{(n+m U)^3 (n+m \bar U)} \gamma_{\sigma,k} \cq_{\sigma,k} \Big]
&& \\
&& \hspace{-9cm} ~=~
-4 \sum_{i,j} N_iN_j \sum_{\vec n=(n,m)^T} \!\!\!\!\!\! ' \; \; \; \frac{1}{(n+m U)^3 (n+m \bar U)}  \Big[
e^{2 \pi i (\vec{a}_i - \vec{a}_j) \cdot \vec{n}} + e^{2 \pi i (-\vec{a}_i + \vec{a}_j) \cdot \vec{n}}  \non
&& \hspace{-1.2cm} - e^{2 \pi i (\vec{a}_i + \vec{a}_j) \cdot \vec{n}} - e^{2 \pi i (-\vec{a}_i - \vec{a}_j) \cdot \vec{n}} \Big] \non
&& \hspace{-8cm}
+~ 2\cdot 32 \sum_i N_i \sum_{\vec n=(n,m)^T} \!\!\!\!\!\! ' \; \; \; \frac{1}{(n+m U)^3 (n+m \bar U)} \Big[ e^{2 \pi i \vec{a}_i \cdot \vec{n}} + e^{-2 \pi i \vec{a}_i \cdot \vec{n}}  \Big] \non
&& \hspace{-8cm}
-~ 4\sum_i N_i \sum_{\vec n=(n,m)^T} \!\!\!\!\!\! ' \; \; \; \frac{1}{(n+m U)^3 (n+m \bar U)} \Big[ e^{4 \pi i \vec{a}_i \cdot \vec{n}} + e^{-4 \pi i \vec{a}_i \cdot \vec{n}}  \Big] \ , \non
&& \hspace{-10.3cm} \sum_\sigma c_\sigma^3 d_\sigma
\sum_{k=0,1} \tr \Big[ \lambda_{s[B]} \sum_{\vec n=(n,m)^T} \!\!\!\!\!\! ' \; \; \; \frac{e^{2 \pi i \vec{\bf A}_\sigma \cdot \vec{n}}}{|n+m U|^4}(n+m \bar U) \gamma_{\sigma,k} \cq_{\sigma,k} \Big] \label{4thtrace} \\
&& \hspace{-9cm} ~=~ -4 \sum_{i,j} N_i N_j \sum_{\vec n=(n,m)^T} \!\!\!\!\!\! ' \; \; \;
\frac{n+m \bar U}{|n+mU|^4} \Big( \delta_{is[B]} \Big[ e^{2 \pi i (\vec{a}_i - \vec{a}_j) \cdot \vec{n}} - e^{2 \pi i (-\vec{a}_i + \vec{a}_j) \cdot \vec{n}}  \non
&& \hspace{-1.5cm} - e^{2 \pi i (\vec{a}_i + \vec{a}_j) \cdot \vec{n}} + e^{2 \pi i (-\vec{a}_i - \vec{a}_j) \cdot \vec{n}} \Big] \non
&& \hspace{-3cm} -~ \delta_{js[B]} \Big[ e^{2 \pi i (\vec{a}_i - \vec{a}_j) \cdot \vec{n}} - e^{2 \pi i (-\vec{a}_i + \vec{a}_j) \cdot \vec{n}}  \non
&& \hspace{-1.5cm} + e^{2 \pi i (\vec{a}_i + \vec{a}_j) \cdot \vec{n}} - e^{2 \pi i (-\vec{a}_i - \vec{a}_j) \cdot \vec{n}} \Big] \Big) \non
&& \hspace{-8cm} +~ 2 \cdot 32 \sum_i N_i \delta_{i s[B]} \sum_{\vec n=(n,m)^T} \!\!\!\!\!\! ' \; \; \;
\frac{n+m \bar U}{|n+mU|^4} \Big[ e^{2 \pi i \vec{a}_i \cdot \vec{n}}
- e^{-2 \pi i \vec{a}_i \cdot \vec{n}} \Big] \non
&& \hspace{-8cm} -~ 8 \sum_i N_i \delta_{i s[B]} \sum_{\vec n=(n,m)^T} \!\!\!\!\!\! ' \; \; \; \frac{n+m \bar U}{|n+mU|^4} \Big[
e^{4 \pi i \vec{a}_i \cdot \vec{n}} - e^{-4 \pi i \vec{a}_i \cdot \vec{n}} \Big] \nonumber \ .
\eeqn
Also both of these traces only get contributions from the 99- and 95-annulus and the 9-brane M\"obius strip. Moreover, the trace in (\ref{4thtrace}) with $\lambda_{s[B]}$ replaced by $\lambda_{s[B]}^\dagger$ gives the same result.

Putting everything together, we finally get
\beqn
\langle V_{U} V_{\bar U} \rangle &=& (p_1 \cdot p_2) \alpha' g_c^2 \alpha'^{-4} \frac{V_4}{(4 \pi^2 \alpha')^2} e^{-\Phi} \Big[ -\frac{3 i}{4 \sqrt{2} (U- \bar U)^2 (S_0' - \bar S_0 ')} \ce_2 (A_k, U)  \label{vuu} \non
&& + \frac{\pi}{4 \sqrt{2} (U - \bar U)(S_0' - \bar S_0 ')} \sum_{\vec n=(n,m)^T} \!\!\!\!\!\! ' \; \; \; \frac{2n+m (U + \bar U)}{|n+mU|^4} \times \non
&& \times \Big( -4 \sum_{i,j} N_i N_j [(A_i - \bar A_i - (A_j - \bar A_j)) (e^{2 \pi i (\vec{a}_i - \vec{a}_j) \cdot \vec{n}} - e^{2 \pi i (-\vec{a}_i + \vec{a}_j) \cdot \vec{n}}) \non
&& -  (A_i - \bar A_i + (A_j - \bar A_j)) (e^{2 \pi i (\vec{a}_i + \vec{a}_j) \cdot \vec{n}} - e^{2 \pi i (-\vec{a}_i - \vec{a}_j) \cdot \vec{n}})] \non
&&+~ 64 \sum_i N_i (A_i - \bar A_i) ( e^{2 \pi i \vec{a}_i \cdot \vec{n}}
- e^{-2 \pi i \vec{a}_i \cdot \vec{n}}) \non
&&  -~ 8 \sum_i N_i (A_i - \bar A_i) ( e^{4 \pi i \vec{a}_i \cdot \vec{n}}
- e^{-4 \pi i \vec{a}_i \cdot \vec{n}}) \Big) \\
&& + \frac{\pi^2}{\sqrt{2} (U-\bar U)^3 (S_0' - \bar S_0 ')} \Big( -4 \sum_{i,j} N_i N_j \non
&&  [(A_i - \bar A_i - (A_j - \bar A_j))^2 (\te_1(A_i - A_j,U)+\te_1(-A_i + A_j,U)) \non
&& - (A_i - \bar A_i + (A_j - \bar A_j))^2 (\te_1(A_i + A_j,U)+\te_1(-A_i - A_j,U))] \non
&& +~ 64 \sum_i N_i (A_i - \bar A_i)^2 (\te_1(A_i,U)+\te_1(-A_i,U)) \non
&& -~ 16 \sum_i N_i (A_i - \bar A_i)^2 (\te_1(2A_i,U)+\te_1(-2A_i,U)) \Big) \Big]\ , \non
\langle V_{A_i} V_{\bar A_j} \rangle &=& (p_1 \cdot p_2) \alpha' g_o^2 \alpha'^{-4} \frac{V_4}{(4 \pi^2 \alpha')^2} e^{-\Phi} \frac{1}{\sqrt{2} (U-\bar U) (S_0' - \bar S_0 ')} \times  \label{vaa} \non
&& \times \Big[ - 8 \delta_{ij} N_i \sum_l N_l [\te_1(A_i - A_l,U)+\te_1(-A_i + A_l,U) \non
&& \hspace{3cm} -\te_1(A_i + A_l,U)-\te_1(-A_i - A_l,U)] \non
&& +~ 8 N_i N_j [\te_1(A_i - A_j,U)+\te_1(-A_i + A_j,U) \\
&& \hspace{2cm} + \te_1(A_i + A_j,U)+\te_1(-A_i - A_j,U)] \non
&& + 64 \delta_{ij} N_i [\te_1(A_i,U) + \te_1(-A_i,U)] \non
&& - 16 \delta_{ij} N_i [\te_1(2 A_i,U) + \te_1(-2 A_i,U)]  \Big] \ , \non
\langle V_{U} V_{S_2'} \rangle &=& (p_1 \cdot p_2) \alpha' g_c^2 \alpha'^{-4} \frac{V_4}{(4 \pi^2 \alpha')^2} e^{-\Phi} \Big[ \frac{i \pi}{8 \sqrt{2} (S_0' - \bar S_0 ')^2} \sum_{\vec n=(n,m)^T} \!\!\!\!\!\! ' \; \; \; \frac{n+m \bar U}{|n+mU|^4} \times \label{vus} \non
&& \times \Big( -4 \sum_{i,j} N_i N_j [(A_i - \bar A_i - (A_j - \bar A_j)) (e^{2 \pi i (\vec{a}_i - \vec{a}_j) \cdot \vec{n}} - e^{2 \pi i (-\vec{a}_i + \vec{a}_j) \cdot \vec{n}}) \non
&& -  (A_i - \bar A_i + (A_j - \bar A_j)) (e^{2 \pi i (\vec{a}_i + \vec{a}_j) \cdot \vec{n}} - e^{2 \pi i (-\vec{a}_i - \vec{a}_j) \cdot \vec{n}})] \non
&&+~ 64 \sum_i N_i (A_i - \bar A_i) ( e^{2 \pi i \vec{a}_i \cdot \vec{n}}
- e^{-2 \pi i \vec{a}_i \cdot \vec{n}}) \non
&&  -~ 8 \sum_i N_i (A_i - \bar A_i) ( e^{4 \pi i \vec{a}_i \cdot \vec{n}}
- e^{-4 \pi i \vec{a}_i \cdot \vec{n}}) \Big) \\
&& - \frac{(U -\bar U)}{8 \sqrt{2} (S_0' - \bar S_0 ')^2}
\Big( -4 \sum_{i,j} N_iN_j \sum_{\vec n=(n,m)^T} \!\!\!\!\!\! ' \; \; \; \frac{1}{(n+m U)^3 (n+m \bar U)} \non
&& [ e^{2 \pi i (\vec{a}_i - \vec{a}_j) \cdot \vec{n}} + e^{2 \pi i (-\vec{a}_i + \vec{a}_j) \cdot \vec{n}} - e^{2 \pi i (\vec{a}_i + \vec{a}_j) \cdot \vec{n}} - e^{2 \pi i (-\vec{a}_i - \vec{a}_j) \cdot \vec{n}} ] \non
&& +~ 64 \sum_i N_i \sum_{\vec n=(n,m)^T} \!\!\!\!\!\! ' \; \; \; \frac{1}{(n+m U)^3 (n+m \bar U)} [ e^{2 \pi i \vec{a}_i \cdot \vec{n}} + e^{-2 \pi i \vec{a}_i \cdot \vec{n}} ] \non
&& -~ 4\sum_i N_i \sum_{\vec n=(n,m)^T} \!\!\!\!\!\! ' \; \; \; \frac{1}{(n+m U)^3 (n+m \bar U)} [ e^{4 \pi i \vec{a}_i \cdot \vec{n}} + e^{-4 \pi i \vec{a}_i \cdot \vec{n}}  ] \Big) \Big]\ , \non
\langle V_{U} V_{\bar A_i} \rangle &=& (p_1 \cdot p_2) \alpha' g_c g_o \alpha'^{-4} \frac{V_4}{(4 \pi^2 \alpha')^2} e^{-\Phi} \Big[ -\frac{1}{4 \sqrt{2} (S_0' - \bar S_0 ')} \sum_{\vec n=(n,m)^T} \!\!\!\!\!\! ' \; \; \; \frac{n+m \bar U}{|n+mU|^4} \times \label{vua} \non
&& \times \Big( -8 \sum_{l} N_i N_l [e^{2 \pi i (\vec{a}_i - \vec{a}_l) \cdot \vec{n}} - e^{2 \pi i (-\vec{a}_i + \vec{a}_l) \cdot \vec{n}} \non
&& \hspace{3cm} -~ e^{2 \pi i (\vec{a}_i + \vec{a}_l) \cdot \vec{n}} + e^{2 \pi i (-\vec{a}_i - \vec{a}_l) \cdot \vec{n}}] \non
&& +~ 64 N_i [ e^{2 \pi i \vec{a}_i \cdot \vec{n}}
- e^{-2 \pi i \vec{a}_i \cdot \vec{n}}] - 8 N_i [ e^{4 \pi i \vec{a}_i \cdot \vec{n}}
- e^{-4 \pi i \vec{a}_i \cdot \vec{n}}] \Big) \\
&&  - \frac{\pi}{\sqrt{2} (U - \bar U)^2 (S_0' - \bar S_0 ')} \Big( - 8 N_i \sum_l N_l [(A_i - \bar A_i - (A_l - \bar A_l))\times \non
&& \times (\te_1(A_i - A_l,U) + \te_1(-A_i +A_l,U)) \non
&& -~ (A_i - \bar A_i + (A_l - \bar A_l))(\te_1(A_i + A_l,U) + \te_1(-A_i - A_l,U))] \non
&& +~ 64 N_i [\te_1(A_i,U) + \te_1(-A_i,U)] - 16 N_i [\te_1(2 A_i,U) + \te_1(-2 A_i,U)] \Big) \Big] \ , \non
\langle V_{A_i} V_{S_2'} \rangle &=& (p_1 \cdot p_2) \alpha' g_c g_o \alpha'^{-4} \frac{V_4}{(4 \pi^2 \alpha')^2} e^{-\Phi} \Big[ -\frac{i (U-\bar U)}{8 \sqrt{2} (S_0' - \bar S_0 ')^2} \sum_{\vec n=(n,m)^T} \!\!\!\!\!\! ' \; \; \; \frac{n+m \bar U}{|n+mU|^4} \times \label{vas} \non
&& \times \Big( -8 \sum_{l} N_i N_l [e^{2 \pi i (\vec{a}_i - \vec{a}_l) \cdot \vec{n}} - e^{2 \pi i (-\vec{a}_i + \vec{a}_l) \cdot \vec{n}} \non
&& \hspace{3cm} -~ e^{2 \pi i (\vec{a}_i + \vec{a}_l) \cdot \vec{n}} + e^{2 \pi i (-\vec{a}_i - \vec{a}_l) \cdot \vec{n}}] \\
&&+~ 64 N_i [ e^{2 \pi i \vec{a}_i \cdot \vec{n}}
- e^{-2 \pi i \vec{a}_i \cdot \vec{n}}] - 8 N_i [ e^{4 \pi i \vec{a}_i \cdot \vec{n}}
- e^{-4 \pi i \vec{a}_i \cdot \vec{n}}] \Big) \Big] \ .\nonumber
\eeqn
To see that this is consistent with (\ref{k1loop}), note that
\beqn \label{derivE}
\partial_U \partial_{\bar U} E_2(A,U) &=& - \frac{2}{(U-\bar U)^2} E_2(A,U) - \frac{2 \pi^2 i (A- \bar A)^2}{(U-\bar U)^3} E_1(A,U) \non
&& - \frac{\pi i}{2} \frac{A - \bar A}{U - \bar U} \sum_{\vec n=(n,m)^T} \!\!\!\!\!\! ' \; \; \; \frac{e^{2 \pi i \vec{a} \cdot \vec{n}}}{|n+mU|^4} (2n+m(U+\bar U))\ , \non
\partial_A \partial_{\bar A} E_2(A,U) &=& \frac{-2\pi^2 i}{U-\bar U} E_1(A,U) \ , \non
\partial_U E_2(A,U) &=& \frac{\pi i (A-\bar A)}{2} \sum_{\vec n=(n,m)^T} \!\!\!\!\!\! ' \; \; \; \frac{e^{2 \pi i \vec{a} \cdot \vec{n}}}{(n+m \bar U) (n+m U)^2} \non
&& -~ \frac{U- \bar U}{2} \sum_{\vec n=(n,m)^T} \!\!\!\!\!\! ' \; \; \; \frac{e^{2 \pi i \vec{a} \cdot \vec{n}}}{(n+m \bar U) (n+m U)^3}\ ,   \non
\partial_U \partial_{\bar A} E_2(A,U) &=& \frac{2 \pi^2 i (A - \bar A)}{(U - \bar U)^2} E_1(A,U) + \frac{\pi i}{2} \sum_{\vec n=(n,m)^T} \!\!\!\!\!\! ' \; \; \; \frac{e^{2 \pi i \vec{a} \cdot \vec{n}}}{(n+m \bar U) (n+m U)^2}\ ,\non
\partial_{A} E_2(A,U) &=& -\frac{\pi i (U-\bar U)}{2} \sum_{\vec n=(n,m)^T} \!\!\!\!\!\! ' \; \; \; \frac{e^{2 \pi i \vec{a} \cdot \vec{n}}}{(n+m \bar U) (n+m U)^2}\ . \label{e2deriv}
\eeqn
To compare this (in combination with (\ref{k1loop})) with
(\ref{vuu})-(\ref{vas}) we have to perform a Weyl rescaling in the
latter. For most cases this just amounts to a rescaling with the
factor given in (\ref{Weyl}). However, as for the case of $S'$
discussed in the main text, also for the kinetic term of $U$ the
correction to the Einstein Hilbert term (\ref{eh1loop}), calculated in
\cite{ABFPT}, has to be taken into account. Again the reason is the
presence of the kinetic term for $U$ already at sphere level, cf.\
(\ref{ktree}). To make this more precise let us have a look at the
relevant terms in the effective action (using $\cv = \cv_{\rm
K3}\sqrt G$)
\beqn \label{uweyl}
&& \hspace{0cm} \frac12 \Big[e^{-2\Phi} \cv + \frac{\tilde c}{\sqrt{G}} \ce_2(A_i,U)  \Big] R
 + \Big[ \frac{e^{-2\Phi} \cv}{(U-\bar U)^2} + \frac{(\tilde c + \tilde c_0) \ce_2(A_i,U) }{\sqrt{G}(U - \bar U)^2} + \ldots \Big]
  \partial_\mu U \partial^\mu \bar U \non
\hspace{0cm}
&\stackrel{\rm Weyl}{\longrightarrow}&
 \frac12 R + \Big[ \frac{1}{(U-\bar U)^2} + \frac{\tilde c_0 \ce_2(A_i,U) e^{2 \Phi}}{(U - \bar U)^2  \sqrt{G} \cv} + \ldots \Big] \partial_\mu U \partial^\mu \bar U \ ,
\eeqn
where we only displayed the part of the one-loop correction to the
kinetic term of $U$ that is proportional to $\ce_2(A_i,U)$ and
omitted terms of order $\co(e^{3 \Phi})$. Furthermore, the
coefficient $\tilde c +
\tilde c_0$ represents a split into the contribution to the correlator
$\langle V_{ZZ}^{(0,0)}V_{\bar Z\bar Z}^{(0,0)} \rangle_\sigma$
coming from fluctuations ($\tilde c$) and zero modes ($\tilde
c_0$), cf.\ (\ref{torustrick}). The correction to the Einstein-Hilbert term only comes from fluctuations (there are no zero modes
along the non-compact directions). This means that
 after the Weyl rescaling, only
zero mode contributions survive in the kinetic term
of $U$ (this is analogous to the case of $S'$, 
discussed in the main text, cf.\
footnote \ref{fluctfn}). This gives exactly the right relative
factor compared to the other contributions in (\ref{vuu}), i.e.\
those that are not proportional to $\ce_2(A_i,U)$, in order to be
consistent with (\ref{e2deriv}) (to be more precise, the factor
$-3i/(4 \sqrt{2})$ in the first line of (\ref{vuu}) becomes
$-i/\sqrt{2}$).

In order to read off the kinetic terms from the amplitudes
(\ref{vuu})-(\ref{vas}) we have to make the replacements
\be \label{substv4}
V_4 ~\rightarrow~ d^4 x \sqrt{-g}
\ee
and
\beqn
(p_1 \cdot p_2) g_c^2 \pi^2 \alpha'^{-4} &\rightarrow&
\partial_\mu U \partial^\mu \bar U \quad , \quad {\rm in \
(\ref{vuu})}\ , \non
(p_1 \cdot p_2) g_o^2 \alpha'^{-4}
&\rightarrow& \partial_\mu A_i \partial^\mu \bar A_j \quad , \quad
{\rm in \ (\ref{vaa})}\ , \non
\frac12 (p_1 \cdot p_2) g_c^2 \pi^2
\alpha'^{-4} &\rightarrow& \partial_\mu U \partial^\mu S_2' \quad
, \quad {\rm in \ (\ref{vus})}\ , \non
(p_1 \cdot p_2) g_c g_o \pi
\alpha'^{-4} &\rightarrow& \partial_\mu U \partial^\mu \bar A_i
\quad , \quad {\rm in \ (\ref{vua})}\ , \non
\frac12 (p_1 \cdot p_2) g_c
g_o \pi \alpha'^{-4} &\rightarrow& \partial_\mu A_i \partial^\mu
S_2' \quad , \quad {\rm in \ (\ref{vas})} \ .  \label{replace}
\eeqn
Again we neglected overall numerical factors. The additional
factor of $\frac12$ on the left hand sides in those cases that
involve an $S_2'$ should ultimately be checked by comparing with
the corresponding normalization at tree level. In order to compare
with the metrics (\ref{k1loop}), one finally has to take into
account that
\be \label{ifactor}
G_{U \bar S'} \partial_\mu U \partial^\mu \bar S' = -i G_{U \bar
S'} \partial_\mu U \partial^\mu S'_2 +  G_{U \bar S'}
\partial_\mu U \partial^\mu S'_1 \ .
\ee
This means that we still have to multiply the results (\ref{vus})
and (\ref{vas}), i.e.\ the correlators involving one $S_2'$,  by
$i$ in order to read off the actual K\"ahler metric. Taking all
this into account one verifies that our results
(\ref{vuu})-(\ref{vas}) are reproduced by the K\"ahler potential
(\ref{tollesK}) (up to a common numerical constant that we did
not determine).

\section{Fixing the constant $c$}
\label{findc}

In section \ref{1loopsection} we have determined the correction to the
K\"ahler potential up to a numerical constant $c$ which we left free
in the final result (\ref{tollesK}). It would be nice to fix it by direct comparison of the 
tree and one-loop contributions to the K\"ahler metric. However, here we 
use an indirect method and make use of the relation (\ref{km-gc}) that relates
the metric to the gauge couplings of D9-brane gauge groups, which reads in our case
\beqn  \label{comparisoneqs}
K_{A_i\bar A_i} \Big|_{A_i=0} &=&
- \frac{1}{4\pi}
\frac{(S-\bar S) - 4\pi c
(U-\bar U) \partial_{A_i}  \partial_{\bar A_i} \ce_2(A_j,U)}{(S-\bar S)(S'-\bar S')(U-\bar U)
}\Big|_{A_i=0} +
\co(e^{3\Phi}) \ , \non
e^K {\rm Re}(f_{{\rm D9}_i})\Big|_{A_i=0}  &=& \frac{1}{2i} \frac{(S-\bar S) + 2i {\rm
Re}(f_{{\rm D9}_i}^{(1)})}{(S-\bar S)(S'-\bar S')(U-\bar U)}\Big|_{A_i=0} +
\co(e^{3\Phi})\ .
\eeqn
This equality, that we expanded
to order $e^{2\Phi}$, is required to hold exactly and thereby fixes the constant $c$.
At order $e^{2\Phi}$ we have
\beqn \label{nicerelation}
2\pi ic\,  (U-\bar U) \partial_{A_i}  \partial_{\bar A_i}
\ce_2(A_j,U)\Big|_{A_i=0} = {\rm Re}(f_{{\rm D9}_i}^{(1)})\Big|_{A_i=0} \ .
\eeqn
Using the relation (\ref{derivE}) one can rewrite the left-hand-side
as a function of $E_1(A,U)$. Thus, (\ref{nicerelation})
becomes
\beqn
4\pi^3c\,  \ce_1^{(i)}(0,U) = {\rm Re}(f_{{\rm D9}_i}^{(1)})\Big|_{A_i=0} \ ,
\eeqn
$\ce_1^{(i)}(A_l,U)$ defined in (\ref{defE1}).
Now we look to our results of \cite{Berg:2004ek,Berg:2004sj} where
we calculated the one-loop corrections to the gauge couplings,
and use them for comparison. One needs to be careful 
about the fact that we
are dealing with the $U(1)_i$ subgroups of the $U(N_i)$ gauge group
factors. Their couplings were computed in section 3.3.2 of \cite{Berg:2004ek}. To compare to
the form of $\ce_1^{(i)}(A_l,U)$ in (\ref{defE1}) it is easiest to
consider the set of equations (36) in \cite{Berg:2004ek} whose sum produces the
one-loop corrections to the gauge kinetic functions up to a factor of $2$ (that can be inferred from formula (13) of \cite{Berg:2004ek}). One only has to replace the theta functions 
by
\beqn \label{KKsums}
\int_0^{\Lambda^2} dl\, \thba{\vec 0}{\vec 0}(\vec a, ie_\sigma lG) e^{-\pi
\chi/(e_\sigma l)} =
\frac{1}{e_\sigma} \left[ e_\sigma \Lambda^2 + \frac{1}{\pi T_2} \tilde E_1
(A,U) + \, ... \right]
\eeqn
in that formula.\footnote{Note that we now use slightly different
conventions compared to \cite{Berg:2004ek} for the modular transformation and the UV
cutoff. The terms proportional to $\Lambda^2$ drop out when summed
over all diagrams.}
With this in mind, one finds that
\beqn
\label{RefD9eq}
{\rm Re}(f_{{\rm D9}_i}^{(1)}) = \frac{1}{32 \pi^3} \ce_1^{(i)}(A_l,U)\ .
\eeqn
This factor can be understood as follows. There is one overall factor $2$ that we mentioned above.
For the annulus $\ca_{95}$ the factor follows immediately from the factor in (36) of 
\cite{Berg:2004ek} and (\ref{KKsums}) (with $e_{(ia)}=1$), whereas for 
$\ca_{99}$ and $\cm_9$, the factor consists of 
\be
2 \frac{e_\sigma^2}{16\pi^2} \frac{1}{4e_\sigma} \frac{1}{e_\sigma\pi}\ ,
\ee
where a factor $-e_\sigma^2/(16\pi^2)$ comes from
(36) of \cite{Berg:2004ek}, the $1/(4e_\sigma)$ from cancelling the factors 
$d_\sigma^2 \cq_{\sigma,k} (= - 4 e_\sigma)$ in (\ref{defE1}) and
the $1/(e_\sigma\pi)$ from (\ref{KKsums}). Plugging 
(\ref{RefD9eq}) 
back into (\ref{comparisoneqs}), we finally find
\beqn \label{tollesc}
c = \frac{1}{128 \pi^6}\ .
\eeqn
An important consistency check of the direct calculation of
\cite{Berg:2004ek} was that the one-loop diagrams provide the
terms in the gauge couplings of D5-branes involving the Wilson
line moduli $A_i$ in the definition of the K\"ahler variable $S'$
compared to $S_0'$. This fixes the relative coefficient of the
tree-level and one-loop gauge couplings uniquely with no
room for other factors.


\section{A 4-point check}
\label{4ptcheck}

In this appendix we would like to check the validity
of relaxation of momentum conservation,
used throughout the main text. We do this by
calculating a specific 4-point function (for
which relaxation is not needed) 
and comparing the result with the one derived
from the 2-point function using the relaxation
of momentum conservation. The simplest
amplitude to consider, that does not require a generalization of
(\ref{abfpttrick}), is the 4-point function of three open string
scalars and one $\bar S'$. More precisely, we will check only the
99 annulus $\ca_{ij}$ with two open string vertices inserted on
the left and one on the right side. 
\begin{figure}[h]
\begin{center}
\psfrag{Ai}[bc][bc][1][0]{$A_i$}
\psfrag{Aj}[bc][bc][1][0]{$A_j$}
\psfrag{Aib}[bc][bc][1][0]{$\bar{A}_i$}
\psfrag{Sp}[bc][bc][1][0]{$S_2'$}
\psfrag{K}[bc][bc][0.6][0]{{\sc K}}
\includegraphics[width=0.4\textwidth]{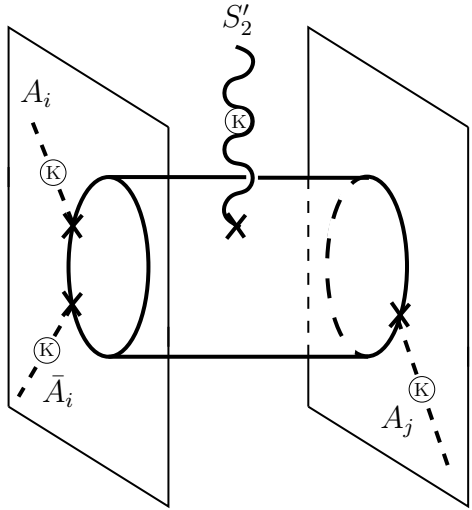}
\caption{The 4-point function $\langle
V_{S_2'}V_{A_j}V_{A_i}V_{\bar{A}_i}\rangle$.
Unlike in fig \ref{fig:disk}, the $A_i$ 
vertex operators are drawn inserted 
on the boundary of the cylinder to show which D-brane
they belong to. Since 
the vertex operators in (\ref{vops}) contain
both open and closed string vertex operators,
it is immaterial whether we draw them with dashed or wiggly lines. }
\label{fig:4pt}
\end{center}
\vspace{-5mm}
\end{figure}
Thus we are concentrating
on\footnote{The notation for the open string vertex operators
deviates slightly from (\ref{nakedvo}) in an obvious way, because
we do not need the general notation here, when we are only
considering the $\ca_{ij}$ diagram.}
\be \label{4pt}
\langle V_{S_2'} V_{A_j}V_{A_i}V_{\bar A_i} \rangle_{\ca_{ij}}  \sim
\frac{i g_o^3 g_c (\alpha')^{-8}}{(S_0' - \bar S_0') (U- \bar U)^{3/2} (T-\bar T)^{3/2}}
\langle V_{Z \bar Z}^{(0,0)} V_Z^{(0) j} V_Z^{(0) i} V_{\bar Z}^{(0) i} \rangle_{\ca_{ij}}\ .
\ee
We will only check the moduli dependence and do not keep track of
the overall factor, except for factors of $\pi$ and $i$. We will
first consider the contribution from the $k=1$ sector and
calculate the term proportional to $p_3 \cdot p_4$. For the
4-point function this does not vanish due to momentum
conservation. We will do the calculation by using the
$\psi$-dependent terms of the last two vertex operators, which
give the desired power in momenta, and then setting the momenta to
zero in the other terms. We will see that this  exactly
reproduces
the moduli dependence that one would get from taking the
derivative of the $A_i \bar A_i$ metric with respect to $A_j$ and
$\bar S'$. Afterwards we will argue that there are no additional
contributions to the term proportional to $p_3 \cdot p_4$, neither
from the $k=0$ nor from the $k=1$ sector.

The aforementioned 
contribution from the $k=1$ sector, i.e.\ the one
derived from taking the $\psi$-dependent terms of the last two
vertex operators and setting the momenta to zero elsewhere, is
given by
\beqn \label{4ptbasic}
\langle V_{Z \bar Z}^{(0,0)} V_Z^{(0) j} V_Z^{(0) i} V_{\bar Z}^{(0) i} \rangle_{\ca_{ij}} & \sim &
(p_3 \cdot p_4) \alpha' \frac{V_4}{(4 \pi^2 \alpha')^2} T_2 \int_0^\infty \frac{dt}{t^4} \int_{\cf_{(ij)}}  d^2 \nu_1 \int_{\partial \ca_{ij}} d \tilde \nu_2 d \tilde \nu_3 d \tilde \nu_4 \non
&& \hspace{-4cm}
\times \sum_{\vec n=(n,m)^T} \Bigg[ {\rm tr} \Big( \lambda_j \lambda_i \lambda_i^\dagger e^{-\pi \vec{n}^{T} G \vec{n} t^{-1}}
e^{2 \pi i \vec{\bf A}_{ij} \cdot \vec{n}} \gamma_{(ij),1}
\sum_{{\alpha\beta}\atop{\rm even}}
\frac{\thba{\alpha}{\beta}(0,\tau)}{\eta^3(\tau)}
 \cz_{\sigma,k}^{\rm int} \zba{\alpha}{\beta} \non
&& \hspace{-.5cm} \times \langle \psi(\nu_3) \psi(\nu_4) \rangle^{\alpha,\beta}
\langle \Psi(\nu_3) \bar \Psi(\nu_4) \rangle^{\alpha,\beta}  \Big) \non
&& \hspace{-3cm}\times
i \alpha'^{-3/2} \Big[ \langle \partial \bar Z (\nu_1) \bar
\partial Z (\bar \nu_1) \dot Z (\nu_2) \rangle + \langle \partial
Z (\nu_1) \bar \partial \bar Z (\bar \nu_1) \dot Z (\nu_2) \rangle
\non
&& \hspace{-1cm}
+ \langle \partial \bar Z (\nu_1) \dot Z (\nu_2) \rangle \langle  \bar \partial Z (\bar \nu_1)\rangle
+ \langle \bar \partial \bar Z (\bar \nu_1) \dot Z (\nu_2) \rangle \langle \partial Z (\nu_1) \rangle
\Big]\Bigg]\ . \non
\eeqn
The bosonic correlators with an odd number of fields in the last
two lines of (\ref{4ptbasic}) only get contributions from the zero
modes (\ref{zeromode}). As in the case of the 2-point functions
discussed before, the sum over spin structures only gives a factor
of $-4$, cf.\ (\ref{spinstructuresums}). Thus the integral over
$\tilde \nu_3$ and $\tilde \nu_4$ is easily performed and gives a
factor of $\pi^2 t^2$. Let us next calculate the bosonic
correlators. They are given by (here we have to keep track of the
exact factors in order to obtain the correct relative factors
between the different contributions of the last and the
penultimate line in (\ref{4ptbasic}))
\beqn
\int_{\cf_{(ij)}}  d^2 \nu_1 \int_{\partial \ca_{ij}} d \tilde \nu_2 \langle \partial \bar Z (\nu_1) \bar \partial Z (\bar \nu_1) \dot Z (\nu_2) \rangle &=& \int_{\cf_{(ij)}}  d^2 \nu_1 \int_{\partial \ca_{ij}} d \tilde \nu_2 \langle \partial Z (\nu_1) \bar \partial \bar Z (\bar \nu_1) \dot Z (\nu_2) \rangle \non
& = & 2 \pi^3 \alpha'^{3/2} \left( \frac{T_2}{2 U_2} \right)^{3/2} \frac{|n+m U|^2 (n+m\bar U)}{t}  \non
\eeqn
and
\beqn
\int_{\cf_{(ij)}}  d^2 \nu_1 \int_{\partial \ca_{ij}} d \tilde \nu_2
\Big[\langle \partial \bar Z (\nu_1) \dot Z (\nu_2) \rangle \langle  \bar \partial Z (\bar \nu_1)\rangle + \langle \bar \partial \bar Z (\bar \nu_1) \dot Z (\nu_2) \rangle \langle \partial Z (\nu_1) \rangle \Big]
&=& \\
&& \hspace{-13cm} = i \alpha'^{1/2} \left( \frac{T_2}{2 U_2} \right)^{1/2} \frac{n+m \bar U}{t} \int_{\cf_{(ij)}}  d^2 \nu_1 \int_{\partial \ca_{ij}} d \tilde \nu_2
 \Big[ \langle \partial \bar Z (\nu_1) \dot Z (\nu_2) \rangle - \langle \bar \partial \bar Z (\bar \nu_1) \dot Z (\nu_2) \rangle \Big] \non
&& \hspace{-13cm} = -\pi^2 \alpha'^{3/2} \left( \frac{T_2}{2 U_2} \right)^{1/2} (n+m \bar U) + 4 \pi^3 \alpha'^{3/2} \left( \frac{T_2}{2 U_2} \right)^{3/2} \frac{|n+m U|^2 (n+m\bar U)}{t} \ ,
\nonumber
\eeqn
where there are two contributions to the 2-point correlators of
$Z$, coming from fluctuations and zero modes, respectively.
Plugging this into (\ref{4ptbasic}) and performing the
$t$-integration in a similar way as in (\ref{intKK}), we finally
end up with\footnote{Note that there is no need for a
regularization in this case because there is no contribution from
a term with $m=n=0$.}
\beqn
\langle V_{Z \bar Z}^{(0,0)} V_Z^{(0) j} V_Z^{(0) i} V_{\bar Z}^{(0) i} \rangle_{\ca_{ij}} &\sim& i \pi^3 \frac{V_4}{(4 \pi^2 \alpha')^2} \sqrt{U_2 T_2} (p_3 \cdot p_4) \alpha'
\\
&& \hspace{2.5cm} \times~{\rm tr} \Big( \lambda_j \lambda_i
\lambda_i^\dagger \gamma_{(ij),1}\sum_{\vec n=(n,m)^T} \!\!\!\!\!\! ' \; \; \; \frac{e^{2 \pi i \vec{\bf A}_{(ij)} \cdot \vec{n}}}{n+m U} \Big)\ . \nonumber
\eeqn
The trace can be evaluated using the matrices of table
\ref{tolletabelle} and formulas (\ref{gammas}) and
(\ref{lambdas}). If we also take into account the factor from the
Weyl rescaling (\ref{Weyl}), we end up with
\beqn \label{4ptresult}
\langle V_{S_2'} V_{A_i}V_{A_i}V_{\bar A_i} \rangle_{\ca_{ij}} &\stackrel{\rm Weyl}{\longrightarrow}&
(p_3 \cdot p_4) \alpha' \frac{V_4}{(4 \pi^2 \alpha')^2} \frac{\pi g_o^3 g_c (\alpha')^{-8}}{(S-\bar S)(S_0' - \bar S_0')^2 (U- \bar U)} N_i N_j \non
&&\times \sum_{\vec n=(n,m)^T} \!\!\!\!\!\! ' \; \; \; \Bigg[\frac{1}{n + m U}
\Big[e^{2 \pi i (\vec{a}_i - \vec{a}_j) \cdot \vec{n}} - e^{2 \pi i (-\vec{a}_i + \vec{a}_j) \cdot \vec{n}} \non
&& \hspace{3cm} +~ e^{2 \pi i (\vec{a}_i + \vec{a}_j) \cdot \vec{n}} - e^{2 \pi i (-\vec{a}_i - \vec{a}_j) \cdot \vec{n}}\Big]\Bigg] \ .
\eeqn
Using
\be
\partial_A \partial_A \partial_{\bar A} E_2(A,U) = \frac{-2\pi^3 i}{U-\bar U} \sum_{\vec n=(n,m)^T} \!\!\!\!\!\! ' \; \; \;
\frac{e^{2 \pi i \vec{a} \cdot \vec{n}}}{n+m U}\ ,
\ee
and substitutions similar to (\ref{substv4}) and (\ref{replace}), i.e.\
\be \label{replace4pt}
i (p_3 \cdot p_4) g_o^3 \pi g_c \alpha'^{-8} ~\rightarrow~ \delta
S_2' \delta A_j \partial_\mu A_i \partial^\mu \bar A_i\ ,
\ee
it is straightforward to see that the moduli dependence of
(\ref{4ptresult}) exactly reproduces the one of the second
derivative of the contribution from $\ca_{ij}$ to the $A_i \bar
A_i$ metric (given in the second line of (\ref{k1loop})) with
respect to $A_j$ and $\bar S'$. The necessary extra factor of $i$
in (\ref{replace4pt}) is as discussed below (\ref{ifactor}).

Next, we would like to argue that there are no further
contributions to the 4-point function coming from either the $k=0$
or $k=1$ sector. Let us begin with the $k=0$ sector. In
order to get a non-vanishing result from the spin structure
summation, one has to contract eight fermionic fields in
(\ref{4pt}). This already leads to a fourth power of momenta,
e.g.\ to a term proportional to $s^2$, where $s \equiv  -(p_1 +
p_2)^2 = -(p_3 + p_4)^2$. In order to get a contribution to the
kinetic term of the $A_i$, the integration over $t$ would have to
give a pole in $s$. This would require the exchange of a massless
particle in the closed string tree channel. However, the massless
particles reside in the sector with $m=n=0$. On the other hand,
contracting eight of the fermionic  world-sheet fields of the
vertex operators in (\ref{4pt}) would leave a bosonic correlator
of the form $\langle
\partial Z \rangle$ or $\langle \bar \partial Z \rangle$, which
only gets contributions from the zero modes (\ref{zeromode}). Thus
the amplitude trivially vanishes for $m=n=0$ and there is no
contribution proportional to $p_3 \cdot p_4$ from the $k=0$
sector.

A similar argument can be put forward for the $k=1$ sector. Again,
an additional contribution could only come from contracting eight
of the fermionic  world-sheet fields of the vertex operators in
(\ref{4pt}) if there were a pole from the $t$-integration. Such a
pole can be excluded in exactly the same way as for the $k=0$
sector. Alternatively, for the $k=1$ sector one could also argue
that the exchanged massless particle would reside in a
hypermultiplet from the twisted sector. Thus the exchange would
require a coupling between hypermultiplets and the vector
multiplets, whose scalars are external states
in the 4-point function (\ref{4pt}). This is, however, forbidden
by supersymmetry. Either way, we conclude that there are no
additional contributions to the 4-point function (\ref{4pt})
proportional to $p_3 \cdot p_4$ and (\ref{4ptresult}) is the final
result.

\end{appendix}



\clearpage
\begin{thebibliography}{99}

\bibitem{Angelantonj:2002ct}
  C.~Angelantonj and A.~Sagnotti,
  {\it Open strings},
  Phys.\ Rept.\  {\bf 371} (2002) 1
  [Erratum-ibid.\  {\bf 376} (2003) 339]
  [hep-th/0204089].

\bibitem{gg2}
  M.~Berg, M.~Haack and B.~K\"ors,
  {\it On volume stabilization by quantum corrections},
  hep-th/0508171.

\bibitem{Kachru:2003aw}
  S.~Kachru, R.~Kallosh, A.~Linde and S.~P.~Trivedi,
  {\it De Sitter vacua in string theory},
  Phys.\ Rev.\ D {\bf 68} (2003) 046005
  [hep-th/0301240].

\bibitem{vonGersdorff:2005bf}
  G.~von Gersdorff and A.~Hebecker,
  {\it K\"ahler corrections for the volume modulus of flux compactifications},
  hep-th/0507131.

\bibitem{Becker:2002nn}
  K.~Becker, M.~Becker, M.~Haack and J.~Louis,
  {\it Supersymmetry breaking and $\alpha'$-corrections to flux induced potentials},
  JHEP {\bf 0206} (2002) 060
  [hep-th/0204254].

\bibitem{Dvali:1998pa}
G.~R.~Dvali and S.~H.~H.~Tye,
{\it Brane inflation},
Phys.\ Lett.\ B {\bf 450} (1999) 72
[hep-ph/9812483].

\bibitem{Dvali:2001fw}
G.~R.~Dvali, Q.~Shafi and S.~Solganik,
{\it D-brane inflation},
hep-th/0105203.

\bibitem{Burgess:2001fx}
C.~P.~Burgess, M.~Majumdar, D.~Nolte, F.~Quevedo, G.~Rajesh and R.~J.~Zhang,
{\it The in\-flation\-ary brane-anti\-brane uni\-verse},
JHEP {\bf 0107} (2001) 047
[hep-th/0105204].

\bibitem{Kachru:2003sx}
  S.~Kachru, R.~Kallosh, A.~Linde, J.~Maldacena, L.~McAllister and S.~P.~Trivedi,
  {\it Towards inflation in string theory},
  JCAP {\bf 0310} (2003) 013
  [hep-th/0308055].

\bibitem{ABFPT}
I.~Antoniadis, C.~Bachas, C.~Fabre, H.~Partouche and T.R.~Taylor,
{\it Aspects of type I - type II - heterotic triality in four dimensions},
Nucl.\ Phys.\ B {\bf 489} (1997) 160
[hep-th/9608012].

\bibitem{Atick:1987gy}
  J.~J.~Atick, L.~J.~Dixon and A.~Sen,
  {\it String calculation of Fayet-Iliopoulos D terms in arbitrary supersymmetric
  compactifications},
  Nucl.\ Phys.\ B {\bf 292} (1987) 109.

\bibitem{Minahan:1987ha}
  J.~A.~Minahan,
  {\it One loop amplitudes on orbifolds and the renormalization of coupling
  constants},
  Nucl.\ Phys.\ B {\bf 298} (1988) 36.

\bibitem{Poppitz:1998dj}
E.~Poppitz,
{\it On the one loop Fayet-Iliopoulos term in chiral four dimensional type I
orbifolds},
Nucl.\ Phys.\ B {\bf 542} (1999) 31
[hep-th/9810010].

\bibitem{Bain:2000fb}
P.~Bain and M.~Berg,
{\it Effective action of matter fields in four-dimensional string
orientifolds},
JHEP {\bf 0004} (2000) 013
[hep-th/0003185].

\bibitem{Antoniadis:2002cs}
I.~Antoniadis, E.~Kiritsis and J.~Rizos,
{\it Anomalous U(1)s in type I superstring vacua},
Nucl.\ Phys.\ B {\bf 637} (2002) 92
[hep-th/0204153].

\bibitem{Antoniadis:2002tr}
I.~Antoniadis, R.~Minasian and P.~Vanhove,
{\it Non-compact Calabi-Yau manifolds and localized gravity},
Nucl.\ Phys.\ B {\bf 648} (2003) 69
[hep-th/0209030].

\bibitem{Witten:1985xb}
  E.~Witten,
  {\it Dimensional reduction of superstring models},
  Phys.\ Lett.\ B {\bf 155} (1985) 151.

\bibitem{Antoniadis:2003sw}
  I.~Antoniadis, R.~Minasian, S.~Theisen and P.~Vanhove,
  {\it String loop corrections to the universal hypermultiplet},
  Class.\ Quant.\ Grav.\  {\bf 20} (2003) 5079
  [hep-th/0307268].

\bibitem{Antoniadis:1997gu}
I.~Antoniadis, H.~Partouche and T.~R.~Taylor,
{\it Duality of N = 2 heterotic-type I compactifications in four dimensions},
Nucl.\ Phys.\ B {\bf 499} (1997) 29
[hep-th/9703076].

\bibitem{Berg:2004ek}
  M.~Berg, M.~Haack and B.~K\"ors,
  {\it Loop corrections to volume moduli and inflation in string theory},
  Phys.\ Rev.\ D {\bf 71} (2005) 026005
  [hep-th/0404087].

\bibitem{Garousi:1996ad}
  M.~R.~Garousi and R.~C.~Myers,
  {\it Superstring scattering from D-branes},
  Nucl.\ Phys.\ B {\bf 475} (1996) 193
  [hep-th/9603194].

\bibitem{Hashimoto:1996bf}
  A.~Hashimoto and I.~R.~Klebanov,
  {\it Scattering of strings from D-branes},
  Nucl.\ Phys.\ Proc.\ Suppl.\  {\bf 55B} (1997) 118
  [hep-th/9611214].

\bibitem{Garousi:1998fg}
  M.~R.~Garousi and R.~C.~Myers,
  {\it World-volume interactions on D-branes},
  Nucl.\ Phys.\ B {\bf 542} (1999) 73
  [hep-th/9809100].

\bibitem{Garousi:2000ea}
  M.~R.~Garousi and R.~C.~Myers,
  {\it World-volume potentials on D-branes},
  JHEP {\bf 0011} (2000) 032
  [hep-th/0010122].

\bibitem{Lust:2004cx}
  D.~L\"ust, P.~Mayr, R.~Richter and S.~Stieberger,
  {\it Scattering of gauge, matter, and moduli fields from intersecting branes},
  Nucl.\ Phys.\ B {\bf 696} (2004) 205
  [hep-th/0404134].

\bibitem{Lust:2004fi}
  D.~L\"ust, S.~Reffert and S.~Stieberger,
  {\it Flux-induced soft supersymmetry breaking in chiral type IIb orientifolds
  with D3/D7-branes},
  Nucl.\ Phys.\ B {\bf 706} (2005) 3
  [hep-th/0406092].

\bibitem{Bianchi:1990yu}
  M.~Bianchi and A.~Sagnotti,
  {\it On the systematics of open string theories},
  Phys.\ Lett.\ B {\bf 247} (1990) 517.

\bibitem{Gimon:1996rq}
  E.~G.~Gimon and J.~Polchinski,
  {\it Consistency conditions for orientifolds and D-manifolds},
  Phys.\ Rev.\ D {\bf 54}, 1667 (1996)
  [hep-th/9601038].

\bibitem{Berkooz:1996iz}
  M.~Berkooz, R.~G.~Leigh, J.~Polchinski, J.~H.~Schwarz, N.~Seiberg and E.~Witten,
  {\it Anomalies, dualities, and topology of D=6 N=1 superstring vacua},
  Nucl.\ Phys.\ B {\bf 475} (1996) 115
  [hep-th/9605184].

\bibitem{Kaplunovsky:1994fg}
V.~Kaplunovsky and J.~Louis,
{\it Field dependent gauge couplings in locally supersymmetric effective quantum
field theories},
Nucl.\ Phys.\ B {\bf 422} (1994) 57
[hep-th/9402005].

\bibitem{deWit:1995zg}
  B.~de Wit, V.~Kaplunovsky, J.~Louis and D.~L\"ust,
  {\it Perturbative couplings of vector multiplets in N=2 heterotic string
  vacua},
  Nucl.\ Phys.\ B {\bf 451} (1995) 53
  [hep-th/9504006].

\bibitem{Polchinski:rr}
J.~Polchinski,
{\it String theory}, 2 vols.,
Cambridge Univ.\ Pr.\ (1998).

\bibitem{Dixon:1990pc}
  L.~J.~Dixon, V.~Kaplunovsky and J.~Louis,
  {\it Moduli dependence of string loop corrections to gauge coupling constants},
  Nucl.\ Phys.\ B {\bf 355} (1991) 649.

\bibitem{Kiritsis:1997hj}
E.~Kiritsis,
{\it Introduction to superstring theory},
hep-th/9709062.

\bibitem{Bachas:1996zt}
  C.~Bachas and C.~Fabre,
  {\it Threshold effects in open-string theory},
  Nucl.\ Phys.\ B {\bf 476} (1996) 418
  [hep-th/9605028].

\bibitem{ABD}
I.~Antoniadis, C.~Bachas and E.~Dudas,
{\it Gauge couplings in four-dimensional type I string orbifolds},
Nucl.\ Phys.\ B {\bf 560} (1999) 93
[hep-th/9906039].

\bibitem{Aldazabal:1998mr}
  G.~Aldazabal, A.~Font, L.~E.~Ibanez and G.~Violero,
  {\it D = 4, N = 1, type IIB orientifolds},
  Nucl.\ Phys.\ B {\bf 536} (1998) 29
  [hep-th/9804026].

\bibitem{Harvey:1995fq}
  J.~A.~Harvey and G.~W.~Moore,
  {\it Algebras, BPS states, and strings},
  Nucl.\ Phys.\ B {\bf 463} (1996) 315
  [hep-th/9510182].

\bibitem{Berkooz:1996dw}
  M.~Berkooz and R.~G.~Leigh,
  {\it A D = 4 N = 1 orbifold of type I strings},
  Nucl.\ Phys.\ B {\bf 483} (1997) 187
  [hep-th/9605049].

\bibitem{Klein:2000qw}
  M.~Klein and R.~Rabadan,
  {\it Z(N) x Z(M) orientifolds with and without discrete torsion},
  JHEP {\bf 0010} (2000) 049
  [hep-th/0008173].

\bibitem{Lust:2005dy}
  D.~L\"ust, S.~Reffert, W.~Schulgin and S.~Stieberger,
  {\it Moduli stabilization in type IIB orientifolds. I: Orbifold limits},
  hep-th/0506090.

\bibitem{Cvetic:2000aq}
  M.~Cvetic and P.~Langacker,
  {\it D = 4 N = 1 type IIB orientifolds with continuous Wilson lines, moving
  branes, and their field theory realization},
  Nucl.\ Phys.\ B {\bf 586} (2000) 287
  [hep-th/0006049].

\bibitem{Cvetic:2000st}
  M.~Cvetic, A.~M.~Uranga and J.~Wang,
  {\it Discrete Wilson lines in N = 1 D = 4 type IIB orientifolds: A  systematic
  exploration for Z(6) orientifold},
  Nucl.\ Phys.\ B {\bf 595} (2001) 63
  [hep-th/0010091].

\bibitem{gg3}
  M.~Berg, M.~Haack and B.~K\"ors,
  work in progress.

\bibitem{Berglund:2005dm}
  P.~Berglund and P.~Mayr,
  {\it Non-perturbative superpotentials in F-theory and string duality},
  hep-th/0504058.

\bibitem{Giddings:2005ff}
  S.~B.~Giddings and A.~Maharana,
  {\it Dynamics of warped compactifications and the shape of the warped
  landscape},
  hep-th/0507158.
  
\bibitem{Antoniadis:1999ge}
  I.~Antoniadis, C.~Bachas and E.~Dudas,
  {\it Gauge couplings in four-dimensional type I string orbifolds},
  Nucl.\ Phys.\ B {\bf 560} (1999) 93
  [hep-th/9906039].

\bibitem{Berg:2000ne}
M.~Berg, C.~DeWitt-Morette, S.~Gwo and E.~Kramer,
  {\it The Pin groups in physics: C, P, and T},
  Rev.\ Math.\ Phys.\  {\bf 13} (2001) 953
  [math-ph/0012006].

\bibitem{Carlip:1988gw}
S.~Carlip and C.~DeWitt-Morette,
  {\it Where The sign of the metric makes a difference},
  Phys.\ Rev.\ Lett.\  {\bf 60} (1988) 1599.

\bibitem{Distler:1992rr}
  J.~Distler,
  {\it A Note on the 3-D Ising model as a string theory},
  Nucl.\ Phys.\ B {\bf 388} (1992) 648
  [hep-th/9205100].

\bibitem{VanNieuwenhuizen:1985be}
P.~Van Nieuwenhuizen,
  {\it An introduction to simple supergravity and the Kaluza-Klein program},

\bibitem{Burgess:1986ah}
  C.~P.~Burgess and T.~R.~Morris,
  {\it Open and unoriented strings a la Polyakov},
  Nucl.\ Phys.\ B {\bf 291} (1987) 256.

\bibitem{Burgess:1986wt}
C.~P.~Burgess and T.~R.~Morris,
{\it Open superstrings a la Polyakov},
Nucl.\ Phys.\ B {\bf 291} (1987) 285.

\bibitem{Epple:2004nh}
  F.~T.~J.~Epple,
  {\it Propagators on one-loop world sheets for orientifolds and intersecting
  branes},
  JHEP {\bf 0501} (2005) 043
  [hep-th/0410177].

\bibitem{Obers:1999um}
N.~A.~Obers and B.~Pioline,
{\it Eisenstein series and string thresholds},
Commun.\ Math.\ Phys.\  {\bf 209} (2000) 275
[hep-th/9903113].

\bibitem{Berg:2004sj}
  M.~Berg, M.~Haack and B.~K\"ors,
  {\it On the moduli dependence of nonperturbative superpotentials in brane
  inflation},
  hep-th/0409282.

\end{thebibliography}
\end{document}